\documentclass[aps,showpacs,preprintnumbers,notitlepage,nofootinbib,superscriptaddress]{revtex4-1}

\usepackage[latin1]{inputenc}
\usepackage{epsfig}
\usepackage{ae}
\usepackage{bm} 
\usepackage{xcolor}
\usepackage{amsmath}
\usepackage{amssymb}
\usepackage{amsfonts}
\usepackage{graphicx}
\usepackage{slashed}
\usepackage{wasysym}
\usepackage{setspace}
\usepackage{multirow}
\usepackage{colortbl}

\newcommand{\Eqref}[1]{Eq.~\eqref{#1}}

\definecolor{lightgray}{gray}{0.8}

\allowdisplaybreaks

\begin{document}

\title{The photon polarization tensor in a homogeneous magnetic or electric field}

\author{Felix Karbstein}
\affiliation{Theoretisch-Physikalisches Institut, Abbe Center of Photonics,
Friedrich-Schiller-Universit\"at Jena, Max-Wien-Platz 1, D-07743 Jena, Germany}
\affiliation{Helmholtz-Institut Jena, Fr\"obelstieg 3, D-07743 Jena, Germany}

\begin{abstract}
 We revisit the photon polarization tensor in a homogeneous external magnetic or electric field.
 The starting point of our considerations is the momentum space representation of the one-loop photon polarization tensor in the presence of a homogeneous electromagnetic field,
 known in terms of a double parameter integral.
 Our focus is on explicit analytical insights for both on- and off-the-light-cone dynamics in a wide range of well-specified physical parameter regimes, ranging from the perturbative to the manifestly nonperturbative strong field regime.
 The basic ideas underlying well-established approximations to the photon polarization tensor are carefully examined and critically reviewed.
 In particular, we systematically keep track of all contributions, both the ones to be neglected and those to be taken into account explicitly, to all orders.
 This allows us to study their ranges of applicability in a much more systematic and rigorous way. 
 We point out the limitations of such approximations and manage to go beyond at several instances.
\end{abstract}

\maketitle

\section{Introduction}

The photon polarization tensor is a central object in quantum electrodynamics (QED).
It contains essential information about the renormalization properties of QED and, accounting for the vacuum fluctuations of the underlying theory, it encodes quantum corrections to Coulomb's force law.
In the presence of an external field, the photon polarization tensor acquires a dependence on the external field, which couples to the quantum fluctuations involving charged particles.
Correspondingly, it gives rise to a variety of dispersive (associated with its real part) and absorptive (associated with its imaginary part) effects affecting photon propagation in electromagnetic fields.
On a more formal level, the finite external field results in a substantially richer tensor structure as compared to the zero field limit,
where the Ward identity immediately constrains the polarization tensor in momentum space to factorize into an overall tensor structure and a single scalar function.

As long as the external field is homogeneous, translational invariance implies that the polarization tensor in momentum space depends only on the transferred four-momentum and the respective field vectors.
In the case of a pure magnetic or electric field, the photon polarization tensor in momentum space can then be decomposed into three independent tensor structures, that can be associated with three distinct polarization modes, and the
corresponding scalar functions.
Thus, the vacuum subject to an external field exhibits medium-like properties.
The difference in the momentum dependence of these modes gives rise to striking observable consequences, such as vacuum birefringence and dichroism \cite{Toll:1952,Baier,BialynickaBirula:1970vy,Adler:1971wn}.

Even for pure and homogeneous fields, the associated scalar functions at one-loop accuracy are highly non-trivial.
They are most conveniently stated in terms of double parameter integrals \cite{BatShab,Urrutia:1977xb,Dittrich:2000zu,Schubert:2000yt} that cannot be tackled analytically in any straightforward way.
One of the integrals is over propertime, and the other one over an additional parameter governing the momentum dependence in the loop.
In case of a pure magnetic field $\vec B$ \cite{Tsai:1974ap}, the entire field dependence of the scalar functions is via trigonometric functions, whose arguments depend multiplicatively on the field amplitude $B=|\vec B|$.
Analogously, for a pure electric field $\vec E$, the dependence is via a factor $E=|\vec E|$ in the arguments of the corresponding hyperbolic functions.

As already discussed on the the level of the Heisenberg-Euler Lagrangian \cite{physics/0605038} (cf., e.g., Ref.~\cite{Jentschura:2001qr}), both situations are related by an electric-magnetic duality.
Whereas the effective action in a pure homogeneous field depends only on the electric charge, the electron mass and the external field\footnote{\label{footnote:1}The effective action is a scalar quantity, and the external field is the only vector in the problem.
Hence, the dependence can be via the field amplitude only.}, the photon polarization tensor in addition features an explicit dependence on the transferred four-momentum.
Thus, besides a mapping of the field, the corresponding duality for the photon polarization tensor involves a transformation of the
components parallel and perpendicular to the external field also \cite{Urrutia:1977xb}.

While the generic analytic properties of the photon polarization tensor in a magnetic field and its different representations: propertime, dispersion-sum, and Landau or spectral sum representation, respectively,
have been studied in great detail by \cite{Shabad:1975ik,Melrose:1976dr,Melrose:1977,Witte:1990}, and very recently again by \cite{Hattori:2012je,Hattori:2012ny},
handy analytical expressions and controlled approximations that hold within certain, well-constrained parameter regimes are still very rare -- even more so beyond on-the-light-cone dynamics. For a recent numerical study, cf. \cite{Ishikawa:2013fxa}.
Results for the photon polarization tensor in other external field configurations like constant crossed fields and plane wave backgrounds are also available \cite{Nikishov:1964zza,Nikishov:1964zz,narozhnyi:1968,ritus:1972,constcrossedfields:1976} 
(for the crossed field case, cf. also \cite{Heinzl:2006pn}).
For recent reviews about strong field QED in the context of high intensity Laser experiments, see \cite{Marklund:2008gj,Dunne:2008kc,DiPiazza:2011tq}.

Triggered by the seminal works of Tsai and Erber \cite{Tsai:1974fa,Tsai:1975iz} in the 1970s,
ongoing efforts have sought to find adequate approximations for the photon polarization tensor in the presence of an external magnetic or electric field (e.g., \cite{Baier:2007dw,Baier:2009it}) in various limits.
However, their derivation in general involves constraints to a certain momentum regime and most of these approximations are tailored to on-the-light-cone dynamics.
While there are some motivations and indications concerning their regimes of applicability, so far more systematic studies of their regimes of validity -- particularly beyond on-the-light-cone dynamics -- have not been performed.
In this paper we aim at going beyond. Our focus is threefold: to thoroughly investigate the regimes of validity of established approximations, to generalize them beyond on-the-light-cone dynamics, and to obtain new analytical results, particularly into the nonperturbative regime.
Correspondingly, our paper can be considered as constituting a viable toolbox, providing approximations to the photon polarization tensor in various well-specified physical parameter regimes.
For clarity and easy reference we summarize the physical parameter regimes studied in this work in Table~\ref{tab:parameterregs}.

\begin{table}
 \center
\begin{spacing}{1.2}
\begin{tabular}{|c||c|c|}
 \hline
 & \multirow{2}{*}{$\bigl(\frac{ef}{m^2}\bigr)^2\frac{k_\perp^2}{4m^2}\ll1$} & \multirow{2}{*}{$\bigl(\frac{ef}{m^2}\bigr)^2\frac{k_\perp^2}{4m^2}\gg1$} \\
 & & \\
 \hline\hline
\multirow{2}{*}{$\frac{ef}{m^2}\ll1$} & {\it perturbative regime}: & {\it weak fields - large momentum}: \\
 & Sec.~\ref{sec:pertweak} & Tsai and Erber \cite{Tsai:1974fa,Tsai:1975iz}; Sec.~\ref{subsec:xito0} \\
\hline
\multirow{2}{*}{$\frac{k_\perp^2}{4m^2}\gg1$} & {\it very weak fields - large momentum}: & \cellcolor{lightgray} \\
 & this work; Sec.~\ref{subsec:xitoinfty}  & \cellcolor{lightgray} \\
\hline
\multirow{2}{*}{$\frac{ef}{k_\perp^2}\ll1$} & \cellcolor{lightgray} & {\it momentum dominance}: \\
 & \cellcolor{lightgray} & this work; Sec.~\ref{subsec:xito0}  \\
\hline\hline
\multirow{2}{*}{$\bigl\{\frac{ef}{m^2},\frac{ef}{k_\perp^2}\bigr\}\gg1$} & \multicolumn{2}{c|}{{\it strong field limit}:} \\
 & \multicolumn{2}{c|}{ this work; Sec.~\ref{sec:strongfield}} \\
\hline
\end{tabular}
\end{spacing}
\caption{In this table we list the physical parameter regimes studied in this work for on-the-light-cone dynamics, $k^2=0$. All regimes are studied for both magnetic, $f=B$, and electric, $f=E$, fields.
We introduce a descriptive label for each regime (written in italics), and reference the corresponding section in this work.
Moreover, we indicate where this work adds major contributions. Obviously, for $k^2=0$ each regime can be characterized by just two different inequalities.
This is no longer the case for $k^2\neq0$, where additional constrains are needed.
We will nevertheless use the same labels for the analogous regimes generalized to $k^2\neq0$ also.
Let us emphasize that the two inequalities chosen to characterize a particular regime are not unique.
Correspondingly, the gray shaded cells can partly overlap with other regimes, 
e.g., the strong field limit overlaps with the regime characterizing the grayish cell in the right column.}
\label{tab:parameterregs}
\end{table}

Our paper is organized as follows. In Sec.~\ref{sec:thepolten} we recall the basic structure of the photon polarization tensor, subject to a pure and homogeneous external field, in momentum space.
In this context, we discuss the corresponding electric-magnetic duality, mapping the polarization tensor in a purely magnetic field onto the corresponding expression in an electric field.
Section~\ref{sec:analyticalins} focuses on several approximations, allowing for explicit analytical insights.
After a short discussion of the perturbative weak field regime, we retrace the approximations of Tsai and Erber, but without restricting ourselves to on-the-light-cone dynamics from the outset.
Thereafter, we study in detail the strong field limit. Again our focus is on handy analytical expressions, applicable in well-specified physical parameter regimes. The paper ends with conclusions in Sec.~\ref{sec:conclusions}.
Extensive appendices provides additional details that have been omitted in the main text.

\section{The photon polarization tensor in a pure and homogeneous field}\label{sec:thepolten}

We focus on the photon polarization tensor at one-loop level, and stick to its representation in the proper-time formalism \cite{Schwinger:1951nm}.
Whereas it is known exactly for arbitrary homogeneous, externally set electromagnetic field configurations in terms of a double parameter integral \cite{Dittrich:2000zu,BatShab,Urrutia:1977xb,Schubert:2000yt},
we here limit ourselves to the special case of a pure, i.e., either magnetic or electric, and homogeneous field.
By this choice we explicitly restrict ourselves to a certain class of reference systems and break Lorentz covariance:
As $\vec{E}$ and $\vec{B}$ are of course not invariant under general Lorentz transformations, only then our notion of discerning $\vec{E}$ and $\vec{B}$ as {\it pure} fields makes sense.
A residual Lorentz covariance remains for boosts along, and rotations around the external field.

Correspondingly, the only two externally set vectors in the problem are the external field and the vector formed by the spatial components of the transferred momentum four-vector.
They govern the entire direction dependence of the photon polarization tensor. Of course, in inhomogeneous fields, the tensor structure can become much more involved.

It is then convenient to decompose the four-vectors $k^{\mu}$ into components parallel and perpendicular to the direction of the external field $\vec f=\{\vec{B},\vec{E}\}$.
Without loss of generality, $\vec f$ is assumed to point in ${\vec e}_1$ direction, and the following decomposition is adopted,
\begin{align}
 k^{\mu}=k_{\parallel}^{\mu}+k_{\perp}^{\mu}\,,\quad\quad k_{\parallel}^{\mu}=(k^0,k^1,0,0)\,,\quad\quad k_{\perp}^{\mu}=(0,0,k^2,k^3)\,.
\end{align}
Our metric convention is $g_{\mu\nu}={\rm diag}(-1,+1,+1,+1)$, such that the four-vector squared reads $k^2={\vec k}^2-(k^0)^2$.
The metric tensor is decomposed as follows, $g^{\mu\nu}=g_{\parallel}^{\mu\nu}+g_{\perp}^{\mu\nu}$, with $g_\parallel^{\mu\nu}={\rm diag}(-1,+1,0,0)$ and $g_\perp^{\mu\nu}={\rm diag}(0,0,+1,+1)$.
In momentum space, the one-loop photon polarization tensor in a pure and homogeneous field can then be written as \cite{BatShab,Shabad:1975ik,Tsai:1974ap,Melrose:1976dr,Urrutia:1977xb},
\begin{multline}
 \Pi^{\mu\nu}(k|\vec{f})=\frac{\alpha}{2\pi}\int_{-1}^{1}\frac{{\rm d}\nu}{2}\int\limits_{0}^{\infty-{\rm i}\eta}\frac{{\rm d} s}{s}\,\biggl\{{\rm e}^{-{\rm i}\Phi_0s}\,\Bigl[N_0\left(g^{\mu\nu}k^2-k^{\mu}k^{\nu}\right)
+(N_1-N_0)\left(g_{\parallel}^{\mu\nu}k_{\parallel}^2-k_{\parallel}^{\mu}k_{\parallel}^{\nu}\right) \\
+(N_2-N_0)\left(g_{\perp}^{\mu\nu}k_{\perp}^2-k_{\perp}^{\mu}k_{\perp}^{\nu}\right)\Bigr]+{\rm c.t.}\biggr\}\,, \label{eq:PIa}
\end{multline}
with contact term
\begin{equation}
 {\rm c.t.}=-(1-\nu^2){\rm e}^{-{\rm i}(m^2-{\rm i}\epsilon)s}\left(g^{\mu\nu}k^2-k^{\mu}k^{\nu}\right),
\end{equation}
where the parameter $s$ denotes the propertime and $\nu$ governs the momentum distribution within the loop,
$\{\epsilon,\eta\}\to0^+$ are infinitesimal parameters, $m$ is the electron mass, $e>0$ the elementary charge, and $\alpha=e^2/4\pi$ is the fine structure constant.
We use units where $c=\hbar=1$.
Whereas $\epsilon$ can be traced back to the {\it Feynman prescription} $m^2\to m^2-{\rm i}\epsilon$ in the propagator,
the parameter $\eta$ is necessary to unambiguously define the propertime integral for a purely magnetic field. It shifts the integration contour slightly below the positive real $s$ axis\footnote{For a pure magnetic field, poles are located on the real $s$ axis at $z=2\pi n$ ($n\in\mathbb{N}$); cf. \Eqref{eq:scalarfcts_B}, below. \label{footnote:2}}.

The {\it phase factor} $\Phi_0$ is given by
\begin{equation}
 \Phi_0=m^2-{\rm i}\epsilon+n_1{k}_{\parallel}^2+n_2{k}_{\perp}^2\,, \label{eq:Phi0}
\end{equation}
and the dependence on the external field is entirely encoded in the {\it scalar functions} $N_0$, $N_{1}$, $N_{2}$, $n_{1}$ and $n_{2}$.
The nonvanishing elementary scalars which involve the external field $\vec{f}$ and remain invariant under boosts along and rotations around $\vec{f}$, constituting the residual Lorentz symmetry of the problem (see above), read
\begin{align}
  F^{\mu\nu}F_{\mu\nu}&=2B^2\,, \nonumber\\
 (k_{\perp}^{\mu}F_{\mu\rho})(k_{\perp}^{\nu}F_{\nu}^{\ \rho})&=B^2k_{\perp}^2\,, \nonumber\\
 -(k_{\parallel}^{\mu}F_{\mu\rho}^*)(k_{\parallel}^{\nu}F_{\nu}^{*\,\rho})&=B^2k_{\parallel}^2\,, \label{eq:LscalarsB}
\end{align}
for a purely magnetic field ($\vec f=\vec B$), and
\begin{align}
 F^{\mu\nu}F_{\mu\nu}&=-2E^2\,, \nonumber\\
 (k_{\parallel}^{\mu}F_{\mu\rho})(k_{\parallel}^{\nu}F_{\nu}^{\ \rho})&=-E^2k_{\parallel}^2\,, \nonumber\\
 -(k_{\perp}^{\mu}F_{\mu\rho}^*)(k_{\perp}^{\nu}F_{\nu}^{*\,\rho})&=-E^2k_{\perp}^2\,, \label{eq:LscalarsE}
\end{align}
in case of an electric field ($\vec f=\vec E$). Here $F_{\mu\nu}^*=\frac{1}{2}\epsilon_{\mu\nu\alpha\beta}F^{\alpha\beta}$ denotes the dual field strength tensor, and $\epsilon_{\mu\nu\alpha\beta}$ is the totally antisymmetric tensor.
In result, the above scalar functions depend on $\vec{f}$ only via its amplitude $f=|\vec{f}|$.
For general constant electromagnetic fields, the independent Lorentz scalars in the problem are $k^2$, $F^{\mu\nu}F_{\mu\nu}$, $F^{\mu\nu}F^*_{\mu\nu}$, and $k^\mu F_{\mu\nu}F^{\nu\rho}k_\rho$ \cite{Shabad:1975ik}.

Specializing to a magnetic field $\vec{f}=\vec{B}$, the explicit expressions for $N_0$, $N_1$, $N_2$, $n_1$ and $n_2$ read
\begin{align}
 N_0(z)&=\frac{z}{\sin z}\left(\cos\nu z-\nu\sin \nu z \cot z\right),& n_{1}(z)&=\frac{1-\nu^2}{4}\,, \nonumber\\
 N_{1}(z)&=z(1-\nu^2)\cot z\,,& n_{2}(z)&=\frac{\cos{\nu z}-\cos{z}}{2z\sin{z}}\,, \nonumber\\
 N_{2}(z)&=\frac{2z\left(\cos \nu z -\cos z\right)}{\sin^3z}\,, \label{eq:scalarfcts_B}
\end{align}
with $z=eBs$.
They exclusively depend on $z$ and $\nu$, and are even in both variables.
The entire $s$ dependence of \Eqref{eq:Phi0} is via $n_2$.

The corresponding expressions for an electric field $\vec{f}=\vec{E}$ can be obtained from those in \Eqref{eq:scalarfcts_B}, substituting $B\,\to\,\pm{\rm i}E$, and at the same time interchanging the labels $1\ \leftrightarrow\ 2$ \cite{Urrutia:1977xb}.
As Eqs.~\eqref{eq:scalarfcts_B} are even in $z$, both signs $\pm{\rm i}E$ are possible and the mapping is not unique.
Identifying the scalar functions in \Eqref{eq:PIa} with the explicit expressions in \Eqref{eq:scalarfcts_B}, on the level of \Eqref{eq:PIa} this {\it formal correspondence} can be rephrased as
\begin{equation}
 B\ \to\ \pm{\rm i}E \quad\quad{\rm and}\quad\quad \parallel\ \leftrightarrow\ \perp\,. \label{eq:formcorresp}
\end{equation}
Equation~\eqref{eq:formcorresp} is also compatible with Eqs.~\eqref{eq:LscalarsB} and \eqref{eq:LscalarsE}.
However, the correspondence~\eqref{eq:formcorresp} does not survive the propertime integration, and thus is not true for the photon polarization tensor on a general level:
Reverting to \Eqref{eq:PIa} and taking into account the location of the poles in the complex $z$ plane (cf. footnote~\ref{footnote:2}), 
an analytical continuation in the field variable $B\to B{\rm e}^{-{\rm i}\delta}$ is viable for $0\leq\delta\leq\frac{\pi}{2}$ but not for $\delta<0$.
As a consequence, only the mapping $B\to B{\rm e}^{-{\rm i}\pi/2}\hat{=}-{\rm i}E$ survives the propertime integration and is valid for the photon polarization tensor on a general level.
This results in the following electric-magnetic duality,
\begin{equation}
\begin{array}{ccc}
 \Pi^{\mu\nu}(k|\vec{B}) &  & \Pi^{\mu\nu}(k|\vec{E})\\
\hline
B & \leftrightarrow & -{\rm i}E\\
\parallel & \leftrightarrow & \perp \\
\end{array}\ . \label{eq:trafo}
\end{equation}
For completeness, we also give the reverse line of argument: Demanding the electric-magnetic duality~\eqref{eq:trafo} to hold, the integration contour of the propertime integral is fixed
to lie slightly below the positive real $s$ axis.
The physical reason for this is that the photon amplitude in an external field can be depleted but not amplified.
Moreover, note that for ${f}={B}$ and ${k}_{\perp}^2=0$ (${f}={E}$ and ${k}_{\parallel}^2=0$)
the phase factor $\Phi_0$ becomes independent of $s$, and the propertime integral simplifies significantly.

Employing projection operators, we finally write the photon polarization tensor, Eq.~\eqref{eq:PIa}, in a slightly different way.
The projection operator onto transversal modes is given by
\begin{equation}
 P^{\mu\nu}_{\rm T}=g^{\mu\nu}-\frac{k^{\mu}k^{\nu}}{k^2}\,.
\end{equation}
At zero field the transferred momentum $k^{\mu}$ is the only externally set four-vector. The Ward identity, $k_{\mu} \Pi^{\mu \nu} =0$, then implies that the photon polarization tensor at zero field is of the form
\begin{equation}
\Pi^{\mu\nu}(k)=P_{\rm T}^{\mu\nu}(k)\,\Pi^{(0)}(k)\,, \label{eq:Pi0}
\end{equation}
where $\Pi^{(0)}(k)$ is a scalar quantity.
In the presence of an external field it is helpful to introduce
\begin{align}
 P^{\mu\nu}_{\parallel}=g_{\parallel}^{\mu\nu}-\frac{k_{\parallel}^{\mu}k_{\parallel}^{\nu}}{k_{\parallel}^2} \quad\quad{\rm and}\quad\quad
 P^{\mu\nu}_{\perp}=g_{\perp}^{\mu\nu}-\frac{k_{\perp}^{\mu}k_{\perp}^{\nu}}{k_{\perp}^2}\,. 
\label{eq:Projs}
\end{align}
Together with
\begin{equation}
 P^{\mu\nu}_{0}\equiv P^{\mu\nu}_{\rm T}-\left(P^{\mu\nu}_{\parallel}+P^{\mu\nu}_{\perp}\right), \label{eq:Proj3}
\end{equation}
$P^{\mu\nu}_{\parallel}$ and $P^{\mu\nu}_{\perp}$ form a set of projection operators that span the transversal subspace.
For a given photon four-momentum $k^{\mu}$, the projectors $P^{\mu\nu}_{p}$
($p=0,\parallel,\perp$) project onto the three independent photon polarization modes in
the presence of the external field.
We denote the angle between the external field and the propagation direction $\vec k$ by $\theta=\varangle(\vec{f},\vec{k})$. 
As the vacuum speed of light in external
fields deviates from its value at zero field, and the {\it polarized vacuum} exhibits medium-like
properties, the occurrence of three (instead of two at zero field) independent polarization modes is not surprising.
As long as $\vec{k}\nparallel\vec{f}$, the projectors $P^{\mu\nu}_{\parallel}$ and
$P^{\mu\nu}_{\perp}$ have an intuitive interpretation. They project onto photon modes with polarization vector parallel and perpendicular to the $(\vec{k},\vec{f})$ plane,
and can be continuously related to polarization modes at zero field.
For the special alignment of $\vec{k}\parallel\vec{f}$ only
one externally set direction is left, and we encounter rotational invariance
around the field axis. Here the modes $0$ and $\perp$ can be continuously related to polarization modes at zero field.

With the help of Eqs.~\eqref{eq:Projs} and \eqref{eq:Proj3}, \Eqref{eq:PIa} can be rewritten as
\begin{equation}
 \Pi^{\mu\nu}(k|\vec{f})=P^{\mu\nu}_{0}\,\Pi_{0}+P^{\mu\nu}_{\parallel}\,\Pi_{\parallel}+P^{\mu\nu}_{\perp}\,\Pi_{\perp}\,, \label{eq:PI_tens}
\end{equation}
where the scalar functions $\Pi_p(k)$, $p\in\{\parallel,\perp,0\}$, are the components of the photon polarization tensor in the respective subspaces, given by
\begin{equation}
\left\{
 \begin{array}{c}
 \Pi_{\parallel}\\
 \Pi_{\perp}\\
 \Pi_{0}
 \end{array}
\right\}
=\frac{\alpha}{2\pi}\int_{-1}^{1}\frac{{\rm d}\nu}{2}\int\limits_{0}^{\infty-{\rm i}\eta}\frac{{\rm d} s}{s}\,\left[{\rm e}^{-{\rm i}\Phi_0s}
\left\{
 \begin{array}{c}
 k_{\parallel}^2 N_{1} + k_{\perp}^2N_0 \\
 k_{\parallel}^2N_0 + k_{\perp}^2 N_{2} \\
 k^2N_0 
 \end{array}
\right\}
 +k^2{\rm c.t.}
\right], \label{eq:PI_comp}
\end{equation}
and the contact term now reads
\begin{equation}
 {\rm c.t.}=-{\rm e}^{-{\rm i}\left(m^2-{\rm i}\epsilon\right)s}(1-\nu^2)\,. \label{eq:ct}
\end{equation}
Here we have pulled out an overall factor $k^2$, such that the contact term, as defined in \Eqref{eq:ct}, does not feature a momentum dependence.
Throughout the calculations performed in this paper, the $\nu$ integration will be reserved to the very end. Thus it is useful to also introduce the abbreviation
\begin{equation}
 \Pi_p(k)=\int_{-1}^{1}\frac{{\rm d}\nu}{2}\,\pi_p(k,\nu)\,. \label{eq:Pipi}
\end{equation}
However, the expressions $\pi_p(k,\nu)$ are not unambiguously defined. They might differ by terms that vanish or can be rearranged by integrations by parts under the $\nu$ integral.

As discussed above, the identification of $N_0$, $N_1$, $N_2$, $n_{1}$ and $n_{2}$ with the explicit functions in \Eqref{eq:scalarfcts_B} results in the mapping~\eqref{eq:trafo}.
For completeness, note that in the limit of vanishing external field, ${z}\to0$,
\begin{equation}
 \{N_0,N_{1},N_{2}\} \ \to\ 1-\nu^2 +{\cal O}({z^{2}}) \quad\quad{\rm and}\quad\quad \{n_{1},n_{2}\} \ \to\  \frac{1-\nu^2}{4} +{\cal O}({z^{2}})\,. \label{eq:scalarfcts_z=0}
\end{equation}
In this particular limit all components $\Pi_p$ in \Eqref{eq:PI_comp} become equal, and
as $P^{\mu\nu}_{\rm T}=\sum_{p} P^{\mu\nu}_{p}$ [cf. \Eqref{eq:Proj3}], the overall tensor structure of $\Pi^{\mu\nu}(k)$ is $\sim P^{\mu\nu}_{\rm T}$.
Thus, at zero field Eqs.~\eqref{eq:PI_tens} and \eqref{eq:PI_comp} reproduce \Eqref{eq:Pi0} with
\begin{equation}
 \Pi^{(0)}(k)
=k^2\frac{\alpha}{2\pi}\int_{-1}^{1}\frac{{\rm d}\nu}{2}\,(1-\nu^2)\int\limits_{0}^{\infty-{\rm i}\eta}\frac{{\rm d} s}{s}\,{\rm e}^{-{\rm i}\left(m^2-{\rm i}\epsilon\right)s}\left({\rm e}^{-{\rm i}\frac{1-\nu^2}{4}{k}^2s}
 -1\right). \label{eq:PI_zerofield}
\end{equation}
Utilizing partial integration in $\nu$ [\Eqref{eq:partint} with $a=0$],
the propertime integral can be performed straightforwardly, and \Eqref{eq:PI_zerofield} can be cast into a convenient representation of the one-loop photon polarization tensor at zero field \cite{Schwinger:1951nm,Dittrich:1985yb},
\begin{align}
 \Pi^{(0)}(k)=(k^2)^2\frac{\alpha}{4\pi}\int_{-1}^{1}\frac{{\rm d}\nu}{2}\,\nu^2\left(\frac{\nu^2}{3}-1\right)\,{\rm i}\int\limits_{0}^{\infty-{\rm i}\eta}{{\rm d} s}\,{\rm e}^{-{\rm i}\phi_0s}=(k^2)^2\frac{\alpha}{4\pi}\int_{-1}^{1}\frac{{\rm d}\nu}{2}\,\nu^2\left(\frac{\nu^2}{3}-1\right)\,\frac{1}{\phi_0}\,, \label{eq:vac0}
\end{align}
where we introduced the zero field analog of the phase factor $\Phi_0$ [cf. \Eqref{eq:Phi0}],
\begin{equation}
 \phi_0=m^2-{\rm i}\epsilon+\frac{1-\nu^2}{4}{k}^2\,. \label{eq:phi0frei}
\end{equation}
This quantity plays an important role in the expansions to be performed in Sec.~\ref{sec:analyticalins}. Apart from the physical parameters $m^2$ and $k^2$, it depends on the integration parameter $\nu$.

Finally, as already noted below \Eqref{eq:trafo}, for vanishing $k_\perp^2$ ($k_\parallel^2$) the expression of the photon polarization tensor in  a magnetic (electric) field simplifies significantly: In this limit the phase factor~\eqref{eq:Phi0} loses any nontrivial
propertime dependence and becomes independent of the external field. In turn, the propertime integration can be performed explicitly. To keep notations simple, we define
\begin{equation}
 \phi_0^\parallel=m^2-{\rm i}\epsilon+\frac{1-\nu^2}{4}k_\parallel^2 \quad\quad \text{and} \quad\quad \phi_0^\perp=m^2-{\rm i}\epsilon+\frac{1-\nu^2}{4}k_\perp^2\,. \label{eq:phi_parperp}
\end{equation}
These expressions resemble \Eqref{eq:phi0frei} with $k^2\to k_\parallel^2$ and $k^2\to k_\perp^2$, respectively.
While the first one is relevant for magnetic fields, the latter one can be associated with electric fields. 
Employing \Eqref{eq:grad1} below, it is straightforward to derive the following identity,
\begin{align}
 \int\limits_{0}^{\infty-{\rm i}\eta}\frac{{\rm d} s}{s}\,\Bigl\{N_i\,{\rm e}^{-{\rm i}\phi_0^\parallel s}+{\rm c.t.}\Bigr\}=(1-\nu^2)\ln\biggl(\frac{m^2-{\rm i}\epsilon}{\phi_0^\parallel}\biggr)+
\int\limits_{0}^{\infty-{\rm i}\eta}\frac{{\rm d} s}{s}\,{\rm e}^{-{\rm i}\phi_0^\parallel s}\Bigl\{N_i\,-(1-\nu^2)\Bigr\}. \label{eq:id}
\end{align}
With its help, we write
\begin{equation}
\left.\left\{
 \begin{array}{c}
 \Pi_{\parallel}\\
 \Pi_{\perp}\\
 \Pi_{0}
 \end{array}
\right\}\right|_{k_\perp^2=0}
=k_{\parallel}^2\,\frac{\alpha}{2\pi}\int_{-1}^{1}\frac{{\rm d}\nu}{2}
\left\{
 \begin{array}{c}
 \eta_1^\parallel(B) \\
 \eta_0^\parallel(B) \\
 \eta_0^\parallel(B)
 \end{array}
\right\}, \label{eq:Pieta}
\end{equation}
where we have introduced
\begin{equation}
 \eta_i^\parallel(B)=(1-\nu^2)\ln\biggl(\frac{m^2-{\rm i}\epsilon}{\phi_0^\parallel}\biggr)+\int\limits_{0}^{\infty-{\rm i}\eta}\frac{{\rm d} s}{s}\,{\rm e}^{-{\rm i}\phi_0^\parallel s}\left[N_i(eBs)-(1-\nu^2)\right], \label{eq:soumgeschr}
\end{equation}
with $i\in\{0,1,2\}$.
The function $\eta_2$ does not contribute at all in the considered limit. As we will need it later, we nevertheless include it here for later reference.
The results for an electric field and $k_\parallel^2=0$ follow from Eqs.~\eqref{eq:id}-\eqref{eq:soumgeschr} by the replacement $\parallel\to\perp$ and $B\to -{\rm i}E$.
Obviously, the polarization tensor becomes degenerate for two polarization modes [cf. the discussion of the projection operators below \Eqref{eq:Proj3}].

Employing integrations by parts, the propertime integrals in \Eqref{eq:soumgeschr} can be reduced to a few basic integrals [cf. also the alternative representation of the scalar functions $N_i(z)$ given in Appendix~\ref{app:seriesNn}, \Eqref{eq:scalarfcts_B_anders}],
\begin{align}
 &\int\limits_{0}^{\infty-{\rm i}\eta}\frac{{\rm d} s}{s}\,{\rm e}^{-{\rm i}\phi_0^\parallel s}\left\{N_0(z)-(1-\nu^2)\right\}
=-\nu^2+
eB\int\limits_{0}^{\infty-{\rm i}\eta}{\rm d} s\,{\rm e}^{-{\rm i}\phi_0^\parallel s}\left[(1-\nu^2)\left(\frac{\cos(\nu z)}{\sinh(z)}-\frac{1}{z}\right)+\nu\frac{{\rm i}\phi_0^\parallel}{eB}\,\frac{\sin(\nu z)}{\sin(z)}\right],\label{eq:intN0} \\
 &\int\limits_{0}^{\infty-{\rm i}\eta}\frac{{\rm d} s}{s}\,{\rm e}^{-{\rm i}\phi_0^\parallel s}\left\{N_1(z)-(1-\nu^2)\right\}
=(1-\nu^2)\,eB\int\limits_{0}^{\infty-{\rm i}\eta}{\rm d} s\,{\rm e}^{-{\rm i}\phi_0^\parallel s}\left[\cot(z)-\frac{1}{z}\right], \label{eq:intN1}\\
&\int\limits_{0}^{\infty-{\rm i}\eta}\frac{{\rm d} s}{s}\,{\rm e}^{-{\rm i}\phi_0^\parallel s}\left\{N_2(z)-(1-\nu^2)\right\}
=-\frac{1+3\nu^2}{2}
+eB\int\limits_{0}^{\infty-{\rm i}\eta}{\rm d} s\,{\rm e}^{-{\rm i}\phi_0^\parallel s} \nonumber\\
&\hspace*{4.2cm}\times\Biggl[(1-\nu^2)\left(\frac{\cos(\nu z)}{\sin(z)}-\frac{1}{z}\right)
+\biggl(\frac{\phi_0^\parallel}{eB}\biggr)^2\!\left(\cot(z)-\frac{\cos(\nu z)}{\sin(z)}\right)+2\nu\frac{{\rm i}\phi_0^\parallel}{eB}\,\frac{\sin(\nu z)}{\sin(z)}\Biggr],\label{eq:intN2}
\end{align}
with $z=eBs$.
Analogous expressions hold for the electric field case; they are obtained from Eqs.~\eqref{eq:intN0}-\eqref{eq:intN2} by the replacement $\perp\to\parallel$, $B\to-{\rm i}E$ and $z\to -{\rm i}z'=-{\rm i}eEs$.

In order to perform the propertime integrations in Eqs.~\eqref{eq:intN0}-\eqref{eq:intN2} we analytically continue the magnetic field to negative imaginary values or -- equivalently -- employ the electric-magnetic duality~\eqref{eq:trafo}. 
Correspondingly, they can be carried out explicitly by resorting to the following identities, obtained from formulae 3.551.2, 3.551.3 and 3.552.1 of \cite{Gradshteyn},
\begin{align}
 &\int_0^{\infty}{\rm d}z'\,z'^{\varepsilon}\,{\rm e}^{-{\rm i}\beta z'}\frac{1}{z'}=\varepsilon^{-1}-\ln({\rm i}\beta)-\gamma+{\cal O}(\varepsilon)\,, \label{eq:grad1} \\ 
 &\int_0^{\infty}{\rm d}z'\,z'^{\varepsilon}\,{\rm e}^{-{\rm i}\beta z'}\coth(z')=\varepsilon^{-1}-\psi\bigl(\tfrac{{\rm i}\beta}{2}\bigr)-({\rm i}\beta)^{-1}-\gamma-\ln(2)+{\cal O}(\varepsilon)\,, \label{eq:grad2} \\
 &\int_0^{\infty}{\rm d}z'\,z'^{\varepsilon}\,{\rm e}^{-{\rm i}\beta z'}\,\frac{\cosh(\nu z')}{\sinh(z')}=\varepsilon^{-1}-\frac{1}{2}\left[\psi\biggl(\frac{1+{\rm i}\beta+\nu}{2}\biggr)+\psi\biggl(\frac{1+{\rm i}\beta-\nu}{2}\biggr)\right]-\gamma-\ln(2)+{\cal O}(\varepsilon)\,, \label{eq:grad3a} \\
 &\int_0^{\infty}{\rm d}z'\,{\rm e}^{-{\rm i}\beta z'}\,\frac{\sinh(\nu z')}{\sinh(z')}=\frac{1}{2}\left[\psi\biggl(\frac{1+{\rm i}\beta+\nu}{2}\biggr)-\psi\biggl(\frac{1+{\rm i}\beta-\nu}{2}\biggr)\right]\,, \label{eq:grad3b}
\end{align}
all valid for Im($\beta$)$<0$ and $\varepsilon\to0^+$. Here $\Psi(\chi)=\frac{\rm d}{{\rm d}\chi}\ln{\Gamma(\chi)}$ denotes the Digamma function.
The integrations over propertime~\eqref{eq:intN0}-\eqref{eq:intN2} yield finite results:
Divergent contributions are grouped such that the divergences encountered in isolated terms [cf. Eqs.~\eqref{eq:grad1}-\eqref{eq:grad3a}] cancel.
Performing the $\nu$ integration, taking into account
\begin{equation}
 \int_{-1}^1{\rm d}\nu\,{\rm e}^{-{\rm i}\phi_0^\perp s}\, f(\nu)=\int_{-1}^1{\rm d}\nu\,{\rm e}^{-{\rm i}\phi_0^\perp s}\, f(-\nu)\,, \label{eq:nusym}
\end{equation}
where $f(\nu)$ denotes an arbitrary function of the parameter $\nu$, the photon polarization tensor
for a magnetic field and $k_\perp^2=0$, \Eqref{eq:Pieta}, can eventually be represented in the following concise form,
\begin{align}
\nonumber\\ 
\left.\Pi_\parallel\right|_{k_\perp^2=0}&=k_\parallel^2\frac{\alpha}{2\pi}\int_{-1}^1\frac{{\rm d}\nu}{2}(1-\nu^2)\Biggl[\ln\biggl(\frac{m^2-{\rm i}\epsilon}{2eB}\biggr)
-\psi\biggl(\frac{\phi_0^\parallel}{2eB}\biggr)-\frac{eB}{\phi_0^\parallel}\Biggr],  \nonumber\\
\left.\Pi_\perp\right|_{k_\perp^2=0}&=\left.\Pi_0\right|_{k_\perp^2=0}=k_\parallel^2\frac{\alpha}{2\pi}\int_{-1}^1\frac{{\rm d}\nu}{2}\Biggl\{(1-\nu^2)\ln\biggl(\frac{m^2-{\rm i}\epsilon}{2eB}\biggr)
-\nu^2-\biggl[1-\nu^2-\frac{\nu\phi_0^\parallel}{eB}\,\biggr]\,\psi\biggl(\frac{\phi_0^\parallel}{2eB}+\frac{1+\nu}{2}\biggr)\Biggr\}, \label{eq:B_kperp=0}
\end{align}
where only the single parameter integration over $\nu$ is still to be performed; cf. also \cite{Cover:1974ij,Tsai:1975tw,Bakshi:1976vd,Cover:1979zz}.
Apart from the limitation to $k_\perp^2=0$, \Eqref{eq:B_kperp=0} is valid for arbitrary values of the momenta and the magnetic field and thus in particular encodes the full field dependence in the nonperturbative regime.
Taking into account \Eqref{eq:trafo}, the analogous expression for an electric field and $k_\parallel=0$ follows straightforwardly from \Eqref{eq:B_kperp=0}.

Resorting to these preparations, in Sec.~\ref{sec:analyticalins} we study the photon polarization tensor as provided in \Eqref{eq:PIa} and Eqs.~\eqref{eq:PI_tens}-\eqref{eq:PI_comp} in detail.
In the explicit calculations, we mostly focus on the case of a pure magnetic field. The corresponding results for an electric field follow straightforwardly via the electric-magnetic duality~\eqref{eq:trafo}.

\section{Analytical insights into the photon polarization tensor}\label{sec:analyticalins}

The representation of the photon polarization tensor in \Eqref{eq:PI_comp} is well-suited for a perturbative small field expansion.
A series expansion in powers of the amplitude of the external field is straightforward.
It effectively amounts to an expansion in powers of $z\to0$ in the integrand of the propertime integral.

\subsection{Perturbative weak field expansion}\label{sec:pertweak}

Let us first note that in the perturbative regime it is permissible to set $\eta\equiv0$ in Eqs.~\eqref{eq:PIa} and \eqref{eq:PI_comp} from the outset.
This comes about as an expansion around $ef=0$ $\leftrightarrow$ ${z}=0$ does not retain any of the corresponding integrands' poles in the complex $s$ plane, and
the propertime integration contour can be shifted onto the real positive $s$ axis.
Hence, evenness in ${z}$ then directly implies evenness in $ef$, and, in full agreement with Furry's theorem, the perturbative expansion of the photon polarization tensor is in even powers of $ef$.
Consequently, the perturbative expansion of \Eqref{eq:PI_comp}, can be written as
\begin{equation}
 \Pi_p^{\rm pert}=\sum\limits_{n=0}^{\infty}\Pi_p^{(2n)}\,,
\label{eq:Pi_p_pertseries}
\end{equation}
where the upper index ${(2n)}$ refers to the order in the perturbative expansion in powers of $ef$, i.e., denotes contributions of order $(ef)^{2n}$.
Formally, the terms in \Eqref{eq:Pi_p_pertseries} are determined as follows,
\begin{equation}
 \Pi_p^{(2n)}=\frac{(ef)^{2n}}{n!}\left[\left(\frac{\partial}{\partial (ef)^2}\right)^{n}\Pi_p\right]_{ef=0} ,
\label{eq:Pi_p_pertterm}
\end{equation}
where the components $\Pi_p$ in \Eqref{eq:PI_comp} act as generators.
The zeroth order term $\Pi_p^{(0)}$ is the same for all polarization modes $p=0,\parallel,\perp$. It corresponds to the photon polarization tensor at zero-field, \Eqref{eq:PI_zerofield}.
In case of a magnetic field, higher order terms with $n\in\mathbb{N}$ straightforwardly follow from
\begin{equation}
\left\{
 \begin{array}{c}
 \Pi_{\parallel}^{(2n)}\\
 \Pi_{\perp}^{(2n)}\\
 \Pi_{0}^{(2n)}
 \end{array}
\right\}
=\frac{\alpha}{2\pi}\int_{-1}^{1}\frac{{\rm d}\nu}{2}\int_{0}^\infty\frac{{\rm d} s}{s}\,{\rm e}^{-{\rm i}\phi_0 s}\,\frac{z^{2n}}{n!}\left[
\left(\frac{\partial}{\partial z^2}\right)^n\left(\left\{
 \begin{array}{c}
 k_{\parallel}^2 N_{1} + k_{\perp}^2N_0 \\
 k_{\parallel}^2N_0 + k_{\perp}^2 N_{2} \\
 k^2N_0 
 \end{array}
\right\}{\rm e}^{-{\rm i}sk_{\perp}^2\tilde{n}_2}\right)
\right]_{z=0},\label{eq:Pi_p_pertterm2}
\end{equation}
where
\begin{equation}
 \tilde{n}_2=n_2-\frac{1-\nu^2}{4}={\cal O}(z^2)\,.
\end{equation}
The nontrivial ${z}$ dependence of the function $n_2$, entering in the exponential, obstructs simple closed form expressions for $\Pi_p^{(2n)}$ for arbitrary field alignments, 
even though the coefficients $\{N_i^{(2n)},n_2^{(2n)}\}\in\mathbb{R}$ with $i=0,1,2$ in
\begin{equation}
 N_i(z)=\sum_{n=0}^{\infty}N_i^{(2n)}z^{2n} \quad\quad{\rm and}\quad\quad n_2(z)=\sum_{n=0}^{\infty}n_2^{(2n)}z^{2n}\,, \label{eq:Ni}
\end{equation}
are known explicitly in terms of sums over a finite number of terms. Nevertheless, \Eqref{eq:Pi_p_pertterm2} can in principle be evaluated for any desired $n$.
The series representations given in \Eqref{eq:Ni} are valid for $|z|<\pi$ (see Appendix~\ref{app:seriesNn}): The original functions in \Eqref{eq:scalarfcts_B} feature inverse powers of $\sin(z)$,
and correspondingly poles at the zeros of the sine function, $z=n\pi$ ($n\in\mathbb{Z}$), with the first pole at $z=\pi$ delimiting the range convergence of the expansion around $z=0$.

All the propertime integrals in \Eqref{eq:Pi_p_pertterm2} can then be performed explicitly, employing
\begin{equation}
 \int_{0}^{\infty}{{\rm d} s}\,s^{l}\,{\rm e}^{-{\rm i}\phi_0s}=\left({\rm i}\,\frac{\partial}{\partial \phi_0}\right)^{l}\int_{0}^{\infty}{{\rm d} s}\,{\rm e}^{-{\rm i}\phi_0s}
=l!\left(\frac{-{\rm i}}{\phi_0}\right)^{l+1}, \label{eq:int0}
\end{equation}
for $l\in\mathbb{N}_0$.
On first sight \Eqref{eq:int0} seems to be incompatible with the radius of convergence of the small $z$ expansion in \Eqref{eq:Ni}.
It is nevertheless possible to reconcile the expansion in $z\to 0$ and the propertime integration~\eqref{eq:int0}:
First, we substitute the dimensionful propertime parameter ``$s$'' for the dimensionless one ``$z$''. Correspondingly, the factor $\phi_0 s$ in the argument of the exponential reads
$\frac{\phi_0}{eB}z$ (or $\frac{\phi_0}{eE}z'$ for an electric field).
Demanding $|\phi_0/(ef)|\gg 1$, such that the integrands of the propertime integral receive their main contribution from the regime $|z|<\pi$,
we can argue that it is permissible to adopt \Eqref{eq:int0} after the expansion around $z=0$.

Keeping track of the various physical parameters, the expansion coefficients~\eqref{eq:Pi_p_pertterm2} of the photon polarization tensor in a magnetic field are of the following structure,
\begin{multline}
\Pi_p^{(2n)} = \frac{\alpha}{2\pi}\int_{-1}^{1}\frac{{\rm d}\nu}{2}\sum_{l=0}^n \left[k_{\parallel}^2c_{p}^{\parallel(n,l)}(\nu^2)+k_{\perp}^2c_{p}^{\perp(n,l)}(\nu^2)\right]\int\limits_{0}^{\infty-{\rm i}\eta}\frac{{\rm d} s}{s}\,z^{2n}\,(-{\rm i}sk_{\perp}^2)^l\,{\rm e}^{-{\rm i}\phi_0s} \\
=\frac{\alpha}{2\pi}\int_{-1}^{1}\frac{{\rm d}\nu}{2}\sum_{l=0}^{n-1}\frac{(2n+l-1)!}{(-1)^{n+l}}\, \left[k_{\parallel}^2c_{p}^{\parallel(n,l)}(\nu^2)+k_{\perp}^2c_{p}^{\perp(n,l)}(\nu^2)\right]\,\left(\frac{eB}{\phi_0}\right)^{2n}\left(\frac{k_{\perp}^2}{\phi_0}\right)^{l} \\
+\frac{4(3n-1)!}{n!\,3^n}\left(\frac{1-\nu^2}{4}\right)^{2n+1}k^2\left(\frac{eB}{\phi_0}\right)^{2n}\left(\frac{k_{\perp}^2}{\phi_0}\right)^{n}\,, \label{Int1}
\end{multline}
with $n\in\mathbb{N}$, and coefficients $c_{p}^{\parallel(n,l)}(\nu^2)$ and $c_{p}^{\perp(n,l)}(\nu^2)$, which also depend on the polarization mode $p$. The corresponding expression for an electric field follows by the electric-magnetic duality~\eqref{eq:trafo}.
From \Eqref{Int1} we can infer that factors $\sim z$ scale as $eB/\phi_0$, while those $\sim sk_\perp^2$ scale as $k_\perp^2/\phi_0$ after having carried out the integration over propertime.
Thus, for the perturbative expansion to yield trustworthy results, in the sense that higher order contributions become less important with increasing order $n$, both conditions,
\begin{equation}
 \left(\frac{eB}{\phi_0}\right)^2\ll1 \quad\quad{\rm and}\quad\quad \left(\frac{eB}{\phi_0}\right)^2\frac{k_\perp^2}{\phi_0}\ll1\,, \label{eq:pertreg}
\end{equation}
have to be fulfilled.
We emphasize that any truncation of the infinite sum in \Eqref{eq:Pi_p_pertseries} is ultimately limited to this parameter regime.

For illustration, we exemplarily provide the leading field dependent ($n=1$) perturbative correction to the photon polarization tensor in a pure magnetic field. It reads
\begin{equation}
\left\{
 \begin{array}{c}
\Pi^{(2)}_\parallel\\ 
 \Pi^{(2)}_\perp\\
 \Pi^{(2)}_0
 \end{array}
\right\}
=-\frac{\alpha}{12\pi}\int_{-1}^{1}\frac{{\rm d}\nu}{2}\,\left(\frac{eB}{\phi_0}\right)^2\left(1-\nu^2\right)^2
\left[
\left\{
\begin{array}{c}
\frac{-2}{1-\nu^2}\\
1\\
1
 \end{array}
\right\}
k_{\parallel}^2
+
\left(\left\{
\begin{array}{c}
1\\
\frac{5-\nu^2}{2(1-\nu^2)}\\
1
\end{array}
\right\}-\frac{\frac{1-\nu^2}{4}k^2}{\phi_0}\right)
k_{\perp}^2
\right].
\label{eq:PI_pert2}
\end{equation}
Specifying to on-the-light-cone dynamics, i.e., setting $k^2={\vec k}^2-\omega^2=0$, where $\omega$ denotes the photon energy, the conditions in \Eqref{eq:pertreg} simplify to
\begin{equation}
 \frac{eB}{m^2}\ll1 \quad\quad{\rm and}\quad\quad \left(\frac{eB}{m^2}\right)^2\frac{\omega^2\sin^2\theta}{m^2}\ll1\,. \label{eq:pertreg-lc}
\end{equation}
In this limit, the polarization tensor at zero field $\Pi^{(0)}$ vanishes and the $\nu$ integration in Eq.~(\ref{eq:PI_pert2}) can be easily performed,
\begin{align}
\left.\left\{
 \begin{array}{c}
 \Pi_\parallel^{(2)}\\
 \Pi_\perp^{(2)}\\
 \Pi_0^{(2)}
 \end{array}
\right\}\right|_{k^2=0}
=-\frac{\alpha}{2\pi}\left(\frac{eB}{m^2}\right)^2\omega^2\sin^2\theta\ \frac{2}{45}
\left\{
 \begin{array}{c}
7\\
4\\
0
 \end{array}
\right\}, \label{eq:pert_lc_LO}
\end{align}
and the index of refraction $n_p$ for on-the-light cone photons, polarized in mode $p=\{\parallel,\perp\}$ \cite{Tsai:1974fa}, 
\begin{equation}
 n_p=1-\frac{1}{2\omega^2}\,\Re(\Pi_p)|_{k^2=0}\,, \label{eq:np}
\end{equation}
can be read off straightforwardly. We obtain
\begin{align}
\left\{
 \begin{array}{c}
 \! n_\parallel \! \\
 \!  n_\perp \!
 \end{array}
\right\}
=1+\frac{\alpha}{4\pi}\left(\frac{eB}{m^2}\right)^2\sin^2\theta\ \frac{2}{45}
\left\{
 \begin{array}{c}
7\\
4
 \end{array}
\right\} + {\cal O}\left((eB)^4\right)\,. \label{eq:refind}
\end{align}
Equation~\eqref{eq:refind} gives rise to the famous result for the velocity shift in weak magnetic fields \cite{Toll:1952}.
Via the definition $v_p=c/n_p$, we find 
\begin{align}
\left\{
 \begin{array}{c}
 \! v_\parallel \! \\
 \! v_\perp \!
 \end{array}
\right\}
=1-\frac{\alpha}{4\pi}\left(\frac{eB}{m^2}\right)^2\sin^2\theta\ \frac{2}{45}
\left\{
 \begin{array}{c}
7\\
4
 \end{array}
\right\} + {\cal O}\left((eB)^4\right)\,.
\end{align}

For completeness, note that in the special case of a constant magnetic field and $k_\perp^2=0$ the structure of \Eqref{Int1} simplifies significantly: All the contributions in \Eqref{Int1} with $l>0$ vanish and $\phi_0$ becomes $\phi_0^\parallel$ [cf. \Eqref{eq:phi_parperp}]. In turn, the expansion~\eqref{Int1} is governed by the single parameter $(eB/\phi_0^\parallel)^2\ll1$; also see \Eqref{eq:pertreg}.
The perturbative weak field expansion in this regime is most conveniently obtained from \Eqref{eq:B_kperp=0} using the asymptotic 
series expansion of the Digamma function for large arguments \cite{Dunne:2004uk},
\begin{equation}
 \Psi(\chi)=\ln(\chi)-\frac{1}{2\chi}-\frac{1}{12\chi^2}+\mathcal{O}\left(
 \frac{B_{2l}}{2l} \frac{1}{\chi^{2l}}\right) \ , 
\label{eq:Psi_asympt}
\end{equation}
with $l = 2,3,4, \dots$ and Bernoulli numbers $B_{2l}$.
Obviously the logarithmic contribution originating in the large-argument expansion~\Eqref{eq:Psi_asympt} cancels with the logarithm in \Eqref{eq:B_kperp=0}, such that the perturbative small field expansion is entirely in even powers of $eB$.
The asymptotic nature of a perturbative expansion in $e B \sim 1/\chi$ is very generic in QED
\cite{Dyson:1952tj,Dunne:2004nc,Dunne:2002rq}.

\subsection{Approximations \`{a} la Tsai and Erber}\label{sec:TsaiErber}

Subsequently our focus is on the photon polarization tensor beyond the perturbative regime.
We start by following a strategy devised by Tsai and Erber for on-the-light-cone dynamics \cite{Tsai:1974fa,Tsai:1975iz}.
The basic idea is to adopt a particular type of expansion of the integrand of the propertime integral in \Eqref{eq:PI_comp}, such that it can eventually be written in terms of Airy functions.

In contrast to \cite{Tsai:1974fa,Tsai:1975iz}, we do not limit ourselves to on-the-light-cone dynamics, and do not perform any substitution in the propertime integration parameter $s$.
Rather, we explicitly keep track of occurrences of the {\it bare} propertime parameter $s$ and the {\it combined} parameter $z$.
This allows for a decisively more controlled expansion, and even grants access to novel parameter regimes beyond the scope of \cite{Tsai:1974fa,Tsai:1975iz}.

We again make use of an expansion in terms of the parameter $z$ under the propertime integral,
but in contrast to Sec.~\ref{sec:pertweak} do not perform a strict series expansion in powers of $z^2$.

Whereas in a strict perturbative weak field expansion only the first term of the phase factor,
\begin{equation}
  \Phi_0 =  \phi_0 + k_{\perp}^2\left[\frac{(1-\nu^2)^2}{48}{z^{2}}+\,{\cal O}({z^{4}})\right], \label{eq:phi_0@z^2}
\end{equation}
is kept in the exponential and all terms proportional to $z^{2n}$ ($n\in\mathbb{N}$) are expanded
to form polynomial contributions in the integrand of the propertime integral, we now keep all the terms written explicitly in \Eqref{eq:phi_0@z^2} in the exponential.
To this end, we rewrite the exponential factor in \Eqref{eq:PI_comp} as follows,
\begin{equation}
 {\rm e}^{-{\rm i}\Phi_0s}={\rm e}^{-{\rm i}\phi_0s - {\rm i} k_{\perp}^2\frac{(1-\nu^2)^2}{48}z^2s}\Biggl[1+\sum_{j=1}^{\infty}\frac{1}{j!}\left(-{\rm i}k_{\perp}^2s\right)^j z^{4j}\biggl(\sum_{n=2}^{\infty}n_2^{(2n)}z^{2(n-2)}\biggr)^j\Biggr]. \label{eq:TsaiErberexp}
\end{equation}
Equation~\eqref{eq:TsaiErberexp} still accounts for the full momentum dependence of the factor ${\rm e}^{-{\rm i}\Phi_0s}$ to all orders.
In particular, this remains also true for truncations of \Eqref{eq:TsaiErberexp} of the form
\begin{equation}
 {\rm e}^{-{\rm i}\Phi_0s}={\rm e}^{-{\rm i}\phi_0s - {\rm i} k_{\perp}^2\frac{(1-\nu^2)^2}{48}z^2s}\Biggl[1+\sum_{j=1
}^{\infty}\frac{(n_2^{(4)})^j}{j!}{\left(-{\rm i}k_{\perp}^2s\right)^j}z^{4j}\left[1+{\cal O}(z^2)\right]\Biggr]. \label{eq:TsaiErberexp_1}
\end{equation}

An expansion as performed in \Eqref{eq:TsaiErberexp} becomes relevant if the contribution $\sim k_{\perp}^2z^2$ in \Eqref{eq:phi_0@z^2} can become as big as, or even surpass, $\phi_0$.
At the same time, the dominant contributions to the propertime integral should stem from small $z$, such that 
terms $\sim k_{\perp}^2{\cal O}({z^{4}})$ remain ``small enough'' to allow for their expansion.
Resorting to our findings in the perturbative weak field regime [cf. below \Eqref{Int1}], 
it is plausible that for $|eB/\phi_0|\ll1$ an expansion of the above type should in particular grant access to the regime $|(\frac{eB}{\phi_0})^2\frac{k_\perp^2}{\phi_0}|\geq1$,
not accessible within an ordinary perturbative weak field expansion.
In addition, the approach should of course still grant access to the perturbative weak field regime~\eqref{eq:pertreg}.
Moreover, we could expect to obtain trustworthy results even for small $\phi_0$ given that simultaneously $k_\perp^2/(eB)\gg1$, such that the propertime integrations still receive their main contribution
from $|z|<\pi$ [cf. below \Eqref{eq:int0}].

Let us now adopt this type of expansion in \Eqref{eq:PI_comp} and thoroughly study its range of applicability.
Substituting the functions $N_i$ with $i\in\{0,1,2\}$ for their series representations in the limit $z\to0$, \Eqref{eq:Ni},
and employing \Eqref{eq:partint} with $a=k_{\perp}^2(eB)^2s^3/48$, we arrive at [cf. \Eqref{eq:Pipi}]
\begin{multline}
\left\{\!\!
 \begin{array}{c}
 \pi_{\parallel}\\
 \pi_{\perp}\\
 \pi_{0}
 \end{array}\!\!
\right\}
=\frac{\alpha}{2\pi}\int\limits_{0}^{\infty-{\rm i}\eta}{\rm d} s\,{\rm e}^{-{\rm i}\phi_0s - {\rm i} k_{\perp}^2\frac{(1-\nu^2)^2}{48}(eB)^2s^3}
\Biggl\{-{\rm i}(k^2)^2\frac{\nu^2(3-\nu^2)}{6}  \\
+\sum_{n=1}^{\infty}\left\{
 \begin{array}{c}
 k_{\parallel}^2 N_{1}^{(2n)} + k_{\perp}^2N_0^{(2n)} \\
 k_{\parallel}^2N_0^{(2n)} + k_{\perp}^2 N_{2}^{(2n)} \\
 k^2N_0^{(2n)}
 \end{array}
\right\}\frac{z^{2n}}{s}\Biggl[\sum_{j=0}^{\infty}\frac{1}{j!}\left(-{\rm i}k_{\perp}^2s\right)^j z^{4j}\biggl(\sum_{l=2}^{\infty}n_2^{(2l)}z^{2(l-2)}\biggr)^j\Biggr] \\
+k^2(1-\nu^2)\frac{1}{s}\Biggl[\frac{\nu^2(3-\nu^2)}{36}(-{\rm i}k_{\perp}^2s)z^2+\sum_{j=1}^{\infty}\frac{1}{j!}\left(-{\rm i}k_{\perp}^2s\right)^j z^{4j}\biggl(\sum_{l=2}^{\infty}n_2^{(2l)}z^{2(l-2)}\biggr)^j\Biggr]\Biggr\}.
 \label{eq:PI_tsai}
\end{multline}

In the next step we aim at carrying out the propertime integration. The integrand of the propertime integral in \Eqref{eq:PI_tsai} consists of an overall exponential factor that multiplies a polynomial in $s$.
Polynomial contributions even in $s$ come with prefactors that are purely imaginary, whereas odd powers of $s$ have real prefactors. 
The integrals to be performed are of the following structure
\begin{equation}
 \int_{0}^{\infty}{\rm d} s\,s^l\,{\rm e}^{-{\rm i}\phi_0s - {\rm i} k_{\perp}^2\frac{(1-\nu^2)^2}{48}(eB)^2s^3}
=\left({\rm i}\,\frac{\partial}{\partial \phi_0}\right)^{l}\int_{0}^{\infty}{\rm d} s\,{\rm e}^{-{\rm i}\phi_0s - {\rm i} k_{\perp}^2\frac{(1-\nu^2)^2}{48}(eB)^2s^3}, \label{eq:gentsaierberint}
\end{equation}
with $l\in\mathbb{N}_0$.
Given that $0\leq\nu<1$ and $|B^2k_{\perp}^2|\neq0$, we can write
\begin{multline}
 \int_{0}^{\infty}{\rm d} s\,{\rm e}^{-{\rm i}\phi_0s - {\rm i} k_{\perp}^2\frac{(1-\nu^2)^2}{48}(eB)^2s^3}
=\frac{1}{\phi_0}\left(\frac{3}{2}\tilde\xi\right)^{2/3}\int_{0}^{\infty}{\rm d} \tau\,{\rm e}^{-{\rm i}\left(\frac{3}{2}\tilde\xi\right)^{2/3}\tau-{\rm i}\frac{1}{3}{\rm sign}\left(B^2k_{\perp}^2\right)\tau^3} \\
=\frac{\pi}{\phi_0}\left(\frac{3}{2}\tilde\xi\right)^{2/3}\biggl[{\rm Ai}\left({\rm sign}\!\left(B^2k_{\perp}^2\right)\!\left(\tfrac{3}{2}\tilde{\xi}\right)^{2/3}\right)
-{\rm i}\,{\rm sign}\!\left(B^2k_{\perp}^2\right){\rm Gi}\left({\rm sign}\!\left(B^2k_{\perp}^2\right)\!\left(\tfrac{3}{2}\tilde\xi\right)^{2/3}\right)\biggr], \label{eq:tsaielm}
\end{multline}
where we substituted $s$ for $\tau$,
\begin{equation}
 \tau=\left(\frac{1-\nu^2}{4}\right)^{2/3}\left|(eB)^2k_{\perp}^2\right|^{1/3}s\,,
\end{equation}
and introduced the dimensionless parameter
\begin{equation}
 \tilde\xi^{2/3}\equiv\left(\frac{2}{3}\frac{4}{1-\nu^2}\right)^{2/3}\frac{\phi_0}{\left|(eB)^2k_{\perp}^2\right|^{1/3}}\,, \label{eq:xi}
\end{equation}
which is real valued apart from the infinitesimal quantity $-{\rm i}\epsilon$, with $\epsilon\to0^+$, encoded in $\phi_0$.
In the last line of \Eqref{eq:tsaielm} we used the definitions 10.4.32 and 10.4.42 of \cite{Abram}.
${\rm Ai}(\chi)$ denotes the Airy function and ${\rm Gi}(\chi)$ is defined via both ${\rm Ai}(\chi)$ and the Airy function of the second kind ${\rm Bi}(\chi)$ (cf. formula 10.4.42 of \cite{Abram}),
\begin{equation}
 {\rm Gi}(\chi)=\frac{1}{3}{\rm Bi}(\chi)+\int_0^{\chi}{\rm d}t\left[{\rm Ai}(\chi){\rm Bi}(t)-{\rm Ai}(t){\rm Bi}(\chi)\right]. \label{eq:Gi}
\end{equation}
For completeness, we note that it is also possible to perform the manipulations in \Eqref{eq:tsaielm} in a slightly different way, such that the final expression on its right-hand side can be evaluated throughout the contour $B\to B{\rm e}^{-{\rm i}\delta}$ with $0\leq\delta\leq\frac{\pi}{2}$ [cf. \Eqref{eq:trafo}]. However, representation~\eqref{eq:tsaielm} is directly applicable to all physically relevant situations and thus completely serves our purposes. Specifying it to the case of an electric field, we just have to substitute
\begin{equation}
 B^2\ \to\ -E^2 \quad\quad{\rm and}\quad\quad k_\perp^2\ \to\ k_\parallel^2\,.
\end{equation}
It is furthermore helpful (see below) to define the following parameter
\begin{equation}
 \xi_\pm\equiv\left[\pm{\rm sign}(B^2k_\perp^2)\tilde{\xi}^{2/3}\right]^{3/2}=\frac{2}{3}\frac{4}{1-\nu^2}\frac{\left[\pm\phi_0\,{\rm sign}(B^2k_\perp^2)\right]^{3/2}}{|(eB)^2k_\perp^2|^{1/2}}\,. \label{eq:xi!}
\end{equation}
For $k^2=0$ the parameter $\xi_+$ in \Eqref{eq:xi!} agrees with the parameter $\xi$ as defined in Eq.~(57) of \cite{Tsai:1974fa}.
As detailed in Appendix~\ref{app:AiryBessel}, \Eqref{eq:tsaielm} can be expressed in terms of an infinite sum of Bessel functions. In turn, the generic propertime integral~\eqref{eq:gentsaierberint} can be written as
\begin{multline}
\int_{0}^{\infty}{\rm d} s\,s^l\,{\rm e}^{-{\rm i}\phi_0s - {\rm i} k_{\perp}^2\frac{(1-\nu^2)^2}{48}(eB)^2s^3}
 =-\frac{\pi}{3}\left({\rm i}\,\frac{3}{2}\frac{\tilde\xi^{2/3}}{\phi_0}\right)^{l+1}\!\!\left({\rm sign}(B^2k_\perp^2)\xi_+^{1/3}\frac{\partial}{\partial\xi_+}\right)^l\xi_+^{1/3}\Biggl\{{\rm i}\frac{\sqrt{3}}{\pi}{\rm K}_{1/3}(\xi_+) \\
+\frac{{\rm sign}(B^2k_{\perp}^2)}{\sqrt{3}}\biggr[{\rm I}_{-1/3}(\xi_+)+{\rm I}_{1/3}(\xi_+)
+4\sum_{n=0}^{\infty}(-1)^n\bigl[{\rm I}_{-1/3}(\xi_+)\,{\rm I}_{2n+4/3}(\xi_+)-{\rm I}_{1/3}(\xi_+)\,{\rm I}_{2n+2/3}(\xi_+)\bigr]\biggr]\Biggr\}, \label{eq:Tsai_int_sj+}
\end{multline}
for $\Re(B^2k_{\perp}^2\phi_0)\geq0$, and 
\begin{multline}
\int_{0}^{\infty}{\rm d} s\,s^l\,{\rm e}^{-{\rm i}\phi_0s - {\rm i} k_{\perp}^2\frac{(1-\nu^2)^2}{48}(eB)^2s^3}
 =-\frac{\pi}{3}\left({\rm i}\,\frac{3}{2}\frac{\tilde \xi^{2/3}}{\phi_0}\right)^{l+1}\!\!\left(-{\rm sign}(B^2k_\perp^2)\xi_-^{1/3}\frac{\partial}{\partial\xi_-}\right)^l\xi_-^{1/3}\Biggl\{{\rm i}\bigl[{\rm J}_{-1/3}(\xi_-)+{\rm J}_{1/3}(\xi_-)\bigr] \\
+\frac{{\rm sign}(B^2k_{\perp}^2)}{\sqrt{3}}\biggl[{\rm J}_{-1/3}(\xi_-)-{\rm J}_{1/3}(\xi_-)
+4\sum_{n=0}^{\infty}\bigl[{\rm J}_{-1/3}(\xi_-)\,{\rm J}_{2n+4/3}(\xi_-)-{\rm J}_{1/3}(\xi_-)\,{\rm J}_{2n+2/3}(\xi_-)\bigr]\biggr]\Biggr\}, \label{eq:Tsai_int_sj-}
\end{multline}
for $\Re(B^2k_{\perp}^2\phi_0)\leq0$.

As $\epsilon\to0^+$, the conditions for $\Re(B^2k_{\perp}^2\phi_0)\gtreqqless0$ basically amount to conditions for ${\rm sign}(B^2k_{\perp}^2\phi_0)\gtreqqless0$.
Correspondingly, the arguments $\xi_\pm$ of the Bessel functions in Eqs.~\eqref{eq:Tsai_int_sj+} and \eqref{eq:Tsai_int_sj-} are essentially real valued and positive; cf. \Eqref{eq:xi!}.

Rewriting the propertime integrals in \Eqref{eq:PI_tsai} in terms of Eqs.~\eqref{eq:Tsai_int_sj+} and \eqref{eq:Tsai_int_sj-} does not allow for immediate insights. The right-hand sides of Eqs.~\eqref{eq:Tsai_int_sj+} and \eqref{eq:Tsai_int_sj-} are still complicated expressions.
However, as for real valued arguments both the ordinary and the modified Bessel functions of the first kind are real valued,
the real and imaginary parts at fixed $l$ are disentangled in Eqs.~\eqref{eq:Tsai_int_sj+} and \eqref{eq:Tsai_int_sj-}, and can thus be inferred straightforwardly.

Subsequently, we focus on limiting cases where further analytical results are accessible: Formally, these cases can be associated with either the limit $\xi\equiv|\tilde\xi|=|\xi_\pm|\to 0$, or $\xi\to\infty$.
It might seem somewhat unusual to include an explicit $\nu$ dependence -- with $\nu$ still to be integrated over -- in the definition of the parameter $\xi$ which is to be sent to zero or infinity, respectively.
While the integrand of the $\nu$ integral could formally of course be expanded in a manifestly $\nu$ independent combination, like \cite{Tsai:1974fa,Tsai:1975iz}
\begin{equation}
 \lambda^{-1}\equiv\frac{1-\nu^2}{4}\xi\left(\frac{m^{2}}{\phi_0}\right)^{3/2}=\frac{2}{3}\frac{m^3}{\left|(eB)^2k_\perp^2\right|^{1/2}}, \label{eq:lambda}
\end{equation}
instead, the {\it true} expansion parameter would still be given by $\xi$. Performing an expansion in $\lambda^{-1}$, the $\nu$ dependent expansion coefficients would rearrange such that the expansion is effectively in $\xi$.   

\subsubsection{Very weak fields - large momentum}  \label{subsec:xitoinfty}

The limit $\xi\to\infty$, or equivalently [cf. \Eqref{eq:xi}]
\begin{equation}
 \xi^{-1}=\frac{3}{2}\frac{1-\nu^2}{4}\left|\frac{(eB)^2k^2_{\perp}}{\phi_0^{3}}\right|^{\frac{1}{2}}\ll 1\,, \label{eq:xitoinfty}
\end{equation}
should be compatible with the perturbative weak field expansion as performed in Sec.~\ref{sec:pertweak}.
In this spirit, we first write
\begin{multline}
 \int_{0}^{\infty}{\rm d} s\,s^l\,{\rm e}^{-{\rm i}\phi_0s - {\rm i} k_{\perp}^2\frac{(1-\nu^2)^2}{48}(eB)^2s^3}=\int_{0}^{\infty}{\rm d} s\,s^l\,{\rm e}^{-{\rm i}\phi_0s - {\rm i}\frac{4}{27}\,{\rm sign}(B^2k_{\perp}^2)\left(\frac{\phi_0}{\tilde\xi^{2/3}}\right)^3s^3}\\
=\sum_{n=0}^{\infty}\frac{(-{\rm i})^n}{n!}\left[\frac{4}{27}\,{\rm sign}(B^2k_{\perp}^2)\frac{\phi^3_0}{\tilde\xi^2}\right]^n\int_{0}^{\infty}{\rm d} s\,s^{3n+l}\,{\rm e}^{-{\rm i}\phi_0s}
=\left(\frac{-{\rm i}}{\phi_0}\right)^{l+1}\sum_{n=0}^{\infty}\frac{(3n+l)!}{n!3^n}\left[\left(\frac{2}{3}\right)^2\frac{{\rm sign}(B^2k_{\perp}^2)}{\tilde\xi^2}\right]^n, \label{eq:sososo}
\end{multline}
where we employed \Eqref{eq:int0} in the last step. One might naively expect that \Eqref{eq:sososo} corresponds to the maximum information attainable in the limit $\xi\to\infty$.
This, however, is not true:
By comparison with Eqs.~\eqref{eq:Tsai_int_sj+} and \eqref{eq:Tsai_int_sj-},
we infer that \Eqref{eq:sososo} only accounts for the real part of the contributions embraced in curly brackets in these expressions.\footnote{\label{footnote:3}As the above analytical approximations are genuinely insensitive to threshold singularities (cf. also the last paragraph of this section), we explicitly exclude the special case $\Re(\phi_0)\equiv0$ from the following considerations.}
The asymptotic behavior of the corresponding imaginary parts can be extracted from Eqs.~\eqref{eq:assymptK} and \eqref{eq:assymptJ+J} in the appendix.
With regard to \Eqref{eq:Tsai_int_sj+}, the respective leading contribution for fixed $l$ is given by
\begin{multline}
 -\frac{\pi}{3}\biggl({\rm i}\,\frac{3}{2}\frac{\tilde\xi^{2/3}}{\phi_0}\biggr)^{l+1}\biggl({\rm sign}(B^2k_\perp^2)\xi_+^{1/3}\frac{\partial}{\partial\xi_+}\biggr)^l\xi_+^{1/3}\,{\rm i}\frac{\sqrt{3}}{\pi}{\rm K}_{1/3}(\xi_+) \\
= \left[(-{\rm i})\,{\rm sign}(B^2k_\perp^2)\right]^l\sqrt{\frac{\pi}{6}}\left(\frac{3}{2}\frac{\tilde\xi^{2/3}}{\phi_0}\xi_+^{1/3}\right)^{l+1}\frac{{\rm e}^{-\xi_+}}{\sqrt{\xi_+}}\biggl[1+{\cal O}\left(\tfrac{1}{\xi_+}\right)\biggr],
\end{multline}
i.e., the term with all the derivatives for $\xi_+$ acting on the factor $\exp(-\xi_+)$ in \Eqref{eq:assymptK}.
For \Eqref{eq:Tsai_int_sj-} we analogously obtain
\begin{multline}
-\frac{\pi}{3}\biggl({\rm i}\,\frac{3}{2}\frac{\tilde\xi^{2/3}}{\phi_0}\biggr)^{l+1}\biggl(-{\rm sign}(B^2k_\perp^2)\xi^{1/3}_-\frac{\partial}{\partial\xi_-}\biggr)^l\xi_-^{1/3}\,{\rm i}\bigl[{\rm J}_{-1/3}(\xi_-)+{\rm J}_{1/3}(\xi_-)\bigr]\\
=\left[(-{\rm i})\,{\rm sign}(B^2k_\perp^2)\right]^{l}\sqrt{\frac{\pi}{3}}\biggl(\frac{3}{2}\frac{\tilde \xi^{2/3}}{\phi_0}\xi_-^{1/3}\biggr)^{l+1}\frac{1}{\sqrt{\xi_-}} \hspace*{5cm} \\
\times\left\{\cos(\xi_-)\Bigl[\cos(\tfrac{\pi l}{2})+\sin(\tfrac{\pi l}{2})+{\cal O}\left(\tfrac{1}{\xi_-}\right)\Bigr]+\sin(\xi_-)\Bigl[\cos(\tfrac{\pi l}{2})-\sin(\tfrac{\pi l}{2})+{\cal O}\left(\tfrac{1}{\xi_-}\right)\Bigr]\right\}.
\end{multline}
Accounting for these contributions in \Eqref{eq:sososo}, we arrive at
\begin{multline}
 \int_{0}^{\infty}{\rm d} s\,s^l\,{\rm e}^{-{\rm i}\phi_0s - {\rm i} k_{\perp}^2\frac{(1-\nu^2)^2}{48}(eB)^2s^3}
=\left(\frac{-{\rm i}}{\phi_0}\right)^{l+1}\Biggl\{\sum_{n=0}^{\infty}\frac{(3n+l)!}{n!3^n}\left[\left(\frac{2}{3}\right)^2\frac{{\rm sign}(B^2k_{\perp}^2)}{\tilde\xi^2}\right]^n \\
+{\rm i}\left[{\rm sign}(B^2k_\perp^2)\right]^l\sqrt{\frac{\pi}{6}}\left(\frac{3}{2}\,\tilde\xi^{2/3}\xi_+^{1/3}\right)^{l+1}\frac{{\rm e}^{-\xi_+}}{\sqrt{\xi_+}}
\biggl[1+{\cal O}\left(\tfrac{1}{\xi_+}\right)\biggr]\Biggr\}, \label{eq:gen_>}
\end{multline}
for $\Re(B^2k_{\perp}^2\phi_0)\geq0$, and
\begin{multline}
 \int_{0}^{\infty}{\rm d} s\,s^l\,{\rm e}^{-{\rm i}\phi_0s - {\rm i} k_{\perp}^2\frac{(1-\nu^2)^2}{48}(eB)^2s^3}
=\left(\frac{-{\rm i}}{\phi_0}\right)^{l+1}\Biggl\{\sum_{n=0}^{\infty}\frac{(3n+l)!}{n!3^n}\left[\left(\frac{2}{3}\right)^2\frac{{\rm sign}(B^2k_{\perp}^2)}{\tilde\xi^2}\right]^n
\\
+{\rm i}\left[{\rm sign}(B^2k_\perp^2)\right]^l\sqrt{\frac{\pi}{3}}\biggl(\frac{3}{2}\,\tilde\xi^{2/3}\xi_-^{1/3}\biggr)^{l+1}\frac{1}{\sqrt{\xi_-}}\biggl[\Bigl((-1)^l\cos(\xi_-)+\sin(\xi_-)\Bigr)\Bigl(\cos(\tfrac{\pi l}{2})-\sin(\tfrac{\pi l}{2})\Bigr)\\
+\cos(\xi_-)\,{\cal O}\!\left(\tfrac{1}{\xi_-}\right)+\sin(\xi_-)\,{\cal O}\!\left(\tfrac{1}{\xi_-}\right)\biggr]\Biggr\}, \label{eq:gen_<}
\end{multline}
for $\Re(B^2k_{\perp}^2\phi_0)\leq0$.
With these preparations, we now aim at analytical insights into the photon polarization tensor, \Eqref{eq:PI_tsai}, in the limit $\xi\to\infty$.
It is convenient to split the various building blocks of \Eqref{eq:PI_tsai} into real and imaginary contributions (cf. footnote~\ref{footnote:3}).
For the real part we obtain
\begin{equation}
 \Re\int_{0}^{\infty}{\rm d} s\,{\rm e}^{-{\rm i}\phi_0s - {\rm i} k_{\perp}^2\frac{(1-\nu^2)^2}{48}(eB)^2s^3}
 \left\{\begin{array}{c}
  \!\!{\rm i}s^0k^2\!\! \\
  \!\!{\rm i}z^2k^2\!\!
 \end{array}\right\}
=\frac{k^2}{\phi_0}\sum_{l=0}^{\infty}\left\{\begin{array}{c}
  \frac{(3l)!}{l!3^l}\\
  -\frac{(3l+2)!}{l!3^l}\left(\frac{eB}{\phi_0}\right)^2
 \end{array}\right\}\!\left[\left(\frac{2}{3}\right)^2\,\frac{{\rm sign}(B^2k_{\perp}^2)}{\tilde\xi^2}\right]^l \label{eq:ints3_1}
\end{equation}
and
\begin{multline}
\Re\int_{0}^{\infty}\frac{{\rm d} s}{s}\,{\rm e}^{-{\rm i}\phi_0s - {\rm i} k_{\perp}^2\frac{(1-\nu^2)^2}{48}(eB)^2s^3}(-{\rm i}k_{\perp}^2s)^{j}z^{2(n+2j)} \\
=(-1)^{n+j}\left(\frac{4}{1-\nu^2}\right)^{2j}\left(\frac{eB}{\phi_0}\right)^{2(n+j)}
\sum_{l=0}^{\infty}\frac{(3l+2n+5j-1)!}{l!\,3^l}\left[\left(\frac{2}{3}\right)^2\,\frac{{\rm sign}(B^2k_{\perp}^2)}{\tilde\xi^2}\right]^{l+j}. \label{eq:ints3_2}
\end{multline}
We emphasize that Eqs.~\eqref{eq:ints3_1} and \eqref{eq:ints3_2} are valid for any value of $\Re(B^2k_{\perp}^2\phi_0)$. The distinction between the regimes with positive and negative $\Re(B^2k_{\perp}^2\phi_0)$ is only relevant for the imaginary parts. In the regime $\Re(B^2k_{\perp}^2\phi_0)\geq0$, the imaginary parts are given by
\begin{equation}
 \Im\int_{0}^{\infty}{\rm d} s\,{\rm e}^{-{\rm i}\phi_0s - {\rm i} k_{\perp}^2\frac{(1-\nu^2)^2}{48}(eB)^2s^3}
 \left\{\begin{array}{c}
  \!\!{\rm i}s^0k^2\!\! \\
  \!\!{\rm i}z^2k^2\!\!
 \end{array}\right\}
=\frac{3}{2}\frac{k^2}{\phi_0}\left\{\begin{array}{c}
  \frac{\tilde\xi^{2/3}\xi_+^{1/3}}{\sqrt{\xi_+}}\\
  -\left(\frac{3}{2}\tilde\xi\frac{eB}{\phi_0}\right)^2\sqrt{\xi_+}
 \end{array}\right\}{\rm e}^{-\xi_+}\sqrt{\frac{\pi}{6}}\left[1+{\cal O}\left(\tfrac{1}{\xi_+}\right)\right] \label{eq:ints2_1}
\end{equation}
and
\begin{multline}
\Im\int_{0}^{\infty}\frac{{\rm d} s}{s}\,{\rm e}^{-{\rm i}\phi_0s - {\rm i} k_{\perp}^2\frac{(1-\nu^2)^2}{48}(eB)^2s^3}(-{\rm i}k_{\perp}^2s)^{j}z^{2(n+2j)} \\
=(-1)^{n+j}\,\left[{\rm sign}(B^2k_{\perp}^2)\right]^{(n+j+1)} \left(\tfrac{4}{1-\nu^2}\right)^{2j}\left(\tfrac{3}{2}\tilde\xi\tfrac{eB}{\phi_0}\right)^{2(n+j)}\left(\tfrac{3}{2}\xi_+\right)^{j-\frac{1}{2}}{\rm e}^{-\xi_+}\sqrt{\tfrac{\pi}{4}}\left[1+{\cal O}\left(\tfrac{1}{\xi_+}\right)\right], \label{eq:ints2_2}
\end{multline}
where we in particular made use of $\xi_\pm^{2/3}=\pm{\rm sign}(B^2k_\perp^2)\tilde\xi^{2/3}$ [cf. \Eqref{eq:xi!}].

In the complementary regime, $\Re(B^2k_{\perp}^2\phi_0)\leq0$, they read
\begin{equation}
 \Im\int_{0}^{\infty}{\rm d} s\,{\rm e}^{-{\rm i}\phi_0s - {\rm i} k_{\perp}^2\frac{(1-\nu^2)^2}{48}(eB)^2s^3}
 \left\{\begin{array}{c}
  \!\!{\rm i}s^0k^2\!\! \\
  \!\!{\rm i}z^2k^2\!\!
 \end{array}\right\}
=\frac{3}{2}\frac{k^2}{\phi_0}\left\{\begin{array}{c}
  \frac{\tilde\xi^{2/3}\xi_-^{1/3}}{\sqrt{\xi_-}}\\
  \left(\frac{3}{2}\tilde\xi\frac{eB}{\phi_0}\right)^2\sqrt{\xi_-}
 \end{array}\right\}\sqrt{\frac{\pi}{3}}\left(\cos\xi_-+\sin\xi_-\right) \label{eq:ints2_1b}
\end{equation}
and
\begin{multline}
\Im\int_{0}^{\infty}\frac{{\rm d} s}{s}\,{\rm e}^{-{\rm i}\phi_0s - {\rm i} k_{\perp}^2\frac{(1-\nu^2)^2}{48}(eB)^2s^3}(-{\rm i}k_{\perp}^2s)^{j}z^{2(n+2j)} \\
=\left[-{\rm sign}(B^2k_\perp^2)\right]^{(n+j+1)}\left(\tfrac{4}{1-\nu^2}\right)^{2j}\left(\tfrac{3}{2}\tilde\xi\tfrac{eB}{\phi_0}\right)^{2(n+j)}\left(\tfrac{3}{2}\xi_-\right)^{j-\frac{1}{2}}\Bigl(\cos\tfrac{5\pi j}{2}+\sin\tfrac{5\pi j}{2}\Bigr)
\sqrt{\tfrac{\pi}{2}}\left(\cos\xi_--(-1)^j\sin\xi_-\right). \label{eq:ints2_2b}
\end{multline}
To keep these expressions compact, we have only included the leading terms in Eqs.~\eqref{eq:ints2_1b} and \eqref{eq:ints2_2b}.
However, the corrections to Eqs.~\eqref{eq:ints2_1b} and \eqref{eq:ints2_2b} can be inferred straightforwardly by setting 
\begin{equation}
 \cos\xi_-\ \to\ \cos\xi_-\left[1+{\cal O}\left(\tfrac{1}{\xi_-}\right)\right],\quad\quad
 \sin\xi_-\ \to\ \sin\xi_-\left[1+{\cal O}\left(\tfrac{1}{\xi_-}\right)\right]. \label{eq:fehler2}
\end{equation}
Noteworthy, the series expansions in Eqs.~\eqref{eq:ints3_1} and \eqref{eq:ints3_2} -- constituting the real part -- are governed by the combinations
\begin{equation}
 \left(\frac{2}{3}\right)^2\frac{{\rm sign}(B^2k_\perp^2)}{\tilde\xi^2}=\left(\frac{1-\nu^2}{4}\right)^2\frac{(eB)^2k_\perp^2}{\phi_0^3}
\end{equation}
and $(eB/\phi_0)^2$, while
Eqs.~\eqref{eq:ints2_1}-\eqref{eq:ints2_2b}  -- the imaginary part -- are governed by $\xi_\pm$ and
\begin{equation}
 \left(\frac{3}{2}\tilde\xi\frac{eB}{\phi_0}\right)^2=\frac{\phi_0}{k_{\perp}^2}\left(\frac{4}{1-\nu^2}\right)^2{\rm sign}(B^2k_{\perp}^2)\,. \label{eq:eBbyphi0_malxi}
\end{equation}

Let us now focus on the real part of the photon polarization tensor~\eqref{eq:PI_tsai}.
In consequence of Eqs.~\eqref{eq:ints3_1} and \eqref{eq:ints3_2}, it can formally be expressed in terms of an infinite series in $(eB)^2$.
Obviously, a truncated version of this series yields trustworthy results, provided that
\begin{equation}
\left|\frac{eB}{\phi_0}\right|^2\ll1 \quad\quad \text{and} \quad\quad \xi^{-1}\ll1\,. \label{eq:reg_Re}
\end{equation}
As expected, these conditions are compatible with those stated in \Eqref{eq:pertreg} for the conventional perturbative weak field expansion.
Hence, $\Re(\Pi_p)$ should reproduce the perturbative weak field expansion, \Eqref{eq:Pi_p_pertseries}.
Employing partial integration with respect to $\nu$, it is straightforward to show explicitly that the contributions $\sim(eB)^0$ and $\sim(eB)^2$ arising in \Eqref{eq:Pipi} agree with Eqs.~\eqref{eq:vac0} and \eqref{eq:PI_pert2}, as derived in Sec.~\ref{sec:pertweak}.

Conversely, the structure of the imaginary part of the photon polarization tensor is manifestly non-perturbative, and thus cannot be inferred from a perturbative weak field expansion.
Having a closer look on the propertime integrals~\eqref{eq:ints2_2} and \eqref{eq:ints2_2b}, we make the following observation:
Focusing only on the terms written explicitly in Eqs.~\eqref{eq:ints2_2} and \eqref{eq:ints2_2b}, the nonnegative integers $n$ and $j$ just appear as powers
-- powers of the dimensionless parameters $z^2$ and $k_{\perp}^2s$ before having carried out the propertime integration, and powers of $(\frac{3}{2}\tilde\xi\frac{eB}{\phi_0})^2$ and $\frac{3}{2}\xi_\pm$ thereafter.
With regard to \Eqref{eq:PI_tsai} this implies that to leading order in $1/\xi_\pm$ the sum over $j$ can be performed straightforwardly resulting in exponential (trigonometric) functions. 
Summing up the leading terms for a given power of the propertime variable $s$, the respective contributions to the imaginary part of \Eqref{eq:PI_tsai} can be concisely represented as
\begin{multline}
\sum_{j=1}^{\infty}\frac{1}{j!}\,\Im\int_{0}^{\infty}\frac{{\rm d} s}{s}\,{\rm e}^{-{\rm i}\phi_0s - {\rm i} k_{\perp}^2\frac{(1-\nu^2)^2}{48}(eB)^2s^3}(-{\rm i}k_{\perp}^2s)^{j}z^{2(n+2j)}\biggl(\sum_{l=2}^{\infty}n_2^{(2l)}z^{2(l-2)}\!\biggr)^j \\
\quad\quad={\rm sign}(B^2k_\perp^2)\sqrt{\tfrac{\pi}{4}}\left[\tfrac{4}{1-\nu^2}\bigl(-\tfrac{\phi_0}{k_\perp^2}\bigr)^{\frac{1}{2}}\right]^{2n}\frac{{\rm e}^{-\xi_+}}{\left(\tfrac{3}{2}\xi_+\right)^{{1}/{2}}}\left[1+{\cal O}\left(\tfrac{1}{\xi_+}\right)\right]
\Biggl[\exp\biggl\{-\tfrac{3}{2}\xi_+\tfrac{k_\perp^2}{\phi_0}\sum_{l=2}^{\infty}n_2^{(2l)}\left[\tfrac{4}{1-\nu^2}\bigl(-\tfrac{\phi_0}{k_\perp^2}\bigr)^{\frac{1}{2}}\right]^{2l}\biggl\}-1\Biggr] \hfill \\
\quad\quad={\rm sign}(B^2k_\perp^2)\sqrt{\tfrac{\pi}{4}}\left[\tfrac{4}{1-\nu^2}\bigl(-\tfrac{\phi_0}{k_\perp^2}\bigr)^{\frac{1}{2}}\right]^{2n}\frac{{\rm e}^{-\xi_+}}{\left(\tfrac{3}{2}\xi_+\right)^{{1}/{2}}}\left({\rm e}^{\Xi_+}-1\right)\left[1+{\cal O}\left(\tfrac{1}{\xi_+}\right)\right], \hfill
\label{eq:sumints1a}
\end{multline}
for $\Re(B^2k_{\perp}^2\phi_0)\geq0$, where we defined
\begin{align}
 \Xi_+\equiv-\tfrac{3}{2}\xi_+\tfrac{k_{\perp}^2}{\phi_0}\biggl[n_2\!\left(\tfrac{4}{1-\nu^2}\bigl(-\tfrac{\phi_0}{k_\perp^2}\bigr)^{\frac{1}{2}}\right)
-\tfrac{1-\nu^2}{4}+\tfrac{1}{3}\tfrac{\phi_0}{k_\perp^2}\biggr]. \label{eq:Xi}
\end{align}
Accounting only for the leading term at a fixed power of $s$ [cf. \Eqref{eq:ints2_2}] directly constrains \Eqref{eq:Xi} to yield trustworthy results for
\begin{equation}
 \Bigl|\frac{\phi_0}{k_\perp^2}\Bigr|\frac{1}{\xi}\ll1\,, \quad\text{and}\quad \Bigr|\frac{\phi_0}{k_\perp^2}\Bigr|\ll1\,.  \label{eq:conds}
\end{equation}
The first condition in \Eqref{eq:conds} is obtained by demanding the subleading contribution in \Eqref{eq:ints2_2} for $n\to n+1$, $j$ fixed, to be substantially smaller than the leading contribution for $n,j$ fixed.
The latter follows analogously by requiring \Eqref{eq:ints2_2} for $j\to j+1$, $n$ fixed, to be substantially smaller than the leading contribution for $n,j$ fixed.
As we have already limited ourselves to $\xi^{-1}\ll1$ from the outset of this section [cf. \eqref{eq:xitoinfty}], the only new condition to be fulfilled is $|\phi_0/k_\perp^2|\ll1$.

Recall that the radius of convergence of the series representation of $n_2(z)$, \Eqref{eq:Ni}, guarantees the above representation to make sense for at least $\left|\bigr(\tfrac{4}{1-\nu^2}\bigl)^2\tfrac{\phi_0}{k_\perp^2}\right|<\pi^2$. 
Sticking to the same assumptions as above, we analogously write
\begin{multline}
\sum_{j=1}^{\infty}\frac{1}{j!}\,\Im\int_{0}^{\infty}\frac{{\rm d} s}{s}\,{\rm e}^{-{\rm i}\phi_0s - {\rm i} k_{\perp}^2\frac{(1-\nu^2)^2}{48}(eB)^2s^3}(-{\rm i}k_{\perp}^2s)^{j}z^{2(n+2j)}\biggl(\sum_{l=2}^{\infty}n_2^{(2l)}z^{2(l-2)}\!\biggr)^j \\
\quad\quad=-{\rm sign}(B^2k_\perp^2)\sqrt{\tfrac{\pi}{2}}\left[\tfrac{4}{1-\nu^2}\bigl(-\tfrac{\phi_0}{k_\perp^2}\bigr)^{\frac{1}{2}}\right]^{2n}\frac{1}{\bigl(\tfrac{3}{2}\xi_-\bigr)^{1/2}}
\sum_{j=1}^{\infty}\frac{1}{j!}\,\left[-\tfrac{3}{2}\xi_-\tfrac{k_\perp^2}{\phi_0}
\biggl(\sum_{l=2}^{\infty}n_2^{(2l)}\left[\tfrac{4}{1-\nu^2}\bigl(-\tfrac{\phi_0}{k_\perp^2}\bigr)^{\frac{1}{2}}\right]^{2l}\biggr)\right]^j \hfill \\
\hspace*{8cm}\times\Bigl(\cos\tfrac{5\pi j}{2}+\sin\tfrac{5\pi j}{2}\Bigr)
\left(\cos\xi_--(-1)^j\sin\xi_-\right)\\
\quad\quad={\rm sign}(B^2k_\perp^2)\sqrt{\tfrac{\pi}{2}}\left[\tfrac{4}{1-\nu^2}\bigl(-\tfrac{\phi_0}{k_\perp^2}\bigr)^{\frac{1}{2}}\right]^{2n}\frac{\cos\xi_-\left(1-\cos\Xi_- -\sin\Xi_-\right)-\sin\xi_-\left(1-\cos\Xi_-+\sin\Xi_-\right)}{\bigl(\tfrac{3}{2}\xi_-\bigr)^{1/2}}\,, \hfill
\label{eq:sumints1b}
\end{multline}
for $\Re(B^2k_{\perp}^2\phi_0)\leq0$, where we introduced
\begin{align}
 \Xi_-&\equiv-\tfrac{3}{2}\xi_-\tfrac{k_{\perp}^2}{\phi_0}\biggl[n_2\!\left(\tfrac{4}{1-\nu^2}\bigl(-\tfrac{\phi_0}{k_\perp^2}\bigr)^{\frac{1}{2}}\right)
-\tfrac{1-\nu^2}{4}+\tfrac{1}{3}\tfrac{\phi_0}{k_\perp^2}\biggr]. \label{eq:tildeXi}
\end{align}
The neglected terms in \Eqref{eq:sumints1b} can be inferred from \Eqref{eq:fehler2}. Also \Eqref{eq:sumints1b} is limited to yield trustworthy results only for $|\phi_0/k_\perp^2|\ll1$; cf. \Eqref{eq:conds}.
Employing exactly the same reasoning for the sum over $n$ also -- thereby reverting to the original representations of the scalar functions in \Eqref{eq:scalarfcts_B} --,
with the help of Eqs.~\eqref{eq:ints2_1}-\eqref{eq:ints2_2b},~\eqref{eq:sumints1a} and \eqref{eq:sumints1b}, the imaginary part of \Eqref{eq:PI_tsai} can eventually be written  in a very compact form. For $\Re(B^2k_{\perp}^2\phi_0)\geq0$ we obtain
\begin{multline}
\Im\left\{\!\!
 \begin{array}{c}
 \pi_{\parallel}\\
 \pi_{\perp}\\
 \pi_{0}
 \end{array}\!\!
\right\}
=\frac{\alpha}{4\sqrt{\pi}}\frac{{\rm sign}(B^2k_\perp^2)}{\left(\tfrac{3}{2}\xi_+\right)^{{1}/{2}}}\Biggl\{
k^2\left[\frac{\nu^2(3-\nu^2)}{6}\left(\frac{2}{3}\frac{4}{1-\nu^2}-\frac{k^2}{\phi_0}\right)\tfrac{3}{2}\xi_+\,-(1-\nu^2)\right] \\
+\left\{
 \begin{array}{c}
 k_{\parallel}^2 N_1\left(\tfrac{4}{1-\nu^2}\bigl(-\tfrac{\phi_0}{k_\perp^2}\bigr)^{\frac{1}{2}}\right) + k_{\perp}^2N_0\left(\tfrac{4}{1-\nu^2}\bigl(-\tfrac{\phi_0}{k_\perp^2}\bigr)^{\frac{1}{2}}\right) \\
 k_{\parallel}^2N_0\left(\tfrac{4}{1-\nu^2}\bigl(-\tfrac{\phi_0}{k_\perp^2}\bigr)^{\frac{1}{2}}\right) + k_{\perp}^2 N_2\left(\tfrac{4}{1-\nu^2}\bigl(-\tfrac{\phi_0}{k_\perp^2}\bigr)^{\frac{1}{2}}\right) \\
 k^2N_0\left(\tfrac{4}{1-\nu^2}\bigl(-\tfrac{\phi_0}{k_\perp^2}\bigr)^{\frac{1}{2}}\right)
 \end{array}
\right\}{\rm e}^{\Xi_+}
 \Biggr\}\,{\rm e}^{-\xi_+}\left[1+{\cal O}\left(\tfrac{1}{\xi_+}\right)\right], \label{eq:Im1durchxi+}
\end{multline}
and for $\Re(B^2k_{\perp}^2\phi_0)\leq0$,
\begin{multline}
\Im\left\{\!\!
 \begin{array}{c}
 \pi_{\parallel}\\
 \pi_{\perp}\\
 \pi_{0}
 \end{array}\!\!
\right\}
=-\frac{\alpha}{2\sqrt{2\pi}}\frac{{\rm sign}(B^2k_\perp^2)}{\left(\tfrac{3}{2}\xi_+\right)^{{1}/{2}}}\Biggl\{k^2\Biggl[\frac{\nu^2(3-\nu^2)}{6}\left(\frac{2}{3}\frac{4}{1-\nu^2}-\frac{k^2}{\phi_0}\right)
\tfrac{3}{2}\xi_-\left(\cos\xi_-+\sin\xi_-\right) \\
-(1-\nu^2)(\cos\xi_--\sin\xi_-)\Biggr] \\
+\left\{
 \begin{array}{c}
 k_{\parallel}^2 N_{1}\left(\tfrac{4}{1-\nu^2}\bigl(-\tfrac{\phi_0}{k_\perp^2}\bigr)^{\frac{1}{2}}\right) + k_{\perp}^2N_0\left(\tfrac{4}{1-\nu^2}\bigl(-\tfrac{\phi_0}{k_\perp^2}\bigr)^{\frac{1}{2}}\right) \\
 k_{\parallel}^2N_0\left(\tfrac{4}{1-\nu^2}\bigl(-\tfrac{\phi_0}{k_\perp^2}\bigr)^{\frac{1}{2}}\right) + k_{\perp}^2 N_{2}\left(\tfrac{4}{1-\nu^2}\bigl(-\tfrac{\phi_0}{k_\perp^2}\bigr)^{\frac{1}{2}}\right) \\
 k^2N_0\left(\tfrac{4}{1-\nu^2}\bigl(-\tfrac{\phi_0}{k_\perp^2}\bigr)^{\frac{1}{2}}\right)
 \end{array}
\right\}\left[\cos(\xi_--\Xi_-)-\sin(\xi_--\Xi_-)\right]\Biggr\}. \label{eq:Im1durchxi-}
\end{multline}
The neglected terms in \Eqref{eq:Im1durchxi-} can again be inferred from \Eqref{eq:fehler2}.
Equations~\eqref{eq:Im1durchxi+} and \eqref{eq:Im1durchxi-} constitute the full analytical expression for the imaginary part of the photon polarization tensor at leading order in a $1/\xi$ expansion and for $|\phi_0/k_\perp^2|\ll1$ as well as $\left|\bigr(\tfrac{4}{1-\nu^2}\bigl)^2\tfrac{\phi_0}{k_\perp^2}\right|<\pi^2$. The latter condition can also be written as
\begin{equation}
 |u|\left|u+\frac{k^2}{4m^2}\right|<\left(\frac{\pi}{2}\right)^2\frac{|k_\perp^2|}{4m^2}\,, \label{eq:radofconv}
\end{equation}
where we introduced $u\equiv\frac{1}{1-\nu^2}$. Substituting $\nu$ for $u$, the integration over the finite $\nu$ interval translates into the integration over an infinite range, 
$\int_{-1}^1\frac{{\rm d}\nu}{2}\to\int_1^\infty\frac{{\rm d}u}{2u^{3/2}\sqrt{u-1}}$.
For given physical parameters the convergence criterion~\eqref{eq:radofconv} provides a condition on the real valued, positive integration parameter $u\geq1$:
The $u$ integration receives a contribution from within the radius of convergence as long as
\begin{equation}
 1\leq u<\sqrt{\left(\frac{\pi}{2}\right)^2\frac{|k_\perp^2|}{4m^2}+\left(\frac{k^2}{8m^2}\right)^2}-\frac{k^2}{8m^2}\,. \label{eq:u<}
\end{equation}
Larger values of $u$ give rise to contributions outside the radius of convergence.
As the evaluation of the photon polarization tensor manifestly requires an integration from $u=1$ to $u\to\infty$,
one might question whether Eqs.~\eqref{eq:Im1durchxi+} and \eqref{eq:Im1durchxi-} constitute trustworthy approximations to the photon polarization tensor.
However, if the main contribution to the integral stems from the $u$ range constrained by \Eqref{eq:u<}, reliable analytical results are still possible.

Taking into account the parameter integration over $\nu$ in Eqs.~\eqref{eq:Im1durchxi+} and \eqref{eq:Im1durchxi-} [cf. \Eqref{eq:Pipi}], for $\Re(B^2k_{\perp}^2\phi_0)\geq0$ we obtain
\begin{multline}
\Im\left\{\!\!
 \begin{array}{c}
 \Pi_{\parallel}\\
 \Pi_{\perp}\\
 \Pi_{0}
 \end{array}\!\!
\right\}
=\frac{{\rm sign}(B^2k_\perp^2)\alpha}{8\sqrt{\pi}}\int_1^\infty\frac{{\rm d}u}{u^{3/2}\sqrt{u-1}}\frac{1}{\left(\tfrac{3}{2}\xi_+\right)^{{1}/{2}}}
\Biggl\{
k^2\left[\frac{(u-1)(u+\frac{1}{2})}{3u^2}\left(\frac{8}{3}u-\frac{k^2}{\phi_0}\right)\tfrac{3}{2}\xi_+\,-\frac{1}{u}\right] \\
+\left\{
 \begin{array}{c}
 k_{\parallel}^2 N_1\left(4u\bigl(-\tfrac{\phi_0}{k_\perp^2}\bigr)^{\frac{1}{2}}\right) + k_{\perp}^2N_0\left(4u\bigl(-\tfrac{\phi_0}{k_\perp^2}\bigr)^{\frac{1}{2}}\right) \\
 k_{\parallel}^2N_0\left(4u\bigl(-\tfrac{\phi_0}{k_\perp^2}\bigr)^{\frac{1}{2}}\right) + k_{\perp}^2 N_2\left(4u\bigl(-\tfrac{\phi_0}{k_\perp^2}\bigr)^{\frac{1}{2}}\right) \\
 k^2N_0\left(4u\bigl(-\tfrac{\phi_0}{k_\perp^2}\bigr)^{\frac{1}{2}}\right)
 \end{array}
\right\}{\rm e}^{\Xi_+}
 \Biggr\}\,{\rm e}^{-\xi_+}\left[1+{\cal O}\left(\tfrac{1}{\xi_+}\right)\right], \label{eq:Im1durchxi+_2}
\end{multline}
and for $\Re(B^2k_{\perp}^2\phi_0)\leq0$,
\begin{multline}
\Im\left\{\!\!
 \begin{array}{c}
 \Pi_{\parallel}\\
 \Pi_{\perp}\\
 \Pi_{0}
 \end{array}\!\!
\right\}
=-\frac{{\rm sign}(B^2k_\perp^2)\alpha}{4\sqrt{2\pi}}\int_1^\infty\frac{{\rm d}u}{u^{3/2}\sqrt{u-1}}\frac{1}{\left(\tfrac{3}{2}\xi_-\right)^{{1}/{2}}}\\
\times\Biggl\{k^2\Biggl[\frac{(u-1)(u+\frac{1}{2})}{3u^2}\left(\frac{8}{3}u-\frac{k^2}{\phi_0}\right)
\tfrac{3}{2}\xi_-\left(\cos\xi_-+\sin\xi_-\right)
-\frac{\cos\xi_--\sin\xi_-}{u}\Biggr] \\
+\left\{
 \begin{array}{c}
 k_{\parallel}^2 N_{1}\left(4u\bigl(-\tfrac{\phi_0}{k_\perp^2}\bigr)^{\frac{1}{2}}\right) + k_{\perp}^2N_0\left(4u\bigl(-\tfrac{\phi_0}{k_\perp^2}\bigr)^{\frac{1}{2}}\right) \\
 k_{\parallel}^2N_0\left(4u\bigl(-\tfrac{\phi_0}{k_\perp^2}\bigr)^{\frac{1}{2}}\right) + k_{\perp}^2 N_{2}\left(4u\bigl(-\tfrac{\phi_0}{k_\perp^2}\bigr)^{\frac{1}{2}}\right) \\
 k^2N_0\left(4u\bigl(-\tfrac{\phi_0}{k_\perp^2}\bigr)^{\frac{1}{2}}\right)
 \end{array}
\right\}
\left[\cos(\xi_--\Xi_-)-\sin(\xi_--\Xi_-)\right]\Biggr\}. \label{eq:Im1durchxi-_2}
\end{multline}
The substitution $\nu\to u$ of course implies that also the parameter $\nu$ contained in the definitions of the scalar functions~\eqref{eq:scalarfcts_B} is expressed in terms of $u$, setting $\nu=(1-\frac{1}{u})^{1/2}$.
Accounting for the respective conditions on $\Re(B^2k_\perp^2\phi_0)$,
both $\xi_+$ in \Eqref{eq:Im1durchxi+_2} and $\xi_-$ in \Eqref{eq:Im1durchxi-_2} [cf. \Eqref{eq:xi!}] can
effectively be identified with the single parameter 
\begin{equation}
 \xi(u)=\frac{4}{3}u\frac{m^2}{|eB|}\frac{2m}{|k_\perp^2|^{1/2}}\left(1+\frac{k^2}{4m^2}\frac{1}{u}\right)^{3/2}
 =\frac{4}{3}u\frac{m^2}{|eB|}\frac{2m}{|k_\perp^2|^{1/2}}\sum_{n=0}^\infty\frac{\Gamma(\frac{5}{2})}{n!\,\Gamma(\frac{5}{2}-n)}\left(\frac{k^2}{4m^2}\frac{1}{u}\right)^n,
 \label{eq:xi_ohne}
\end{equation}
where --  in the second step -- we made use of Newton's generalized binomial theorem.

To allow for further analytical insights, we now explicitly limit ourselves to small $|k^2|\ll m^2$.
In this limit, the parameter $\xi$ scales as
\begin{equation}
 \xi(u)\approx\frac{4}{3}u\frac{m^2}{|eB|}\frac{2m}{|k_\perp^2|^{1/2}}\,, \label{eq:xi(u)approx}
\end{equation}
and condition~\eqref{eq:u<} effectively amounts to $1\leq u<\frac{\pi}{2}\bigl|\frac{k_\perp^2}{4m^2}\bigr|^{1/2}$.
Correspondingly, at the upper limit of the radius of convergence the parameter $\xi$ is then approximately given by
\begin{equation}
 \xi\left(\frac{\pi}{2}\Bigl|\frac{k_\perp^2}{4m^2}\Bigr|^{1/2}\right)\approx\frac{2\pi}{3}\frac{m^2}{|eB|}\,, \label{eq:xi(arg)}
\end{equation}
such that -- most obviously for $\Re(B^2k_\perp^2\phi_0)\geq0$, featuring an exponential suppression with increasing $u$ -- 
we may expect to obtain trustworthy results under the above constraints as long as $\frac{|eB|}{m^2}\ll1$.

In the next step we aim at carrying out the $u$ integration.
Therefore, recall that for $\Re(B^2k_\perp^2\phi_0)\geq0$ all contributions within the outermost curly brackets in \Eqref{eq:Im1durchxi+_2} -- most explicitly before formal resummation into trigonometric and additional exponential functions --
can be viewed in terms of an overall exponential factor $\sim\exp(-\xi_+)$ multiplying an infinite series in the parameter $u$ and its inverse.
Given that $\bigl(\frac{m^2}{|eB|}\frac{2m}{|k_\perp^2|^{1/2}}\bigr)\frac{|k^2|}{m^2}\ll1$, we can moreover expand the overall exponential factor as follows [cf. \Eqref{eq:xi_ohne}],
\begin{equation}
 {\rm e}^{-\xi_+}={\rm e}^{-\frac{4}{3}u\frac{m^2}{|eB|}\frac{2m}{|k_\perp^2|^{1/2}}}\sum_{l=0}^\infty\frac{1}{l!}\left[-\frac{4}{3}u\frac{m^2}{|eB|}\frac{2m}{|k_\perp^2|^{1/2}}\sum_{n=1}^\infty\frac{\Gamma(\frac{5}{2})}{n!\,\Gamma(\frac{5}{2}-n)}\left(\frac{k^2}{4m^2}\frac{1}{u}\right)^n\right]^l, \label{eq:expxi}
\end{equation}
such that the argument of the remaining exponential is linear in $u$. 
With the help of the following identity (formulae 3.383.4 and 9.232.1 of \cite{Gradshteyn})
\begin{equation}
 \int_1^\infty\frac{{\rm d}u}{u^2\sqrt{u-1}}\,u^\sigma\,{\rm e}^{-\beta u}=\sqrt{\pi}\,\beta^{-\frac{2\sigma-1}{4}}{\rm e}^{-\frac{\beta}{2}}\,{\rm W}_{\frac{2\sigma-3}{4},\frac{2\sigma-3}{4}}(\beta),
\end{equation}
valid for $\Re(\beta)>0$, the integrals over $u$ can be expressed in terms of the Whittaker hypergeometric function ${\rm W}_{.,.}(.)$, whose asymptotic expansion for large $|\beta|$, $|\arg\beta|<\pi$ reads (formula 9.227 of \cite{Gradshteyn})
\begin{equation}
 {\rm W}_{\frac{2\sigma-3}{4},\frac{2\sigma-3}{4}}(\beta)=\beta^{\frac{2\sigma-3}{4}}{\rm e}^{-\frac{\beta}{2}}\left[1+{\cal O}\bigl(\tfrac{1}{\beta}\bigr)\right].
\end{equation}
Hence, we have
\begin{equation}
 \int_1^\infty\frac{{\rm d}u}{u^2\sqrt{u-1}}\,u^\sigma\,{\rm e}^{-\beta u}=\sqrt{\frac{\pi}{\beta}}\,{\rm e}^{-\beta}\left[1+{\cal O}\bigl(\tfrac{1}{\beta}\bigr)\right]. \label{eq:assW}
\end{equation}
Most notably, the leading term of the asymptotic expansion in \Eqref{eq:assW} does not at all depend on the parameter $\sigma$.
Moreover, adopting \Eqref{eq:assW} to the calculations to be performed here, the parameter $\beta$ can be identified with $\frac{4}{3}\frac{m^2}{|eB|}\frac{2m}{|k_\perp^2|^{1/2}}$ [cf. \Eqref{eq:expxi}].
Correspondingly, the subleading terms in \Eqref{eq:assW} can be safely neglected:
They are of the same order as the subleading contributions to \Eqref{eq:Im1durchxi+_2}, already not explicitly taken into account.
Thus, for  $\bigl(\frac{m^2}{|eB|}\frac{2m}{|k_\perp^2|^{1/2}}\bigr)\frac{|k^2|}{m^2}\ll1$ the $u$ integral in \Eqref{eq:Im1durchxi+_2} can be performed by formally expanding the integrand to all orders in $u$, keeping only terms linear in $u$ in the exponential.
All other contributions are expanded to form polynomials in $u$ (cf. Appendix~\ref{app:seriesNn}). After performing the integration with the help of \Eqref{eq:assW} the result is resummed again.
This should be permissible if the main contribution to the integral stems from $2\frac{2m}{|k_\perp|}u<\pi$, or equivalently, $1\leq u<\frac{\pi}{2}\frac{|k_\perp|}{2m}$; cf. below \Eqref{eq:xi(u)approx}.
More practically, for \Eqref{eq:Im1durchxi+_2} this amounts to the following recipe: Replace 
\begin{equation}
 \int_1^\infty\frac{{\rm d}u}{u^2\sqrt{u-1}} \quad \to \quad \sqrt{\pi\frac{3}{4}\frac{|eB|}{m^2}\frac{|k_\perp^2|^{1/2}}{2m}}\,,
\end{equation}
and set $u\equiv1$ ($\leftrightarrow\nu=0$) in the remaining terms.
To keep notations compact we still write the result in terms of the scalar functions $N_i$ with $i\in\{0,1,2\}$ and $n_2$.
However, these expressions now have to be understood to be evaluated at $\nu=0$, i.e., here [cf. \Eqref{eq:scalarfcts_B}]
\begin{align}
 N_0(z)=\frac{z}{\sin z}\,,\quad\quad N_{1}(z)=z\cot z\,, \quad\quad N_{2}(z)&=\frac{2z\left(1 -\cos z\right)}{\sin^3z}\,, \quad\quad\text{and}\quad\quad n_{2}(z)=\frac{1-\cos{z}}{2z\sin{z}}\,.  \label{eq:scalarfcts_B_nu0}
\end{align}
As the additional constraint $\bigl(\frac{m^2}{|eB|}\frac{2m}{|k_\perp^2|^{1/2}}\bigr)\frac{|k^2|}{m^2}\ll1$, implies $\Re(\phi_0|_{u=1})>0$, for \Eqref{eq:Im1durchxi+_2} this procedure results in
\begin{multline}
\Im\left\{\!\!
 \begin{array}{c}
 \Pi_{\parallel}\\
 \Pi_{\perp}\\
 \Pi_{0}
 \end{array}\!\!
\right\}
=\frac{\alpha}{16}\sqrt{\frac{3}{2}}\frac{|eB|}{m^2}\frac{|k_\perp^2|^{1/2}}{2m}\left(1+\frac{k^2}{4m^2}\right)^{-3/4}\,{\rm e}^{-\frac{1}{3}\frac{\left(4m^2+k^2\right)^{3/2}}{|eB||k_\perp^2|^{1/2}}}\Biggl\{
-k^2 \\
+\left\{
 \begin{array}{c}
 k_{\parallel}^2 N_1\left(2\bigl(-\tfrac{4m^2-{\rm i}\epsilon+k^2}{k_\perp^2}\bigr)^{\frac{1}{2}}\right) + k_{\perp}^2N_0\left(2\bigl(-\tfrac{4m^2-{\rm i}\epsilon+k^2}{k_\perp^2}\bigr)^{\frac{1}{2}}\right) \\
 k_{\parallel}^2N_0\left(2\bigl(-\tfrac{4m^2-{\rm i}\epsilon+k^2}{k_\perp^2}\bigr)^{\frac{1}{2}}\right) + k_{\perp}^2 N_2\left(2\bigl(-\tfrac{4m^2-{\rm i}\epsilon+k^2}{k_\perp^2}\bigr)^{\frac{1}{2}}\right) \\
 k^2N_0\left(2\bigl(-\tfrac{4m^2-{\rm i}\epsilon+k^2}{k_\perp^2}\bigr)^{\frac{1}{2}}\right)
 \end{array}
\right\}\\
\times{\rm e}^{-2\frac{k_{\perp}^2}{|eB|}\frac{\left(4m^2+k^2\right)^{1/2}}{|k_\perp^2|^{1/2}}\biggl[n_2\!\left(2\bigl(-\tfrac{4m^2-{\rm i}\epsilon+k^2}{k_\perp^2}\bigr)^{\frac{1}{2}}\right)
-\frac{1}{4}+\frac{1}{12}\frac{4m^2+k^2}{k_\perp^2}\biggr]}
 \Biggr\}\left[1+{\cal O}\left(\tfrac{|eB|}{m^2}\tfrac{|k_\perp^2|^{1/2}}{2m}\right)\right], \label{eq:Im1durchxi+_2ausint}
\end{multline}
now applicable for $B^2k_{\perp}^2\geq0$.
An analogous expression can be derived from \Eqref{eq:Im1durchxi-_2}, adopting the same reasoning as above, and employing the following identities,
\begin{align}
 \int_1^\infty\frac{{\rm d}u}{u^2\sqrt{u-1}}\,u^\sigma\,{\rm e}^{-\epsilon u^\kappa}\cos(bu)
 &=\sqrt{\frac{\pi}{|b|}}\,\frac{\cos|b|-\sin|b|}{\sqrt{2}}\left[1+{\cal O}\bigl(\tfrac{1}{b}\bigr)\right]\,, \\
 \int_1^\infty\frac{{\rm d}u}{u^2\sqrt{u-1}}\,u^\sigma\,{\rm e}^{-\epsilon u^\kappa}\sin(bu)
 &=\sqrt{\frac{\pi}{|b|}}\,{\rm sign}(b)\frac{\sin|b| + \cos|b|}{\sqrt{2}}\left[1+{\cal O}\bigl(\tfrac{1}{b}\bigr)\right]\,,
\end{align}
with $b\in\mathbb{R}$, $\kappa>0$ and $\epsilon\to0^+$, which can be derived straightforwardly from \Eqref{eq:assW}.
The presence of the convergence ensuring exponential factor is most clearly visible in the second expression in the first line of \Eqref{eq:tsaielm},
where an overall term $\sim{\rm e}^{-\epsilon(1-\nu^2)^{-2/3}}={\rm e}^{-\epsilon u^{2/3}}$ can be factored out.
Hence, \Eqref{eq:Im1durchxi-_2} gives rise to
\begin{multline}
\Im\left\{\!\!
 \begin{array}{c}
 \Pi_{\parallel}\\
 \Pi_{\perp}\\
 \Pi_{0}
 \end{array}\!\!
\right\}
=-\frac{\alpha}{8}\sqrt{\frac{3}{2}}\frac{|eB|}{m^2}\frac{|k_\perp^2|^{1/2}}{2m}\left(1+\frac{k^2}{4m^2}\right)^{-3/4}\Biggl\{-k^2\,\sin\left(\frac{1}{3}\frac{\left(4m^2+k^2\right)^{3/2}}{|eB||k_\perp^2|^{1/2}}\right)\\
+\left\{
 \begin{array}{c}
 k_{\parallel}^2 N_{1}\left(2\bigl(-\tfrac{4m^2-{\rm i}\epsilon+k^2}{k_\perp^2}\bigr)^{\frac{1}{2}}\right) + k_{\perp}^2N_0\left(2\bigl(-\tfrac{4m^2-{\rm i}\epsilon+k^2}{k_\perp^2}\bigr)^{\frac{1}{2}}\right) \\
 k_{\parallel}^2N_0\left(2\bigl(-\tfrac{4m^2-{\rm i}\epsilon+k^2}{k_\perp^2}\bigr)^{\frac{1}{2}}\right) + k_{\perp}^2 N_{2}\left(2\bigl(-\tfrac{4m^2-{\rm i}\epsilon+k^2}{k_\perp^2}\bigr)^{\frac{1}{2}}\right) \\
 k^2N_0\left(2\bigl(-\tfrac{4m^2-{\rm i}\epsilon+k^2}{k_\perp^2}\bigr)^{\frac{1}{2}}\right)
 \end{array}
\right\}\\
\times\sin\left(\frac{1}{3}\frac{\left(4m^2+k^2\right)^{3/2}}{|eB||k_\perp^2|^{1/2}}+2\frac{k_{\perp}^2}{|eB|}\frac{\left(4m^2+k^2\right)^{1/2}}{|k_\perp^2|^{1/2}}\biggl[n_2\!\left(2\bigl(-\tfrac{4m^2-{\rm i}\epsilon+k^2}{k_\perp^2}\bigr)^{\frac{1}{2}}\right)
-\tfrac{1}{4}+\tfrac{1}{12}\tfrac{4m^2+k^2}{k_\perp^2}\biggr]\right)\Biggr\}, \label{eq:Im1durchxi-_2ausint}
\end{multline}
valid for $B^2k_{\perp}^2\leq0$.
Eqs.~\eqref{eq:Im1durchxi+_2ausint} and \eqref{eq:Im1durchxi-_2ausint} constitute our result for the imaginary part of the photon polarization tensor in the parameter regime
\begin{equation}
 \frac{|k^2|}{m^2}\ll\frac{ef}{m^2}\frac{|k_\perp^2|^{1/2}}{2m}\ll 1\,, \quad\text{and}\quad \frac{2m}{|k_\perp^2|^{1/2}}\ll 1\,. \label{eq:valin}
\end{equation}
Let us stress again that in order to derive the above results it was essential to systematically keep track of the orders of all contributions, both the neglected ones and the ones taken into account explicitly.

Finally, we turn to on-the-light-cone dynamics, where the expressions are less complicated. As
\begin{equation}
 k^2=0 \quad\leftrightarrow\quad k_{\perp}^2=-k_{\parallel}^2=\omega^2\sin^2\theta\geq0\,, \label{eq:olcd}
\end{equation}
implies $B^2k_{\perp}^2\geq0$ and $-E^2k_{\parallel}^2\geq0$,
in this limit the entire information for both the magnetic and electric field cases is contained in \Eqref{eq:Im1durchxi+_2ausint}.
Moreover, note that
\begin{equation}
 \left.\bigl(-\tfrac{4m^2-{\rm i\epsilon+k^2}}{k_\perp^2}\bigr)^{\frac{1}{2}}\right|_{k^2=0}=\tfrac{2m}{|k_\perp|}
 \begin{cases}
  {\rm i} \quad&\text{for}\quad k_\perp^2>0\,, \\
  1 \quad&\text{for}\quad k_\perp^2<0\,.
 \end{cases}
\end{equation}
Focusing on $k^2=0$, the $p=0$ component in \Eqref{eq:Im1durchxi+_2ausint} vanishes, $\Im(\Pi_0)|_{k^2=0}=0$, while
the imaginary part of the other components, $p\in\{\parallel,\perp\}$, of the photon polarization tensor can finally be written as
\begin{multline}
\left.\Im\left\{\!\!
 \begin{array}{c}
 \Pi_{\parallel}\\
 \Pi_{\perp}
 \end{array}\!\!
\right\}\right|_{k^2=0}
=-\frac{\alpha}{8}\sqrt{\frac{3}{2}}\,\frac{eB}{m^2}\,k_{\perp}^2
\tanh\bigl(\tfrac{2m}{|k_\perp|}\bigr)
\left\{\!\!
 \begin{array}{c}
   1\\
   \frac{1}{2\cosh^2\bigl(\frac{2m}{|k_\perp|}\bigr)}
 \end{array}\!\!
\right\} \\
\times{\rm e}^{-2\,\frac{m^2}{eB}\frac{2m}{|k_\perp|}+\frac{m|k_\perp|}{eB}\left(1-\frac{|k_\perp|}{2m}\tanh\bigl(\tfrac{2m}{|k_\perp|}\bigr)\right)}
\left[1+{\cal O}\left(\tfrac{eB}{m^2}\tfrac{|k_\perp|}{2m}\right)\right] \label{eq:Im1durchxi+_Bk^2=0_uausint}
\end{multline}
in case of a magnetic field, and
\begin{multline}
\left.\Im\left\{\!\!
 \begin{array}{c}
 \Pi_{\perp}\\
 \Pi_{\parallel}
 \end{array}\!\!
\right\}\right|_{k^2=0}
=-\frac{\alpha}{8}\sqrt{\frac{3}{2}}\,\frac{eE}{m^2}\,k_{\perp}^2\tan\bigl(\tfrac{2m}{|k_\perp|}\bigr)
\left\{\!\!
 \begin{array}{c}
   1 \\
   \frac{1}{2\cos^2\bigl(\tfrac{2m}{|k_\perp|}\bigr)}
 \end{array}\!\!
\right\} \\
\times{\rm e}^{-2\,\frac{m^2}{eE}\frac{2m}{|k_\perp|}-\frac{m|k_\perp|}{eE}\left(1-\frac{|k_\perp|}{2m}\tan\bigl(\tfrac{2m}{|k_\perp|}\bigr)\right)}
\left[1+{\cal O}\left(\tfrac{eE}{m^2}\tfrac{|k_\perp|}{2m}\right)\right] \label{eq:Im1durchxi+_Ek^2=0_uausint}
\end{multline}
for an electric field. To arrive at these expressions we have employed double and half angle formulas. 
The terms written explicitly in Eqs.~\eqref{eq:Im1durchxi+_Bk^2=0_uausint} and \eqref{eq:Im1durchxi+_Ek^2=0_uausint} constitute 
the full result at leading order in the $1/\xi$ expansion for on-the-light-cone dynamics. They are expected to grant access to the regime characterized by [cf. \Eqref{eq:valin}]
\begin{equation}
 \frac{ef}{m^2}\frac{|k_\perp|}{2m}\ll1\,, \quad\text{and} \quad \frac{2m}{|k_\perp|}\ll1\,, \label{eq:valin_k^2=0}
\end{equation}
or equivalently
\begin{equation}
  \frac{ef}{m^2}\ll\frac{2m}{|k_\perp|}\ll1 \,,
\end{equation}
i.e., for {\it very weak fields}, but a {\it large} transversal {\it momentum}.
The latter condition in \Eqref{eq:valin_k^2=0} implies an explicit restriction to kinematics allowing for the creation of real electron positron pairs in a magnetic field from on-the-light-cone photons, i.e., photons fulfilling $k^2=0$: To see this, it is illustrative to perform a Lorentz transformation along the direction of the magnetic field, 
such that the parallel momentum component $\vec k_{\parallel}$ of the photons becomes zero,
i.e., $\vec k_{\parallel}'=0$. Vector components orthogonal to the direction of the Lorentz boost remain unaltered.
In this particular reference system (denoted by $'$), the particles are still subject to a homogeneous external magnetic field. This comes about as a Lorentz boost
in the direction of the magnetic field does not induce an electric field.
However, the light-cone condition for photons now reads $k^2=\vec k_{\perp}^2-(k'^0)^2=0$,
and the on-set condition for real pair creation becomes
\begin{equation}
 k'^0=|\vec k_{\perp}|=2m \quad\leftrightarrow\quad \omega\sin\theta=2m \,.
\label{eq:pair}
\end{equation}
Equations~\eqref{eq:Im1durchxi+_Bk^2=0_uausint} and \eqref{eq:Im1durchxi+_Ek^2=0_uausint} are intimately related to the absorption coefficient $\kappa_p$ of on-the-light-cone photons, polarized in mode $p=\{\parallel,\perp\}$ and propagating in a magnetic or electric field, respectively, given by \cite{Tsai:1974fa}
\begin{equation}
 \kappa_p=-\frac{1}{\omega}\,\Im(\Pi_p)|_{k^2=0}\,. \label{eq:kappa_p}
\end{equation}
Throughout this paper we neglect any backreaction effects due to the production of real electron positron pairs; for some general considerations in this direction, cf.  \cite{Fradkin:1981sc,Barashev:1985zm,Gavrilov:1987,Fradkin:1991zq,Gavrilov:2007hq}.

With the help of the following identities,
\begin{equation}
 \frac{ef}{m^2}=\left(\frac{ef}{m^2}\frac{|k_\perp|}{2m}\right)\frac{2m}{|k_\perp|}\,, \quad\text{and}\quad \frac{ef}{m|k_\perp|}=\frac{1}{2}\left(\frac{ef}{m^2}\frac{|k_\perp|}{2m}\right)\left(\frac{2m}{|k_\perp|}\right)^2, \label{eq:h_ids}
\end{equation}
Eqs.~\eqref{eq:Im1durchxi+_Bk^2=0_uausint} and \eqref{eq:Im1durchxi+_Ek^2=0_uausint} -- besides the overall factor of $k_\perp^2$ -- can be entirely written in terms of the two independent parameters $\frac{ef}{m^2}\frac{|k_\perp|}{2m}$ and $\frac{2m}{|k_\perp|}$.
Using this replacement in Eqs.~\eqref{eq:Im1durchxi+_Bk^2=0_uausint} and \eqref{eq:Im1durchxi+_Ek^2=0_uausint} and
performing an expansion in powers of $\frac{2m}{|k_\perp|}$, we straightforwardly obtain
\begin{multline}
\left.\Im\left\{\!\!
 \begin{array}{c}
 \Pi_{\parallel}\\
 \Pi_{\perp}
 \end{array}\!\!
\right\}\right|_{k^2=0}
=-\frac{\alpha}{4}\sqrt{\frac{3}{2}}\,eB\,\frac{|k_\perp|}{2m}
\left\{\!\!
 \begin{array}{c}
   2\\
   1
 \end{array}\!\!
\right\}
{\rm e}^{-\tfrac{4}{3}\tfrac{m^2}{eB}\tfrac{2m}{|k_\perp|}}
 \left\{1+{\cal O}\left(\bigl(\tfrac{2m}{|k_\perp|}\bigr)^2\right)\left[1+{\cal O}\left(\tfrac{4}{3}\tfrac{m^2}{eB}\tfrac{2m}{|k_\perp|}\right)\right]+{\cal O}\left(\tfrac{3}{4}\tfrac{eB}{m^2}\tfrac{|k_\perp|}{2m}\right)\right\} \label{eq:Im1durchxi+_Bk^2=0_TsaiErber}
\end{multline}
in case of a magnetic field, and
\begin{multline}
\left.\Im\left\{\!\!
 \begin{array}{c}
 \Pi_{\perp}\\
 \Pi_{\parallel}
 \end{array}\!\!
\right\}\right|_{k^2=0}
=-\frac{\alpha}{4}\sqrt{\frac{3}{2}}\,eE\,\frac{|k_\perp|}{2m}
\left\{\!\!
 \begin{array}{c}
   2\\
   1
 \end{array}\!\!
\right\}
{\rm e}^{-\tfrac{4}{3}\tfrac{m^2}{eE}\tfrac{2m}{|k_\perp|}}
 \left\{1+{\cal O}\left(\bigl(\tfrac{2m}{|k_\perp|}\bigr)^2\right)\left[1+{\cal O}\left(\tfrac{4}{3}\tfrac{m^2}{eE}\tfrac{2m}{|k_\perp|}\right)\right]+{\cal O}\left(\tfrac{3}{4}\tfrac{eE}{m^2}\tfrac{|k_\perp|}{2m}\right)\right\} \label{eq:Im1durchxi+_Ek^2=0_TsaiErber}
\end{multline}
for an electric field.
By inspection of the subleading terms, we infer that Eqs.~\eqref{eq:Im1durchxi+_Bk^2=0_TsaiErber} and \eqref{eq:Im1durchxi+_Ek^2=0_TsaiErber} should yield trustworthy results, given the following hierarchy of scales,
\begin{equation}
 \frac{4m^2}{|k_\perp^2|}\ll\frac{ef}{m^2}\frac{|k_\perp|}{2m}\ll 1\,. \label{eq:trueTsaiErberregime}
\end{equation} 
The terms written explicitly in \Eqref{eq:Im1durchxi+_Bk^2=0_TsaiErber} agree with an expression derived for homogeneous magnetic fields by Tsai and Erber \cite{Tsai:1974fa}; their Eq. (59b).
If, e.g., $\frac{4m^2}{|k_{\perp}^2|}\gtrsim\frac{eB}{m^2}\frac{|k_\perp|}{2m}$, inequality~\eqref{eq:trueTsaiErberregime} is violated (cf. also \cite{Baier:2007dw}) and one should rather work with \Eqref{eq:Im1durchxi+_Bk^2=0_uausint} derived in this work.

\begin{figure}[h]
\center
\includegraphics[width=0.65\textwidth]{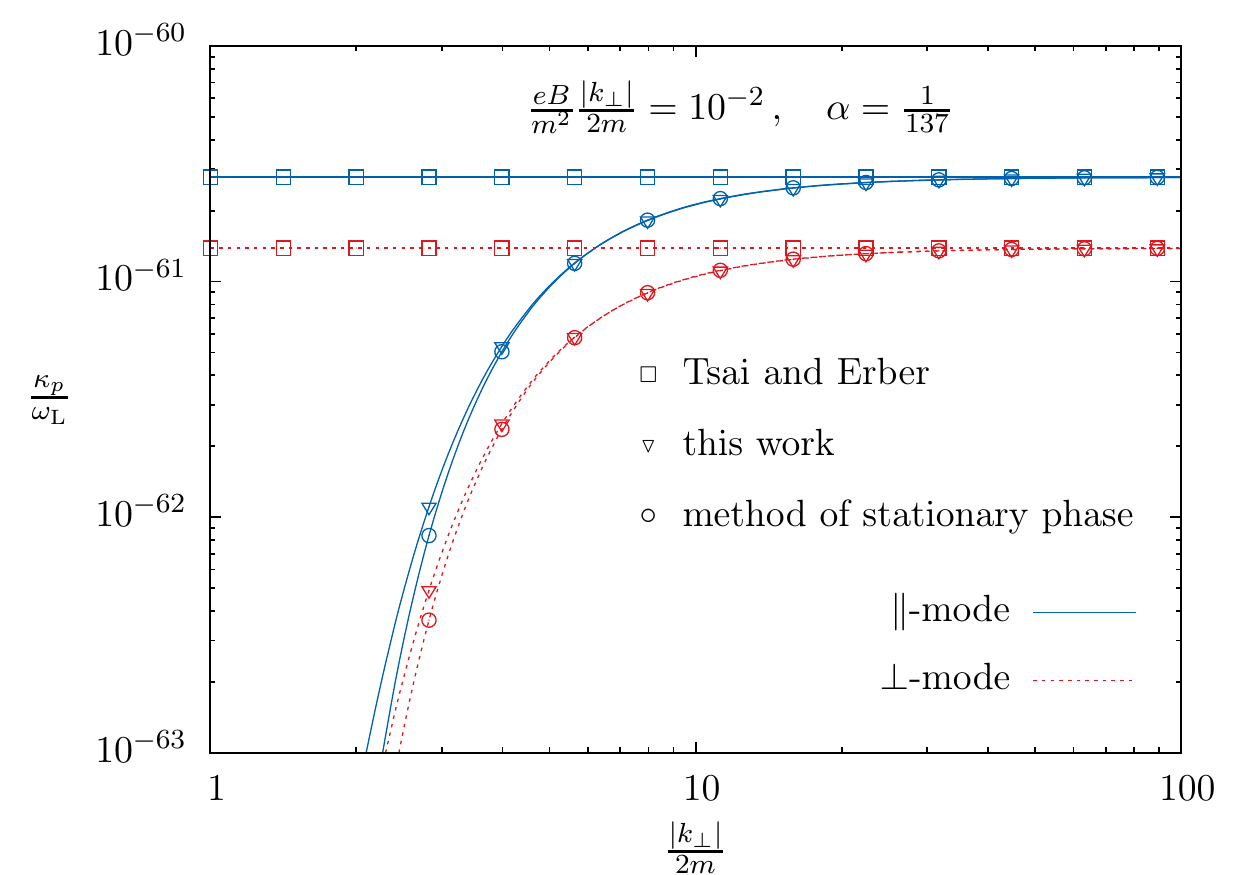} 
\caption{Photon absorption coefficient $\kappa_{\parallel,\perp}$ in a magnetic field in units of the Larmor frequency $\omega_{\rm L}=\frac{eB}{m}$
plotted as a function of the dimensionless ratio $\frac{|k_\perp|}{2m}$; the photon propagation direction is assumed to be orthogonal to the magnetic field, i.e., $\theta=\frac{\pi}{2}$, such that $|k_\perp|=\omega$:
Results as obtained from \Eqref{eq:Im1durchxi+_Bk^2=0_TsaiErber} originally derived by Tsai and Erber (curves marked with squares),
the new approximation~\eqref{eq:Im1durchxi+_Bk^2=0_uausint} devised in this work (marked with triangles), and the method of stationary phase~\eqref{eq:methstat_B} (marked with circles).
While -- for the explicit parameters adopted here -- the approximation of Tsai and Erber is only applicable for large values of $\frac{|k_\perp|}{2m}\gtrsim25$,
the results of our new approximation are in good agreement with those of the method of stationary phase for $\frac{|k_\perp|}{2m}\gtrsim4$.}
\label{fig:ImB}
\end{figure}

\begin{figure}[h]
\center
\includegraphics[width=0.65\textwidth]{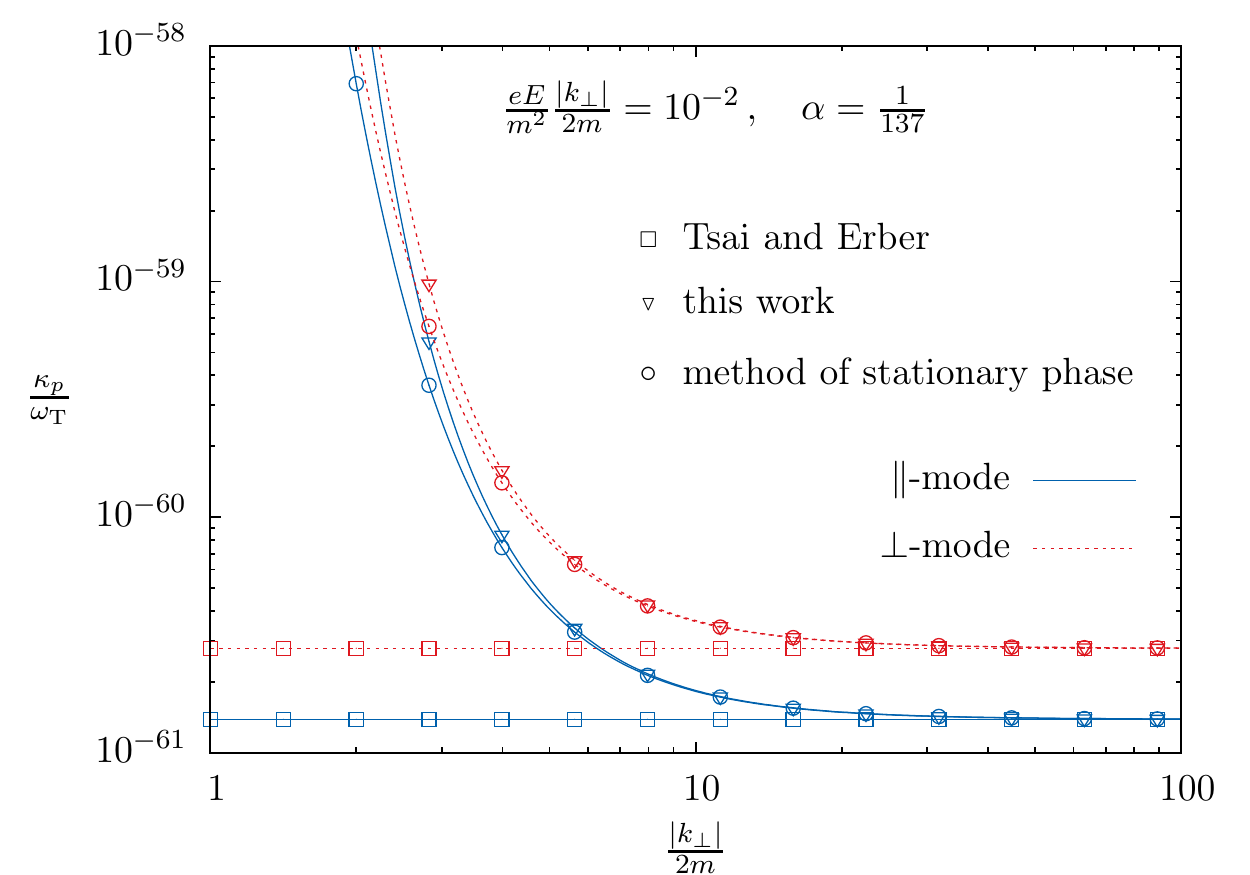} 
\caption{Absorption coefficient $\kappa_{\parallel,\perp}$ in an electric field in units of the tunneling frequency $\omega_{\rm T}=\frac{eE}{m}$ plotted as a function of the dimensionless ratio $\frac{|k_\perp|}{2m}$;
the photon propagation direction is assumed to be orthogonal to the electric field, i.e., $\theta=\frac{\pi}{2}$, such that $|k_\perp|=\omega$:
Results as obtained from \Eqref{eq:Im1durchxi+_Ek^2=0_TsaiErber} (curves marked with squares),
the new approximation~\eqref{eq:Im1durchxi+_Ek^2=0_uausint} devised in this work (marked with triangles), and the method of stationary phase~\eqref{eq:methstat_E} (marked with circles).
For the explicit parameters adopted here, \Eqref{eq:Im1durchxi+_Ek^2=0_TsaiErber} yields trustworthy results only for large values of $\frac{|k_\perp|}{2m}\gtrsim25$.
In full agreement with Fig.~\ref{fig:ImB}, the results of our new approximation are compatible with those of the method of stationary phase for $\frac{|k_\perp|}{2m}\gtrsim4$.}
\label{fig:ImE}
\end{figure}

Equation~\eqref{eq:h_ids} implies that the dimensionless ratios
\begin{itemize}
 \item $\frac{\kappa_p}{\omega_{\rm L}}$ for a magnetic field, with Larmor frequency $\omega_{\rm L}=\frac{eB}{m}$, and
 \item $\frac{\kappa_p}{\omega_{\rm T}}$ for an electric field, with ``tunneling frequency'' $\omega_{\rm T}=\frac{eE}{m}$,
\end{itemize}
besides an overall factor of $\sin\theta$, are fully governed by just two independent physical parameters, e.g.,
$\frac{|k_\perp|}{2m}$ and the combined parameter $\frac{ef}{m^2}\frac{|k_\perp|}{2m}$.
In Figs.~\ref{fig:ImB} and \ref{fig:ImE} we plot these quantities for a fixed value of $\frac{ef}{m^2}\frac{|k_\perp|}{2m}$ as a function of the parameter $\frac{|k_\perp|}{2m}$. The photon propagation direction is assumed to be orthogonal to the external field, i.e., $\theta=\frac{\pi}{2}$. 
Alternatively, the horizontal axis in Figs.~\ref{fig:ImB} and \ref{fig:ImE} can be rescaled into the dimensionless field strength, employing $\frac{ef}{m^2}=(\frac{ef}{m^2}\frac{|k_\perp|}{2m})(\frac{|k_\perp|}{2m})^{-1}$.
The curves marked with squares are obtained from Eqs.~\eqref{eq:Im1durchxi+_Bk^2=0_TsaiErber} and \eqref{eq:Im1durchxi+_Ek^2=0_TsaiErber}. 
For fixed values of the combination $\frac{ef}{m^2}\frac{|k_\perp|}{2m}$, they turn out to be completely independent of $\frac{|k_\perp|}{2m}$.
For $f=B=E$, \Eqref{eq:Im1durchxi+_Bk^2=0_TsaiErber} can be mapped onto \Eqref{eq:Im1durchxi+_Ek^2=0_TsaiErber} by exchanging the polarization components, $\Pi_\parallel\leftrightarrow\Pi_\perp$. Hence, apart from this change,
the curves marked with squares in Figs.~\ref{fig:ImB} and \ref{fig:ImE} fall on top of each other.
The curves marked with triangles in Fig.~\ref{fig:ImB} are obtained from \Eqref{eq:Im1durchxi+_Bk^2=0_uausint} and those marked with triangles in Fig.~\ref{fig:ImE} from \Eqref{eq:Im1durchxi+_Ek^2=0_uausint}.
For comparison, we additionally depict the results obtained with the method of stationary phase \cite{Baier:2007dw,Baier:2009it,Dunne:2009gi}, also expected to yield reliable results in the parameter regime under consideration (cf. Appendix~\ref{sec:methstat}).
Equations~\eqref{eq:methstat_B} and \eqref{eq:methstat_E} result in the curves marked with circles.

As expected, for a given polarization mode and for large values of $\frac{|k_\perp|}{2m}$, compatible with \Eqref{eq:trueTsaiErberregime}, the curves in Figs.~\ref{fig:ImB} and \ref{fig:ImE} basically fall on top of each other;
for the explicit value of the combined parameter $\frac{ef}{m^2}\frac{|k_\perp|}{2m}=10^{-2}$ adopted in Figs.~\ref{fig:ImB} and \ref{fig:ImE}, \Eqref{eq:trueTsaiErberregime} results in the condition $\frac{|k_\perp|}{2m}\gg10$.
For $\frac{|k_\perp|}{2m}\gtrsim4$ the curves marked with triangles corresponding to our new approximation
are in good agreement with those marked with circles obtained by the method of stationary phase, derived in a completely different way. 
This is perfectly compatible with the regime of applicability of our new approximation, $\frac{ef}{m^2}\frac{|k_\perp|}{2m}\ll1$ and $\frac{2m}{|k_\perp|}\ll1$.
Outside their range of applicability, i.e., for $\frac{4m^2}{|k_\perp^2|}\gtrsim\frac{ef}{m^2}\frac{|k_\perp|}{2m}$, the Tsai and Erber type approximations~\eqref{eq:Im1durchxi+_Bk^2=0_TsaiErber} and \eqref{eq:Im1durchxi+_Ek^2=0_TsaiErber}
generically seem to overestimate the imaginary part of the photon polarization tensor for magnetic fields, while they tend to underestimate it in case of an electric field.

Let us finally emphasize that the results obtained in this section give rise to smooth analytic curves in Figs.~\ref{fig:ImB} and \ref{fig:ImE}, even though the original expression of the photon polarization tensor in a magnetic field~\eqref{eq:PIa} is known to exhibit singularities at each threshold of electron-positron pair creation in the state with the given Landau quantum numbers \cite{shabad,Shabad:1972rg,Shabad:1975ik,Witte:1990} [cf. also \Eqref{eq:Bpairthresh} below].
These singularities are commonly denoted as threshold, root or cyclotron resonances in the literature, and give rise to a {\it sawtooth pattern} in actual numerical evaluations \cite{Daugherty:1984tr,Baier:2007dw,Ishikawa:2013fxa}.
As demonstrated by \cite{Baier:2007dw}, the above types of approximations can be considered as providing us with the smoothed out limit of the original sawtooth pattern averaged over a characteristic distance of the order of the separation between two subsequent peaks.

\subsubsection{Weak fields - large momentum, and momentum dominance} \label{subsec:xito0}

We now focus on the limit $\xi\to0$, i.e., the regime where
\begin{equation}
 \xi=\frac{2}{3}\frac{4}{1-\nu^2}\left|\frac{\phi_0^3}{(eB)^2k_{\perp}^2}\right|^{\frac{1}{2}}\ll1\,. \label{eq:xito0}
\end{equation}
While the analogous condition for $\xi\to\infty$, \Eqref{eq:xitoinfty}, could be fulfilled throughout the interval of integration over the parameter $\nu$,
this is not true for \Eqref{eq:xito0} which is obviously violated for $\nu\to\pm1$. 

Employing the identities in Appendix~\ref{app:xi->0} and using the Cauchy product, Eqs.~\eqref{eq:Tsai_int_sj+} and \eqref{eq:Tsai_int_sj-} can be written in terms of an infinite series representation in powers of $\tilde\xi^{2/3}$,
\begin{multline}
 \int_{0}^{\infty}{\rm d} s\,s^l\,{\rm e}^{-{\rm i}\phi_0s - {\rm i} k_{\perp}^2\frac{(1-\nu^2)^2}{48}(eB)^2s^3}
=-\pi\left(\frac{\rm i}{\phi_0}\right)^{l+1}\Biggl\{\left({\rm i}+\tfrac{{\rm sign}(B^2k_{\perp}^2)}{\sqrt{3}}\right) 
\sum_{n=\lceil\frac{l}{3}\rceil}^{\infty} \frac{\frac{(3n)!}{(3n-l)!}2^{-2n-\frac{2}{3}}}{n!\,\Gamma \left(n+\frac{2}{3}\right)}\bigl(\tilde\xi^{2/3}\bigr)^{3n+1} \\
-\left({\rm i}-\tfrac{{\rm sign}(B^2k_{\perp}^2)}{\sqrt{3}}\right) \sum_{n=\lceil\frac{l-1}{3}\rceil}^{\infty} \frac{\frac{(3n+1)!}{(3n+1-l)!}2^{-2n-\frac{4}{3}}}{n!\,\Gamma \left(n+\frac{4}{3}\right)}\bigl(\tilde\xi^{2/3}\bigr)^{3n+2}
+\tfrac{{\rm sign}(B^2k_{\perp}^2)}{\sqrt{3}} \sum_{n=\lceil\frac{l-2}{3}\rceil}^{\infty} \frac{\frac{(3n+2)!}{(3n+2-l)!}\rho(n)}{2^{2n}}\bigl(\tilde\xi^{2/3}\bigr)^{3n+3}
\Biggr\}, \label{eq:Tsai_smallxi}
\end{multline}
where $\lceil\chi\rceil={\rm min}\{n\in\mathbb{Z}\,|\,n\geq\chi\}$ denotes the smallest integer no less than $\chi$, and
\begin{equation}
 \rho(n)=\sum_{m=0}^n\frac{(-1)^{n+m}\,\Gamma(2n+2)}{(m!)\,\Gamma(2n+2-m)}\Biggl[\frac{1}{\Gamma\!\left(m+\frac{2}{3}\right)\,\Gamma\!\left(2n-m+\frac{7}{3}\right)} -\frac{1}{\Gamma\!\left(m+\frac{4}{3}\right)\,\Gamma\!\left(2n-m+\frac{5}{3}\right)}\Biggr].
\end{equation}
Let us emphasize that \Eqref{eq:Tsai_smallxi} -- which only depends on $\tilde\xi^{2/3}$ and not on the parameters $\xi_\pm$ sensitive to the sign of $\Re(B^2k_{\perp}^2\phi_0)$ -- is true for any value of $\Re(B^2k_{\perp}^2\phi_0)$.
More schematically, \Eqref{eq:Tsai_smallxi} has the following structure,
\begin{equation}
 \int_{0}^{\infty}{\rm d} s\,s^l\,{\rm e}^{-{\rm i}\phi_0s - {\rm i} k_{\perp}^2\frac{(1-\nu^2)^2}{48}(eB)^2s^3}\sim\left(\frac{\tilde\xi^{2/3}}{\phi_0}\right)^{l+1}\left[1+{\cal O}(\tilde\xi^{2/3})\right]. \label{eq:Tsai_smallxiLO}
\end{equation}
The leading contribution to \Eqref{eq:Tsai_smallxiLO} in the limit $\xi\to0$ is independent of $\phi_0$, which drops out in the ratio $\tilde\xi^{2/3}/\phi_0$.
As a direct consequence, also the following dimensionless ratios
\begin{align}
 \biggl(\frac{\tilde\xi^{2/3}eB}{\phi_0}\biggr)^2&=\left(\frac{2}{3}\frac{4}{1-\nu^2}\right)^{\frac{4}{3}}\left(\frac{|(eB)^2|}{|k_{\perp}^2|^{2}}\right)^{\frac{1}{3}}{\rm sign}(B^2)\,,\\
 \frac{\tilde\xi^{2/3}k_{\perp}^2}{\phi_0}&=\left(\frac{2}{3}\frac{4}{1-\nu^2}\right)^{\frac{2}{3}}\left(\frac{|k_{\perp}^2|^2}{|(eB)^2|}\right)^{\frac{1}{3}}{\rm sign}(k_{\perp}^2)\,, \\
 \biggl(\frac{\tilde\xi^{2/3}eB}{\phi_0}\biggr)^2\frac{\tilde\xi^{2/3}k_{\perp}^2}{\phi_0}&=\left(\frac{2}{3}\frac{4}{1-\nu^2}\right)^2{\rm sign}(B^2k_{\perp}^2)\,,
\end{align}
which are naturally induced when adopting \Eqref{eq:Tsai_smallxi} in \Eqref{eq:PI_tsai}, do not depend on $\phi_0$.
Contrarily, the divergence for $\nu\to\pm1$ encountered in $\tilde\xi^{2/3}\sim\phi_0(1-\nu^2)^{-2/3}$ [cf. \Eqref{eq:xi}] is not diminished in any of these elementary ratios.
Hence, the divergence for $\nu\to\pm 1$ contained in $\tilde\xi^{2/3}$ can be seen as opposing the expansion in \Eqref{eq:Tsai_smallxi}:
The higher the order in the expansion, the worse the divergence.
Of course, the photon polarization tensor~\eqref{eq:PI_tsai} features additional $\nu$ dependencies besides those contained in the above propertime integrals.
However, even though the above contributions get multiplied with polynomials in $\nu^2$, divergences for $\nu\to\pm1$ persist.
After rewriting the propertime integral in \Eqref{eq:PI_tsai} in terms of an infinite series in powers of $\tilde\xi^{2/3}$, the $\nu$ dependence of \Eqref{eq:PI_tsai} is encoded in terms of the form $(\nu^2)^{\rho}/(1-\nu^2)^{\sigma}$,
with exponents $\rho\geq0$ and $\sigma>0$.
The expansion~\eqref{eq:Tsai_smallxi} generically induces contributions with $\sigma>1$.
The integration of such terms over the full $\nu$ regime in \Eqref{eq:Pipi} yields a finite result for $\sigma<1$, 
\begin{equation}
 \int_{-1}^{1}{\rm d}\nu\,\frac{(\nu^2)^{\rho}}{(1-\nu^2)^{\sigma}}=\frac{\Gamma(1-\sigma)\,\Gamma(\frac{1}{2}+\rho)}{\Gamma(\frac{3}{2}+\rho-\sigma)}\,,\label{eq:nuint}
\end{equation}
and diverges for $\sigma\geq1$.
Let us emphasize that the original propertime expression was completely well-behaved for all values of $\nu$. The convergence problems for $\nu\to\pm1$ are a direct consequence of the expansion performed here.
Of course, one could think about adopting \Eqref{eq:Tsai_smallxi} in a certain range of the $\nu$ interval only, while treating the remainder, e.g., numerically.
This is however outside the scope of the present paper which aims at analytical insights into parameter regimes, where unambiguous, overall expansion parameters can be identified.
Moreover, in Refs.~\cite{Tsai:1974fa,Tsai:1975iz} Tsai and Erber have presented analytical results for the regime $\xi\to0$, by effectively resorting to a leading order expansion in $\tilde\xi^{2/3}$.
Our goal is to carefully rederive and confirm their results, while -- at the same time -- pointing out possible limitations.

To circumvent the divergence problems, -- and in order to make the subsequent discussion most transparent -- we substitute $\nu$ for $\tilde u=1-\nu^2$,
such that $\int_{-1}^1{\rm d}\nu\to\int_0^1\frac{{\rm d}\tilde u}{\sqrt{1-\tilde u}}$.
Thereafter we split the integration into two intervals, namely $\int_0^1{\rm d}\tilde u\to\int_0^{c/\rho}{\rm d}\tilde u+\int_{c/\rho}^1{\rm d}\tilde u$,
where we defined
\begin{equation}
 \rho^{2/3}\equiv1/(\tilde u\tilde\xi)^{2/3}=\frac{1}{\phi_0(c/\rho)}\left|\frac{9}{64}(eB)^2k_\perp^2\right|^{1/3}, \label{eq:b}
\end{equation}
and introduced the additional -- for the moment unspecified -- dimensionless parameter $c$, which is chosen to fulfill
\begin{equation}
 \frac{c}{\rho}\ll 1\,, \quad\text{while}\quad |c|\gg 1\,. \label{eq:c}
\end{equation}
The generic propertime integral -- written in terms of the `new' variables $\tilde u$ and $\rho$ -- reads  [cf. \Eqref{eq:sososo}]
\begin{equation}
 \int_{0}^{\infty}{\rm d} s\,s^l\,{\rm e}^{-{\rm i}\phi_0s - {\rm i} k_{\perp}^2\frac{(1-\nu^2)^2}{48}(eB)^2s^3}
=\int_{0}^{\infty}{\rm d} s\,s^l\,{\rm e}^{-{\rm i}\phi_0(\tilde u)s - {\rm i}\frac{4}{27}\,{\rm sign}(B^2k_{\perp}^2)[\phi_0(c/\rho)]^3(\tilde u\rho)^2s^3}, \label{eq:I1}
\end{equation}
with $\phi_0(\tilde u)=m^2-{\rm i}\epsilon+\frac{k^2}{4}\tilde u^2$. 
While the first condition in \Eqref{eq:c} will allow us to approximate $\int_0^{c/\rho}{\rm d}\tilde u\, f(\tilde u)\approx\frac{c}{\rho}[f(c/\rho)-f(0)]$,
where $f(\tilde u)$ denotes the integrand of the $\tilde u$ integral,
the second one ensures the parameter $\tilde\xi^{2/3}=1/(\tilde u\rho)^{2/3}$ to be small for $\tilde u\in[c/\rho\ldots1]$ and thus \Eqref{eq:xito0} to hold throughout the respective integration interval.
Equation~\eqref{eq:c} and (the right-hand side of) \Eqref{eq:I1} imply that the relevant propertime integrations can still be performed with \Eqref{eq:Tsai_smallxi},
replacing $\phi_0\to\phi_0(\tilde u)$ and $\tilde\xi^{2/3}\to \frac{1}{(\tilde u\rho)^{2/3}}\frac{\phi_0(\tilde u)}{\phi_0(c/\rho)}$.
For the two different $\tilde u$ intervals \Eqref{eq:Tsai_smallxiLO} implies
\begin{multline}
 \int_0^{c/\rho}\frac{{\rm d}\tilde u}{\sqrt{1-\tilde u}} \, \tilde u^n \int_{0}^{\infty}{\rm d} s\,s^l\,{\rm e}^{-{\rm i}\phi_0s - {\rm i} \frac{\tilde u^2}{48}k_{\perp}^2(eB)^2s^3} \\
\sim\frac{c}{\rho}\left\{\left(\frac{c}{\rho}\right)^n \left(\frac{1}{c^{2/3}\phi_0(c/\rho)}\right)^{l+1}\left[1+{\cal O}(\tfrac{c}{\rho})+{\cal O}(\tfrac{1}{c^{2/3}})\right]+\delta_{n0}\left(\frac{1}{m^2}\right)^{l+1}{\cal O}(1)\right\}, \label{eq:Tsai_smallxiLO_1}
\end{multline}
where we also used \Eqref{eq:int0}, and
\begin{align}
 &\int_{c/\rho}^1\frac{{\rm d}\tilde u}{\sqrt{1-\tilde u}} \, \tilde u^n \int_{0}^{\infty}{\rm d} s\,s^l\,{\rm e}^{-{\rm i}\phi_0s - {\rm i}\frac{\tilde u^2}{48} k_{\perp}^2(eB)^2s^3} \nonumber\\
&\quad\quad\sim\left(\frac{1}{\rho^{2/3}\phi_0(c/\rho)}\right)^{l+1}\int_0^{\sqrt{1-c/\rho}}{\rm d}\nu\, \left(\frac{1}{1-\nu^2}\right)^{\frac{2}{3}(l+1)-n}\left[1+{\cal O}(\tfrac{1}{((1-\nu^2)\rho)^{2/3}}\tfrac{\phi_0(1-\nu^2)}{\phi_0(c/\rho)})\right] \nonumber\\
&\quad\quad\approx\left(\frac{1}{\rho^{2/3}\phi_0(c/\rho)}\right)^{l+1}\int_0^{1-\frac{c}{2\rho}}{\rm d}\nu\, \left(\frac{1}{2(1-\nu)}\right)^{\frac{2}{3}(l+1)-n}\left[1+{\cal O}(\tfrac{1}{(2(1-\nu)\rho)^{2/3}})\right] \nonumber\\
&\quad\quad=\left(\frac{1}{\phi_0(c/\rho)}\right)^{l+1}\frac{1}{\rho^{2/3}}\left[\frac{2^{n+1}}{2^{\frac{2}{3}(l+1)}}\left(\frac{1}{\rho^{2/3}}\right)^{l}\left(1+{\cal O}(\tfrac{1}{\rho^{2/3}})\right)-
\left(\frac{c}{\rho}\right)^{n+\frac{1}{3}}\left(\frac{1}{c^{2/3}}\right)^{l}\left(1+{\cal O}(\tfrac{1}{c^{2/3}})\right)\right].
 \label{eq:Tsai_smallxiLO_2}
\end{align}
Adopting Eqs.~\eqref{eq:Tsai_smallxiLO_1} and \eqref{eq:Tsai_smallxiLO_2} to the basic building blocks of \Eqref{eq:PI_tsai}, we can infer their scaling under the above assumptions.
The corresponding explicit expressions are relegated to Appendix~\ref{sec:basicbuild}, Eqs.~\eqref{eq:x1to0_1}-\eqref{eq:x1to0_2b}.
Let us emphasize that in Appendix~\ref{sec:basicbuild} we carefully keep track of the various powers of the parameter $\tilde u=1-\nu^2$ in \Eqref{eq:PI_tsai}: We in particular make use of the fact that -- as outlined in detail in Appendix~\ref{app:seriesNn} -- a global factor $\sim(1-\nu^2)$ can be factored out of the functions $N_0^{(2n)}$, $N_1^{(2n)}$ and $N_2^{(2n)}$, with $n\in\mathbb{N}$, and a factor $\sim(1-\nu^2)^2$ out of the function $n_2^{(2n)}$, with $n\geq2$. 
This turns out to be absolutely essential in singling out the correct scaling behavior of the leading contribution of \Eqref{eq:PI_tsai} for large values of $\rho$. 
Moreover, note that
\begin{equation}
 k_\perp^2\left(\frac{eB}{\rho^{2/3}\phi_0}\right)^2\sim\rho^{2/3}\phi_0\,, \quad\quad
 \text{while} \quad\quad k_\parallel^2\left(\frac{eB}{\rho^{2/3}\phi_0}\right)^2\sim k^2\left(\frac{eB}{\rho^{2/3}\phi_0}\right)^2-\rho^{2/3}\phi_0\,. \label{eq:soja}
\end{equation}
As $\phi_0$ drops out in the product $\rho^{2/3}\phi_0$ [cf. below \Eqref{eq:Tsai_smallxiLO}], the argument of $\phi_0$ is not relevant in combinations of the form $\rho^{2/3}\phi_0$, which thus are independent of $c$. 

In a first step we treat the auxiliary parameter $c$ as a dimensionless numerical constant, chosen to fulfill \Eqref{eq:c}.
Correspondingly, specializing to on-the-light-cone dynamics from the outset, i.e., setting $k^2\equiv0$, the leading contribution to Eqs.~\eqref{eq:x1to0_1}-\eqref{eq:x1to0_2b} in the limit characterized by both
\begin{equation}
 \frac{1}{\rho_0}\ll1 \quad\text{and}\quad\frac{|eB|}{m^2}\ll1, \label{eq:cond_lc}
\end{equation}
with [cf. \Eqref{eq:b}]
\begin{equation}
 \rho_0^{2/3}\equiv\left.\rho^{2/3}\right|_{k^2=0}=\frac{|9k_\perp^2(eB)^2|^{1/3}}{4m^2}\,,
\end{equation}
is expected to scale $\sim m^2\rho^{2/3}_0$.
In particular, it is in general not sufficient to demand $|1/\rho|\ll1$ alone as also contributions $\sim(eB/\phi_0)^{2l}$ with $l\in\mathbb{N}_0$, which are unscreened by inverse powers of $\rho$, are induced.
Condition~\eqref{eq:cond_lc} implicitly contains a restriction to $|k_\perp|/(2m)\gg1$.
Alternatively, it can be represented as
\begin{equation}
 \frac{2m}{|k_\perp|}\ll\frac{|eB|}{m^2}\ll1\,.
\end{equation}
Obviously this ordering of scales holds for {\it weak fields} and a {\it large} transversal {\it momentum}.
Off the light cone the situation is slightly more involved: From Eqs.~\eqref{eq:x1to0_1}-\eqref{eq:x1to0_2b} it is straightforward to infer that in order to guarantee the leading contribution to scale as $\rho^{2/3}$, besides
\begin{equation}
 \left|\frac{1}{\rho}\right|\ll1\quad \text{and} \quad \left|\frac{eB}{\phi_0(c/\rho)}\right|\ll1 \label{eq:cond1}
\end{equation}
we have to demand [cf. also \Eqref{eq:soja}]
\begin{equation}
 \left|\frac{k^2}{\rho^{2/3}\phi_0}\right|\ll1\,,\quad \left|\frac{k^2}{\rho^{2/3}\phi_0}\frac{k^2}{\rho m^2}\right|\ll1, \quad\text{and}\quad \left|\left(\frac{eB}{\rho^{2/3}\phi_0}\right)^2\frac{k^2}{\rho^{2/3}\phi_0}\right|\ll1\,. \label{eq:cond2}
\end{equation}
However, the parameter regime for the leading contribution to scale $\sim\rho^{2/3}$ could in principle even be characterized by fewer constraints, as certain contributions in our
decomposition~\eqref{eq:Tsai_smallxiLO_1}-\eqref{eq:Tsai_smallxiLO_2} could vanish or cancel, and thereby render some constraints irrelevant. Correspondingly, the conditions stated here are all sufficient but not mandatorily necessary.
Expanding the fraction $\frac{1}{\phi_0(c/\rho)}$ as follows,
\begin{equation}
 \frac{1}{\phi(c/\rho)}=\frac{1}{m^2+\frac{k^2}{4}\frac{c^2}{\rho^2}}=\frac{1}{m^2}\left[1+{\cal O}\left(\tfrac{k^2}{4\rho^2m^2}\right)\right], \label{eq:1/phi0cbyrho}
\end{equation}
it is easy to see that in particular for
\begin{equation}
 \frac{1}{\rho_0}\ll1\,,\quad \frac{|eB|}{m^2}\ll1 \quad\text{and}\quad \left(\frac{1}{\rho_0}\right)^{2/3}\frac{|k^2|}{m^2}\ll1 \label{eq:cond3}
\end{equation}
all conditions in Eqs.~\eqref{eq:cond1} and \eqref{eq:cond2} can be met simultaneously. Focusing on $k^2=0$, the last condition is trivially fulfilled and \Eqref{eq:cond_lc} is retained.

By inspection of the building blocks in Appendix~\ref{sec:basicbuild} we find another viable choice of the dimensionless parameter $c$, compatible with \Eqref{eq:c} and the expansions performed in Eqs.~\eqref{eq:x1to0_1}-\eqref{eq:x1to0_2b}, namely
\begin{equation}
 c^{2/3}\equiv\left(\frac{9}{64}\right)^{1/3}\frac{|eB|}{\phi_0(c/\rho)}\,, \quad\text{and thus}\quad \left(\frac{c}{\rho}\right)^{2/3}=\left|\frac{eB}{k_\perp^2}\right|^{1/3}\,.
\end{equation}
For this choice, the ratio ${|eB|}/(c^{2/3}\phi_0)$ corresponds to a purely numeric value.
For the leading contribution to scale as $\rho^{2/3}$ we now have to demand
\begin{multline}
 \left|\frac{1}{\rho}\right|\ll1\,, \quad \left|\frac{eB}{k_\perp^2}\right|\ll1\,,\quad\left|\frac{k^2}{\rho^{2/3}\phi_0}\right|\ll1\,, \quad \left|\frac{eB}{\rho^{2/3}\phi_0}\right|\ll1\,,\\
 \left|\frac{k^2}{\rho^{2/3}\phi_0}\frac{k^2}{eB}\right|^2\left|\frac{eB}{k_\perp^2}\right|\ll1\,,\quad\text{and} \quad\left|\frac{k^2}{\rho^{2/3}\phi_0}\frac{k^2}{m^2}\right|^2\left|\frac{eB}{k_\perp^2}\right|\ll1\,. \label{eq:offlc_mcps}
\end{multline}
Specializing to on-the-light-cone dynamics from the outset, we are only left with the following two conditions,
\begin{equation}
 \frac{1}{\rho_0}\ll1\,, \quad\text{and}\quad \left|\frac{eB}{k_\perp^2}\right|\ll\frac{9}{64}\approx0.14\,, \label{eq:onlc_mcps}
\end{equation}
which imply that the transversal momentum is the dominant scale, i.e., {\it momentum dominance}. 
Let us emphasize that conditions~\eqref{eq:onlc_mcps} are also compatible with the limiting case $m\to0$, which is, e.g., of relevance in the search for minicharged particles \cite{Ahlers:2006iz,Dobrich:2012sw,Dobrich:2012jd}.

Thus, in particular in the two regimes~\eqref{eq:cond1}-\eqref{eq:cond3} and \eqref{eq:offlc_mcps}-\eqref{eq:onlc_mcps} the infinite series in \Eqref{eq:PI_tsai} can be reliably truncated and the leading term, which scales as $\rho^{2/3}$, stems from the following contribution of \Eqref{eq:PI_tsai} (cf. Appendix~\ref{sec:basicbuild}),
\begin{equation}
  \left\{\!\!
 \begin{array}{c}
 \Pi_{\parallel}\\
 \Pi_{\perp}\\
 \Pi_{0}
 \end{array}\!\!
\right\}
=\frac{\alpha}{2\pi}k_{\perp}^2(eB)^2\int_{-1}^1\frac{{\rm d}\nu}{2}\int_{0}^{\infty}{\rm d} s\,s\,{\rm e}^{-{\rm i}\phi_0s - {\rm i} k_{\perp}^2\frac{(1-\nu^2)^2}{48}(eB)^2s^3}
\left\{
 \begin{array}{c}
  N_0^{(2)}-N_{1}^{(2)} \\
  N_{2}^{(2)}-N_0^{(2)}  \\
  0
 \end{array}
\right\}. \label{eq:soho}
\end{equation}
Most notably, this particular contribution can be evaluated directly with the help of Eqs.~\eqref{eq:Tsai_smallxi} and \eqref{eq:nuint}:
As the $\nu$ integration converges, a decomposition of the $\nu$ integral as introduced below \Eqref{eq:nuint} is not necessary, thereby rendering the contribution $\sim\rho^{2/3}$ manifestly independent of any auxiliary parameter $c$.
Let us however emphasize again that such decomposition becomes important in the determination of higher order contributions, and was absolutely essential to establish a consistent truncation scheme and constrain its range of applicability.

Carrying out the integrations in \Eqref{eq:soho}, we finally obtain [cf. also App.~\ref{app:seriesNn}]
\begin{equation}
\left\{\!\!
 \begin{array}{c}
 \Pi_{\parallel}\\
 \Pi_{\perp}\\
 \Pi_{0}
 \end{array}\!\!
\right\}
=\frac{2}{3}\frac{\alpha}{\pi}\left|(eB)^2k_{\perp}^2\right|^{1/3}\left(\frac{1}{6}\right)^{\frac{1}{3}} \frac{9\sqrt{\pi}\,\Gamma^2(\frac{2}{3})}{7\Gamma(\frac{1}{6})}\left(1-{\rm i}\sqrt{3}\,{\rm sign}(B^2k_\perp^2)\right)
\left\{
 \begin{array}{c}
  3 \\
  2 \\
  0
 \end{array}
\right\}.
 \label{eq:soho1}
\end{equation}
The analogous result for an electric field is obtained straightforwardly, employing the electric-magnetic duality~\eqref{eq:trafo}.
Equation~\eqref{eq:soho1} comprises the results derived by Tsai and Erber for homogeneous magnetic fields: the real part corresponds to Eq. (10) of \cite{Tsai:1975iz}, while the imaginary part amounts to Eq. (59a) of \cite{Tsai:1974fa}.
Our results both complement and go beyond those of Tsai and Erber: We now have explicitly shown that \Eqref{eq:soho1} is only applicable in the regimes as characterized by Eqs.~\eqref{eq:cond1}-\eqref{eq:cond3} and Eqs.~\eqref{eq:offlc_mcps}-\eqref{eq:onlc_mcps}, respectively.

\subsection{Strong field limit}\label{sec:strongfield}

Aiming at the strong field limit, i.e., the regime where the scale $ef$, $f\in\{B,E\}$ dominates all other physical scales available in the problem,
it is helpful to introduce the dimensionless parameter $y\equiv{\rm e}^{-{\rm i}z}={\rm e}^{-{\rm i}eBs}$, which transforms into ${\rm e}^{-z'}={\rm e}^{-eEs}$ under the electric-magnetic duality. Given that the propertime integration contour lies slightly below the real positive $s$ axis [cf. the discussion in the context of \Eqref{eq:trafo}] this parameter obviously fulfills $0\leq|y|<1$ for $eBs\neq0$, and $y=1$ for $eBs=1$.
The results to be discussed in the current section will then arise from formal expansions in this parameter and in $k_\perp^2/(2eB)$ [for electric fields: $k_\parallel^2/(2eE)$].
Expansions of this type were pioneered by Shabad~\cite{Shabad:1975ik} (see also \cite{Witte:1990}). Here, our focus ultimately is on compact analytical expressions, with basically all integrations carried out, thereby allowing for immediate insights into the physical parameters dependencies.
For the subsequent considerations it is convenient to write the phase factor as [cf. \Eqref{eq:phi_parperp}] $\Phi_0=\phi_0^{\parallel}+n_2\,k_{\perp}^2$.
With the help of the following identity,
\begin{equation}
 -{\rm i}2zn_2=-1+y^{1+\nu}+y^{1-\nu}-\left(2-y^{1+\nu}-y^{1-\nu}\right)\sum_{n=1}^{\infty}y^{2n}\,,
\end{equation}
we write
\begin{align}
 {\rm e}^{-{\rm i}n_2k_{\perp}^2s}={\rm e}^{-{\rm i}2zn_2\frac{k_\perp^2}{2eB}}={\rm e}^{-\frac{k_{\perp}^2}{2eB}}{\rm e}^{\frac{k_{\perp}^2}{2eB}\left(y^{1+\nu}+y^{1-\nu}\right)}\,\exp\!\left\{-\frac{k_{\perp}^2}{2eB}\left(2-y^{1+\nu}-y^{1-\nu}\right)\sum_{n=1}^{\infty}y^{2n}\right\},
\end{align}
and as $-1\leq\nu\leq1$, obtain
\begin{equation}
 {\rm e}^{-{\rm i}n_2k_{\perp}^2s}
=1+\sum_{l=1}^{\infty}\frac{1}{l!}\left(\frac{k_{\perp}^2}{2eB}\right)^l\Biggl\{(-1)^l
+{\rm e}^{-\frac{k_\perp^2}{2eB}}\left[y^{l(1+\nu)}+y^{l(1-\nu)}\right]\Biggr\}+{\cal O}(\tfrac{k_{\perp}^2}{2eB}){\cal O}(y^2)\,. \label{eq:teil1}
\end{equation}
In \Eqref{eq:teil1} we have neglected all the terms which are at least of ${\cal O}(y^2)$, irrespective of the value of $\nu$.
Given the original $\nu$ interval, $\nu\in[-1\ldots1]$, the expansion is not strict in the sense that for $l\geq2$ the term $y^{l(1+\nu)}$ also contributes beyond ${\cal O}(y^2)$, namely for $\nu\in[\frac{2-l}{l}\ldots1]$.
Analogously, the term $y^{l(1-\nu)}$ is of ${\cal O}(y^2)$ for $\nu\in[-1\ldots\frac{l-2}{l}]$.
An explicit restriction to ${\cal O}(y^2)$ could, e.g., be implemented by an adequate restriction of the $\nu$ integration interval.
Here we stick to the full $\nu$ interval. This will allow us to identify and explicitly evaluate certain generic building blocks
of the photon polarization tensor in the {\it Landau level representation} to be obtained below. It will in particular enable us to exactly account for logarithmic contributions $\sim(eB)^{-2}\ln(eB)$ also.

Moreover, we write
\begin{align}
\frac{N_0}{s}&={\rm i}eB\sum_{n=0}^{\infty}\left\{\left[1+\nu(2n+1)\right]y^{1+\nu}+\left[1-\nu(2n+1)\right]y^{1-\nu}\right\}y^{2n}, \nonumber\\
\frac{N_1}{s}&={\rm i}eB(1-\nu^2)\left(1+2\sum_{n=1}^{\infty}y^{2n}\right), \nonumber\\
\frac{N_2}{s}&=4{\rm i}eB\sum_{n=1}^{\infty}n\left[2n-(n+1)(y^{1+\nu}+y^{1-\nu})\right]y^{2n}\,. \label{eq:teil2}
\end{align}
The contact term was originally given in terms of a field independent integral representation [cf. Eqs.~\eqref{eq:PI_comp} and \eqref{eq:ct}].
To guarantee its correct inclusion when aiming at results in the strong field limit,
we do not naively insert the above series representations~\eqref{eq:teil1} and \eqref{eq:teil2} into the original expression of the photon polarization tensor~\eqref{eq:PI_comp}, but first rewrite \Eqref{eq:PI_comp} in the following form [cf. \Eqref{eq:Pieta}]
\begin{equation}
\left\{
 \begin{array}{c}
 \!\Pi_{\parallel}\!\\
 \!\Pi_{\perp}\!\\
 \!\Pi_{0}\!
 \end{array}
\right\}
=\frac{\alpha}{2\pi}\int_{-1}^{1}\frac{{\rm d}\nu}{2}
\left\{
 \begin{array}{c}
 \!k_\parallel^2\eta_1+k_{\perp}^2\eta_0 \\
 \!k_\parallel^2\eta_0+k_{\perp}^2\eta_2 \\
 k^2\eta_0
 \end{array}
\right\}, \label{eq:PI_comp_largeB}
\end{equation}
where we have introduced the shortcut notation
\begin{equation}
 \eta_i=\eta_i^\parallel+\int\limits_{0}^{\infty-{\rm i}\eta}\frac{{\rm d} s}{s}\,
{\rm e}^{-{\rm i}\phi_0^\parallel s}\left({\rm e}^{-{\rm i}n_2k_\perp^2s}-1\right)N_i\,, \label{eq:eta_i}
\end{equation}
with $i\in\{0,1,2\}$.
Equation~\eqref{eq:PI_comp_largeB} has the advantage that the UV divergence to be canceled by the contact term already cancels on the level of the integrand of the $\nu$ integral [cf. Eqs.~\eqref{eq:intN0}-\eqref{eq:intN2}].
Correspondingly, the $\nu$ integration does not have to be performed to arrive at a manifestly finite expression. This was not true for the original representation~\eqref{eq:PI_comp}.
A naive expansion of \Eqref{eq:PI_comp} would immediately result in spurious contributions that can be attributed to an inadequate treatment and erroneous cancellation of the UV divergence.
Conversely, the terms in \Eqref{eq:PI_comp_largeB} are manifestly finite and perfectly amenable for a strong field expansion:
The second term in \Eqref{eq:eta_i} can be straightforwardly converted into a series in $y$ with the help of the above identities, yielding
\begin{align}
 \int_{-1}^1\frac{{\rm d}\nu}{2}\,\frac{N_0}{s}\left({\rm e}^{-{\rm i}n_2k_\perp^2s}-1\right)& \nonumber\\
&\hspace*{-2cm}=2{\rm i}eB\int_{-1}^1\frac{{\rm d}\nu}{2}\biggl\{(1+\nu)y^{1+\nu}\sum_{n=1}^{\infty}\frac{1}{n!}\left(\frac{k_\perp^2}{2eB}\right)^n\biggl[(-1)^n+y^{n(1+\nu)}{\rm e}^{-\frac{k_\perp^2}{2eB}}\biggr] +{\cal O}(\tfrac{k_{\perp}^2}{2eB}){\cal O}(y^2)\biggr\}, \label{eq:sf_block1}\\
 \int_{-1}^1\frac{{\rm d}\nu}{2}\,\frac{N_1}{s}\left({\rm e}^{-{\rm i}n_2k_\perp^2s}-1\right)&={\rm i}eB\int_{-1}^1\frac{{\rm d}\nu}{2}\,(1-\nu^2) \nonumber\\
&\hspace*{+1.55cm}\times\biggl\{\sum_{n=1}^{\infty}\frac{1}{n!}\left(\frac{k_\perp^2}{2eB}\right)^n
\biggl[(-1)^n+2\,y^{n(1+\nu)}{\rm e}^{-\frac{k_\perp^2}{2eB}}\biggr]+{\cal O}(\tfrac{k_{\perp}^2}{2eB}){\cal O}(y^2)\biggr\}, \label{eq:sf_block2} \\
 \int_{-1}^1\frac{{\rm d}\nu}{2}\,\frac{N_2}{s}\left({\rm e}^{-{\rm i}n_2k_\perp^2s}-1\right)&={\rm i}eB\int_{-1}^1\frac{{\rm d}\nu}{2}\,{\cal O}(\tfrac{k_{\perp}^2}{2eB}){\cal O}(y^2). \label{eq:sf_block3}
\end{align}
Here we have performed a double expansion in both $k_\perp^2/(2eB)$ and $y$, and have neglected terms which are {\it suppressed} by at least a factor of $y^2k_\perp^2/(2eB)$. 
Conversely, we have in particular kept all contributions with a $y$ dependence of the form $y^{n(1+\nu)}$, with $n\in\mathbb{N}$, that in the vicinity of $\nu\to-1$ approach $y^0$, and whose field dependence becomes increasingly less pronounced.
However, note the overall multiplicative factor $(1+\nu)$ counteracting this behavior by diminishing corresponding contributions.
An analogous representation for the first term in \Eqref{eq:eta_i} can be obtained from Eqs.~\eqref{eq:soumgeschr}-\eqref{eq:intN2},
employing
\begin{equation}
 \frac{\cos(\nu z)}{\sin(z)}={\rm i}\left(y^{1+\nu}+y^{1-\nu}\right)\sum_{n=0}^\infty y^{2n}\,, \quad\quad
 \frac{\sin(\nu z)}{\sin(z)}
=\left(y^{1-\nu}-y^{1+\nu}\right)\sum_{n=0}^\infty y^{2n}\,, \quad\quad
 \cot(z)={\rm i}\Bigl[1+2\sum_{n=1}^\infty y^{2n}\Bigr], \label{eq:zeugs}
\end{equation}
and the following identity
\begin{equation}
 \int\limits_{0}^{\infty-{\rm i}\eta}\,\frac{{\rm d} s}{s}\,{\rm e}^{-{\rm i}\phi_0^\parallel s}
=\psi\biggl(\frac{\phi_0^\parallel}{2eB}\biggr)-\ln\biggl(\frac{\phi_0^\parallel}{2eB}\biggr)
+\sum_{n=0}^\infty \int\limits_{0}^{\infty-{\rm i}\eta}\,{\rm d}s\,{\rm e}^{-{\rm i}\phi_0^\parallel s} y^n
=\biggl(\sum_{n=1}^\infty\frac{1}{n}-\gamma\biggr)-\ln\biggl(\frac{\phi_0^\parallel}{2eB}\biggr)\,, \label{eq:ct_anders}
\end{equation}
where $\gamma$ denotes the Euler-Mascheroni constant.  
Equation~\eqref{eq:ct_anders} can be derived straightforwardly from formula 8.361.8 of \cite{Gradshteyn},
\begin{equation}
 \int_{0}^{\infty}\frac{{\rm d} z'}{z'}\,{\rm e}^{-{\rm i}\beta z'}=\psi({\rm i}\beta)-\ln({\rm i}\beta)+\int_{0}^{\infty}{\rm d} z'\,{\rm e}^{-{\rm i}\beta z'}\frac{1}{1-{\rm e}^{-z'}}\,,
\end{equation}
valid for ${\rm Im}(\beta)<0$, and the exact series representation of the Digamma function, formula 8.362 of \cite{Gradshteyn},
\begin{equation}
 \psi(\chi)=-\gamma-\frac{1}{\chi}+\sum_{n=1}^{\infty}\frac{\chi}{n(\chi+n)}=-\gamma-\frac{1}{\chi}+\sum_{n=1}^{\infty}\left(\frac{1}{n}-\frac{1}{\chi+n}\right) .
\label{eq:Psi_exact}
\end{equation}
Note that, even though the original propertime integral expression on the left-hand side of \Eqref{eq:ct_anders} seems {\it a priori} independent of the external field, its right-hand side
 -- constituting a (strong field) series, or equivalently Landau level representation of the original propertime integral -- features an explicit field dependence. This comes about as follows:
The propertime parameter $s$ is a dimensionful quantity of dimension mass-squared, to be rendered dimensionless by an additional physical mass scale. Aiming at a strong field expansion,the dominant scale is $\sim eB$.
We argue that the natural reference scale is in fact given by $\Delta m=m_{n+1}-m_n=2eB$, corresponding to the ``mass difference'' between two consecutive Landau levels, labeled by integers $n$ and $n+1$, and featuring magnetic field dependent masses $m_n^2=m^2+2eBn$.
Thus, the quantity $2eB$ constitutes the reference scale to render $s$ dimensionless, which explains the occurrence of the ratio $\phi_0^\parallel/(2eB)$ on the right-hand side of \Eqref{eq:ct_anders}.
By means of the electric-magnetic duality~\eqref{eq:trafo} the above reasoning also holds for the case of an electric field. Here, the corresponding dimensionless ratio in the argument of the logarithm is given by $\phi_0^\parallel/(2eE)$.

Let us emphasize that the above \textit{ad hoc} assumption can be explicitly confirmed and verified, noting that the results of Eqs.~\eqref{eq:intN0_s}-\eqref{eq:intN2_s} below, utilizing this assumption in their derivation, can alternatively be derived from the functions $\eta_i^\parallel(B)$ with the propertime integrations already carried out [cf. \Eqref{eq:B_kperp=0}].
Following this alternative approach, no ambiguity arises; cf. below \Eqref{eq:intN2_s} and also \cite{Karbstein:2011ja}.
Employing Eqs.~\eqref{eq:zeugs} and \eqref{eq:ct_anders} in \Eqref{eq:soumgeschr}, we obtain
\begin{align}
\int_{-1}^1\frac{{\rm d}\nu}{2}\,\eta_0^\parallel(B)
&=\int_{-1}^1\frac{{\rm d}\nu}{2}\biggl\{\,{\rm c.t.}-\nu^2+2{\rm i}
eB\int\limits_{0}^{\infty-{\rm i}\eta}\!{\rm d} s\,{\rm e}^{-{\rm i}\phi_0^\parallel s}
\biggl(1-\nu^2-\nu\frac{\phi_0^\parallel}{eB}\biggr)\sum_{n=0}^\infty y^{1+\nu+2n}\biggr\}, \label{eq:eta0series}\\
 \int_{-1}^1\frac{{\rm d}\nu}{2}\,\eta_1^\parallel(B)
&=\int_{-1}^1\frac{{\rm d}\nu}{2}\biggl\{\,{\rm c.t.}
+{\rm i}eB\int\limits_{0}^{\infty-{\rm i}\eta}\!{\rm d} s\,{\rm e}^{-{\rm i}\phi_0^\parallel s}\Bigl(1+2\sum_{n=1}^\infty y^{2n}\Bigr)\biggr\},\label{eq:eta1series} \ \\
\int_{-1}^1\frac{{\rm d}\nu}{2}\,\eta_2^\parallel(B)
&=\int_{-1}^1\frac{{\rm d}\nu}{2}\biggl\{\,{\rm c.t.}-\frac{1+3\nu^2}{2}+2{\rm i}eB\int\limits_{0}^{\infty-{\rm i}\eta}\!{\rm d} s\,{\rm e}^{-{\rm i}\phi_0^\parallel s} \nonumber\\
&\hspace*{3cm}
\times\Biggl[\biggl[1-\biggl(\nu+\frac{\phi_0^\parallel}{eB}\biggr)^2\biggr]\sum_{n=0}^\infty y^{1+\nu+2n}
+\biggl(\frac{\phi_0^\parallel}{eB}\biggr)^2\Bigl(\frac{1}{2}+\sum_{n=1}^\infty y^{2n}\Bigr)\Biggr]\biggr\}, \label{eq:eta2series}
\end{align}
where we introduced
\begin{equation}
 {\rm c.t.}=(1-\nu^2)\biggl[\ln\biggl(\frac{m^2-{\rm i}\epsilon}{2eB}\biggr)-\biggl(\sum_{n=1}^\infty\frac{1}{n}-\gamma\biggr)\biggr].
\end{equation}

The propertime integrations can now be performed straightforwardly: All contributions in Eqs.~\eqref{eq:sf_block1}-\eqref{eq:sf_block3} and \eqref{eq:eta0series}-\eqref{eq:eta2series} are proportional to powers of $y={\rm e}^{-{\rm i}z}$, and thus only depend linearly on the propertime parameter $s$ in the exponential.
Correspondingly, the basic building blocks of \Eqref{eq:PI_comp_largeB} can be written as
\begin{align}
 \int_{-1}^1\frac{{\rm d}\nu}{2}\int\limits_{0}^{\infty-{\rm i}\eta}\frac{{\rm d} s}{s}{\rm e}^{-{\rm i}\phi_0^\parallel s}\left({\rm e}^{-{\rm i}n_2k_\perp^2s}-1\right)N_0&=2eB\int_{-1}^1\frac{{\rm d}\nu}{2}\,\biggl\{(1+\nu)\sum_{n=1}^{\infty}\frac{1}{n!}\left(\frac{k_{\perp}^2}{2eB}\right)^n\nonumber\\
&\hspace*{-1.6cm}\times\biggl[\frac{(-1)^n}{\phi_0^\parallel+eB(1+\nu)}+\frac{{\rm e}^{-\frac{k_\perp^2}{2eB}}}{\phi_0^\parallel+eB(n+1)(1+\nu)}\biggr] +{\cal O}(\tfrac{k_{\perp}^2}{2eB}){\cal O}\Bigl(\tfrac{1}{\phi_0^\parallel+2eB}\Bigr)\biggr\}, \label{eq:sf_block1_2}\\
 \int_{-1}^1\frac{{\rm d}\nu}{2}\int\limits_{0}^{\infty-{\rm i}\eta}\frac{{\rm d} s}{s}{\rm e}^{-{\rm i}\phi_0^\parallel s}\left({\rm e}^{-{\rm i}n_2k_\perp^2s}-1\right)N_1&=eB\int_{-1}^1\frac{{\rm d}\nu}{2}\,(1-\nu^2)\biggl\{\sum_{n=1}^{\infty}\frac{1}{n!}\left(\frac{k_{\perp}^2}{2eB}\right)^n  \nonumber\\
&\quad\quad\times\biggl[\frac{(-1)^n}{\phi_0^\parallel}+\frac{2\,{\rm e}^{-\frac{k_\perp^2}{2eB}}}{\phi_0^\parallel+eBn(1+\nu)}\biggr]+{\cal O}(\tfrac{k_{\perp}^2}{2eB}){\cal O}\Bigl(\tfrac{1}{\phi_0^\parallel+2eB}\Bigr)\biggr\}, \label{eq:sf_block2_2} \\
 \int_{-1}^1\frac{{\rm d}\nu}{2}\int\limits_{0}^{\infty-{\rm i}\eta}\frac{{\rm d} s}{s}{\rm e}^{-{\rm i}\phi_0^\parallel s}\left({\rm e}^{-{\rm i}n_2k_\perp^2s}-1\right)N_2&=eB\int_{-1}^1\frac{{\rm d}\nu}{2}\,
{\cal O}(\tfrac{k_{\perp}^2}{2eB}){\cal O}\Bigl(\tfrac{1}{\phi_0^\parallel+2eB}\Bigr), \label{eq:sf_block3_2}
\end{align}
where ${\cal O}\Bigl(\tfrac{1}{\phi_0^\parallel+2eB}\Bigr)$ is to be understood as denoting terms of the structure $\tfrac{1}{\phi_0^\parallel+\kappa eB}$ with $\kappa\geq2$,
and
\begin{align}
\int_{-1}^1\frac{{\rm d}\nu}{2}\,\eta_0^\parallel(B)
&=\int_{-1}^1\frac{{\rm d}\nu}{2}\biggl\{\,{\rm c.t.}-\nu^2+
2eB\sum_{n=0}^\infty \frac{1+\nu(2n+1)}{\phi_0^\parallel+eB(1+\nu+2n)}\biggr\}, \label{eq:intN0_s} \\
\int_{-1}^1\frac{{\rm d}\nu}{2}\,\eta_1^\parallel(B)
&=\int_{-1}^1\frac{{\rm d}\nu}{2}\biggl\{\,{\rm c.t.}
+eB\,(1-\nu^2)\Biggl[\frac{1}{\phi_0^\parallel}+\sum_{n=1}^\infty\frac{2}{\phi_0^\parallel+2neB} \Biggr]\biggr\}, \label{eq:intN1_s} \\
\int_{-1}^1\frac{{\rm d}\nu}{2}\,\eta_2^\parallel(B)
&=\int_{-1}^1\frac{{\rm d}\nu}{2}\biggl\{\,{\rm c.t.}-\frac{1+3\nu^2}{2}+\sum_{n=0}^\infty2-\frac{\phi_0^\parallel}{eB}+8eB\,\sum_{n=1}^\infty\Biggl[\frac{n^2}{\phi_0^\parallel+2neB}-\frac{n(n+1)}{\phi_0^\parallel+eB(1+\nu+2n)}\Biggr]\biggl\}. \label{eq:intN2_s}
\end{align}
Note that the right-hand sides of Eqs.~\eqref{eq:intN0_s}-\eqref{eq:intN2_s} correspond to exact series representations of the functions on their left-hand sides. They can alternatively be obtained by employing the exact series representation of the Digamma function~\eqref{eq:Psi_exact} in the functions $\eta_i^\parallel(B)$ with the propertime integrations already carried out, as derived in Sec.~\ref{sec:thepolten}: While the corresponding expressions for $\eta_0^\parallel(B)$ and $\eta_1^\parallel(B)$ can be read off from Eqs.~\eqref{eq:Pieta} and \eqref{eq:B_kperp=0}, the one for $\eta_2^\parallel(B)$ can be derived analogously and reads [cf. Eqs.~\eqref{eq:soumgeschr} and \eqref{eq:intN2}-\eqref{eq:grad3b}]
\begin{align}
\int_{-1}^{1}\frac{{\rm d}\nu}{2}\,\eta_2^\parallel(B)&=\int_{-1}^1\frac{{\rm d}\nu}{2}\Biggl\{(1-\nu^2)\ln\biggl(\frac{m^2-{\rm i}\epsilon}{2eB}\biggr)-\frac{1+3\nu^2}{2}-\frac{\phi_0^\parallel}{eB}\biggl[1+\frac{\phi_0^\parallel}{eB}\,\psi\biggl(\frac{\phi_0^\parallel}{2eB}\biggr)\biggr] \nonumber\\
&\hspace*{4.5cm}-\biggl[1-\nu^2 -\frac{2\nu\phi_0^\parallel}{eB}-\biggl(\frac{\phi_0^\parallel}{eB}\biggr)^2\,\biggr]\psi\biggl(\frac{\phi_0^\parallel}{2eB}+\frac{1+\nu}{2}\biggr) \Biggr\}. \label{eq:tildeN2}
\end{align}

So far we have not really specified the parameter regime where the particular truncations as performed in Eqs.~\eqref{eq:sf_block1_2}-\eqref{eq:sf_block3_2} yield trustworthy results:
The neglect of terms of order $k_\perp^2/(2eB)$ suggests the limitation to $k_\perp^2/(2eB)\ll1$. This condition involves only physical parameters.
Conversely, the situation is not so clear for the second expansion, where we neglected contributions $\sim[\phi_0^\parallel+\kappa eB]^{-1}=[m^2+\kappa eB+\frac{1-\nu^2}{4}k_\parallel^2]^{-1}$ with $\kappa\geq2$.
Apart from the physical parameters $k_\parallel^2$, $m^2$ and $eB$, these terms depend on the integration parameter $\nu$.
Hence, definitive statements about the range of applicability of the latter expansion can only be given after the $\nu$ integration has been carried out.

In a next step we explicitly perform the $\nu$ integrations. The corresponding results will then be expanded to allow for analytical insights and a better understanding of the parameter dependencies and the range of applicability of Eqs.~\eqref{eq:sf_block1_2}-\eqref{eq:intN2_s}.
We emphasize again that all $\nu$ integrations will be performed over the full $\nu$ interval, $\nu\in[-1\ldots1]$; cf. the corresponding comment below \Eqref{eq:teil1}.

The $\nu$ integrals to be performed in Eqs.~\eqref{eq:sf_block1_2}-\eqref{eq:intN2_s} are of the simple structure
\begin{equation}
  \int{\rm d}\nu\,\frac{\nu^j}{\phi_0^\parallel+2eBn+eBl(1+\nu)} = \int{\rm d}\nu\,\frac{\nu^j}{a+2b\nu+c\nu^2}\,, \label{eq:nuintegral}
\end{equation}
with $j\in\{0,1,2\}$ and coefficients
\begin{equation}
  a=m^2-{\rm i}\epsilon+eB(2n+l)+\frac{k_\parallel^2}{4}\,, \quad b=\frac{eBl}{2}\,, \quad c=-\frac{k_\parallel^2}{4}\,, \label{eq:abc}
\end{equation}
featuring the momentum and external field dependence. As detailed in Appendix~\ref{app:Bint}, the integrals with $j=1,2$ can be expressed in terms of the most basic one with $j=0$.
For completeness, we also provide the explicit expressions of the definite integral for $ac>b^2$,
\begin{equation}
 \int_{-1}^1{\rm d}\nu\,\frac{1}{a+2b\nu+c\nu^2}=\frac{1}{\sqrt{ac-b^2}}\biggl[\arctan\biggl(\frac{b+c}{\sqrt{ac-b^2}}\biggr)-\arctan\biggl(\frac{b-c}{\sqrt{ac-b^2}}\biggr)\biggr], \label{eq:ac>b^2_-11}
\end{equation}
and in the complementary regime, $ac<b^2$,
\begin{multline}
 \int_{-1}^1{\rm d}\nu\,\frac{1}{a+2b\nu+c\nu^2}
=\frac{1}{2}\frac{1}{\sqrt{b^2-ac}}\biggl\{\ln\biggl|\frac{c-a-2\sqrt{b^2-ac}}{c-a+2\sqrt{b^2-ac}}\biggr| \\
 +2\pi{\rm i}\,\Theta\Bigl(-k_\parallel^2-\left(\sqrt{m^2+2eBn}+\sqrt{m^2+2(n+l)eB}\right)^2\Bigr)\biggr\}, \label{eq:ac<b^2ReIm}
\end{multline}
[cf. Appendix~\ref{app:Bint}].
Together with Eqs.~\eqref{eq:lala}-\eqref{eq:lalalala} in the Appendix, Eqs.~\eqref{eq:ac>b^2_-11} and \eqref{eq:ac<b^2ReIm}
explicitly confirm that the photon polarization tensor in the presence of a magnetic field develops an imaginary part only above the threshold,
\begin{equation}
 -k_\parallel^2\geq\left(\sqrt{m^2+2eBn}+\sqrt{m^2+2(n+l)eB}\right)^2, \label{eq:Bpairthresh>}
\end{equation}
to create an electron on the $n$th Landau level and a positron on the $(n+l)$th level (or vice versa) \cite{Shabad:1975ik,shabad}.
The necessary condition for an imaginary part to occur in \Eqref{eq:ac<b^2ReIm} is $-k_\parallel^2\geq(2m)^2$,
obtained by setting $n=l=0$ in \Eqref{eq:Bpairthresh>}, i.e., it increases step-like from zero to a finite value at $-k_\parallel^2=(2m)^2$.
Above this lowest threshold, the next discrete increase of the imaginary part occurs at
\begin{equation}
 -k_\parallel^2=\left(m+\sqrt{m^2+2eB}\right)^2, \label{eq:cond_pc1}
\end{equation}
corresponding to $n=0$ and $l=1$.
Focusing on the strong field limit, i.e., the regime where $eB$ is assumed to dominate all other dimensionful parameters, $eB\gg\{m^2,|k_\parallel^2|,k_\perp^2\}$, it is instructive to rewrite the condition~\eqref{eq:cond_pc1} as
\begin{equation}
 -\frac{k_\parallel^2}{2eB}=1+\frac{m^2}{eB}+\sqrt{2\,\frac{m^2}{eB}\left(1+\frac{m^2}{2eB}\right)}\,. \label{eq:cond_pc2}
\end{equation}
Obviously this condition can never be fulfilled in the strong field limit. Correspondingly, the contribution from $n=l=0$ encodes the full imaginary part of the photon polarization tensor in this particular limit.

For electric fields $E>0$ and $l\neq0$, the parameter $a$ is genuinely complex, $b$ is purely imaginary and only $c$ is real valued (cf. Appendix~\ref{app:strongEfield}). In this case we thus perform the $\nu$ integral as follows,
\begin{align}
 \int{\rm d}\nu\,\frac{1}{a+2b\nu+c\nu^2}=\frac{1}{2}\frac{{\rm i}}{\sqrt{ac-b^2}} \biggl[\ln\biggl(1-\frac{{\rm i}(c\nu+b)}{\sqrt{ac-b^2}}\biggr) - \ln\biggl(1+\frac{{\rm i}(c\nu+b)}{\sqrt{ac-b^2}}\biggr)\biggr]\ +\ C, \label{eq:basicnuint_E}
\end{align}
with integration constant $C$,
irrespective of the particular values of $a$, $b$ and $c$. Equation~\eqref{eq:basicnuint_E} genuinely features both real and imaginary parts -- in contrast to the magnetic field there is no threshold condition, and pair production occurs for arbitrarily weak electric fields.

In the following we organize the results of the above integrations in terms of an expansion in inverse powers of the parameter $ef\gg\{m^2,|k_\parallel^2|,k_\perp^2\}$.
In case of a magnetic field, the situation $ac<b^2$ is of particular interest, as -- at least for $l\neq0$ -- it is compatible with the strong field limit, $eB\gg\{m^2,|k_\parallel^2|\}$ [cf. \Eqref{eq:abc}].
For $\{n,l\}>0$, we find for the real part of the $\nu$ integration
\begin{equation}
 \Re\int_{-1}^1{\rm d}\nu\,\frac{1}{a+2b\nu+c\nu^2}
\left\{\begin{array}{c}
1\\ \nu\\ \nu^2
\end{array}\right\}
=
\frac{1}{leB}
\left\{\begin{array}{c}
\ln\left(\frac{n+l}{n}\right)\\ 2-\left(1+\frac{2n}{l}\right)\ln\left(\frac{n+l}{n}\right)\\ \left(1+\frac{2n}{l}\right)\left[-2+\left(1+\frac{2n}{l}\right)\ln\left(\frac{n+l}{n}\right)\right]
\end{array}\right\}
+{\cal O}\Bigl(\frac{1}{(eB)^{2}}\Bigr), \label{eq:i_start}
\end{equation}
while for $l=0,\ n>0$ we obtain
\begin{equation}
 \Re\int_{-1}^1{\rm d}\nu\,\frac{1}{a+2b\nu+c\nu^2}
\left\{\begin{array}{c}
1\\ \nu\\ \nu^2
\end{array}\right\}
=
\frac{1}{neB}\left\{\begin{array}{c}
1\\ 0\\ 1/3
\end{array}\right\}
+{\cal O}\Bigl(\frac{1}{(eB)^{2}}\Bigr), \label{eq:i_mitte}
\end{equation}
and for $n=0,\ l>0$,
\begin{multline}
 \Re\int_{-1}^1{\rm d}\nu\,\frac{1}{a+2b\nu+c\nu^2}
\left\{\begin{array}{c}
1\\ \nu\\ \nu^2
\end{array}\right\} \\
=
\frac{1}{leB}\left[
\left\{\begin{array}{c}
\ln\left(\frac{2leB}{m^2}\right)\\ 2-\ln\left(\frac{2leB}{m^2}\right)\\ \ln\left(\frac{2leB}{m^2}\right)-2
\end{array}\right\}
+\frac{1}{leB}\left\{\begin{array}{c}
k_\parallel^2+\frac{1}{2}m^2-\frac{1}{2}k_\parallel^2\ln\left(\frac{2leB}{m^2}\right)\\ \left(\frac{1}{2}k_\parallel^2-m^2\right)\ln\left(\frac{2leB}{m^2}\right)-\frac{3}{2}k_\parallel^2-\frac{1}{2}m^2\\ \frac{5}{3}k_\parallel^2-\frac{3}{2}m^2 + \left(2m^2-\frac{1}{2}k_\parallel^2\right)\ln\left(\frac{2leB}{m^2}\right)
\end{array}\right\}\right]
+{\cal O}\Bigl(\frac{1}{(eB)^{3}}\Bigr). \label{eq:i_end}
\end{multline}
Of course, the corresponding integrals for $n=l=0$ are independent of the field strength.
All contributions in Eqs.~\eqref{eq:i_start}-\eqref{eq:i_end} are suppressed by an overall factor of $(eB)^{-1}$.
Equations~\eqref{eq:i_start}, \eqref{eq:i_mitte} can be written as strict power series in $(eB)^{-1}$.
Contrarily, for $n=0$ but arbitrary values of $l\in\mathbb{N}$ also logarithmic contributions in $eB$ show up.
For extremely large values of $eB/m^2\gg1$ these logarithms can be sizable, i.e., $\ln(eB/m^2)\gg1$, and are expected to constitute the dominant contributions at a given order in the expansion in $(eB)^{-1}$.

For $n=0$, while $l>1$, we in particular obtain [cf. \Eqref{eq:i_end}]
\begin{equation}
 \Re\int_{-1}^1{\rm d}\nu\,\frac{1}{a+2b\nu+c\nu^2}
\left\{\begin{array}{c}
1+\nu \\ 1-\nu^2
\end{array}\right\}
=
\frac{1}{leB}\left[\left\{\begin{array}{c}
2 \\ 2
\end{array}\right\} + \frac{1}{leB}\left\{\begin{array}{c}
-\frac{1}{2}k_\parallel^2-m^2\ln\left(\frac{2leB}{m^2}\right) \\ 2m^2-\frac{2}{3}k_\parallel^2-2m^2\ln\left(\frac{2leB}{m^2}\right)
\end{array}\right\}\right]
+{\cal O}\Bigl(\frac{1}{(eB)^{3}}\Bigr). \label{eq:1+nu/-nu^2}
\end{equation}
Equation~\eqref{eq:1+nu/-nu^2} implies that, even though \Eqref{eq:i_end} starts contributing at order $(eB)^{-1}\ln(eB)$, the leading field dependent logarithmic contribution in Eqs.~\eqref{eq:sf_block1_2}-\eqref{eq:intN2_s} is suppressed by an additional factor of $(eB)^{-1}$ and thus is proportional to $(eB)^{-2}\ln(eB)$:
All terms with $l=0$ are multiplied by an additional factor of $1+\nu$ or $1-\nu^2$, respectively,
and the contribution at order $(eB)^{-1}\ln(eB)$ completely cancels out in \Eqref{eq:1+nu/-nu^2}.
We emphasize that the terms written explicitly in Eqs.~\eqref{eq:sf_block1_2}-\eqref{eq:intN2_s} in fact give rise to all logarithmic contributions of the structure $\sim(eB)^{-n}\ln(eB)$, with $n\geq2$
in the photon polarization tensor.

The calculation for the corresponding imaginary parts is almost trivial: As already noted below \Eqref{eq:ac<b^2ReIm}, in the strong field limit an imaginary part can only arise from the contribution with $n=l=0$.
Hence, the argument of the Heaviside function in \Eqref{eq:ac<b^2ReIm} becomes independent of the magnetic field strength and reads $-k_\parallel^2-(2m)^2$.
This yields
\begin{equation}
 \left.\Im\int_{-1}^1{\rm d}\nu\,\frac{1}{a+2b\nu+c\nu^2}
\left\{\begin{array}{c}
1\\ \nu\\ \nu^2
\end{array}\right\}\right|_{n=l=0}
=\frac{4\pi}{\sqrt{(k_\parallel^2+4m^2)k_\parallel^2}}\,\Theta(-k_\parallel^2-4m^2)
\left\{\begin{array}{c}
1\\
0\\
1+\frac{4m^2}{k_\parallel^2}
\end{array}\right\}. \label{eq:ImpartsBf}
\end{equation}

Employing Eqs.~\eqref{eq:i_start}-\eqref{eq:1+nu/-nu^2} in Eqs.~\eqref{eq:sf_block1_2}-\eqref{eq:sf_block3_2}, we obtain
\begin{equation}
 \Re\int_{-1}^1\frac{{\rm d}\nu}{2}\int\limits_{0}^{\infty-{\rm i}\eta}\frac{{\rm d} s}{s}{\rm e}^{-{\rm i}\phi_0^\parallel s}\left({\rm e}^{-{\rm i}n_2k_\perp^2s}-1\right)N_1(eBs)=-\frac{k_\perp^2}{2}\,\int_{-1}^1\frac{{\rm d}\nu}{2}\,\frac{1-\nu^2}{\phi_0^\parallel}
+ {\cal O}\Bigl(\tfrac{1}{eB}\Bigr),
\end{equation}
while for $i\in\{0,2\}$ we find
\begin{equation}
 \Re\int_{-1}^1\frac{{\rm d}\nu}{2}\int\limits_{0}^{\infty-{\rm i}\eta}\frac{{\rm d} s}{s}{\rm e}^{-{\rm i}\phi_0^\parallel s}\left({\rm e}^{-{\rm i}n_2k_\perp^2s}-1\right)N_i(eBs)={\cal O}\Bigl(\tfrac{1}{eB}\Bigr).
\end{equation}
This directly implies [cf. \Eqref{eq:eta_i}]
\begin{equation}
 \Re\int_{-1}^1\frac{{\rm d}\nu}{2}\,\eta_i(B)=\Re\int_{-1}^1\frac{{\rm d}\nu}{2}\,\eta^\parallel_i(B)+{\cal O}\Bigl(\tfrac{1}{eB}\Bigr), \label{eq:sf_idd}
\end{equation}
for $i\in\{0,2\}$, and
\begin{equation}
 \Re\int_{-1}^1\frac{{\rm d}\nu}{2}\,\eta_1(B)=\Re\int_{-1}^1\frac{{\rm d}\nu}{2}\,\eta^\parallel_1(B)-\frac{k_\perp^2}{2}\,\Re\int_{-1}^1\frac{{\rm d}\nu}{2}\,\frac{1-\nu^2}{\phi_0^\parallel}+{\cal O}\Bigl(\tfrac{1}{eB}\Bigr). \label{eq:sf_idd3}
\end{equation}
Utilizing these findings as well as Eqs.~\eqref{eq:i_start}-\eqref{eq:1+nu/-nu^2} in Eqs.~\eqref{eq:intN0_s}-\eqref{eq:intN2_s}, it is straightforward to derive
\begin{align}
 \Re\int_{-1}^1\frac{{\rm d}\nu}{2}\,\eta_0(B)
&=\left(\frac{2}{3}+\frac{m^2-{\rm i}\epsilon}{eB}\right)\ln\biggl(\frac{m^2-{\rm i}\epsilon}{2eB}\biggr)+\frac{2\gamma}{3}+\frac{2}{3}+\Sigma+{\cal O}\Bigl(\tfrac{1}{eB}\Bigr), \label{eq:Reeta0} \\
 \Re\int_{-1}^1\frac{{\rm d}\nu}{2}\,\eta_1(B)
&=\left(eB-\frac{k_\perp^2}{2}\right)\,\Re\int_{-1}^1\frac{{\rm d}\nu}{2}\frac{1-\nu^2}{\phi_0^\parallel}+\frac{2}{3}\ln\biggl(\frac{m^2-{\rm i}\epsilon}{2eB}\biggr)+\frac{2\gamma}{3}+{\cal O}\Bigl(\tfrac{1}{eB}\Bigr), \\
 \Re\int_{-1}^1\frac{{\rm d}\nu}{2}\,\eta_2(B)
&=\frac{2}{3}\ln\biggl(\frac{m^2-{\rm i}\epsilon}{2eB}\biggr)+\frac{2\gamma}{3}+\Sigma+{\cal O}\Bigl(\tfrac{1}{eB}\Bigr), \label{eq:Reeta2}
\end{align}
where we defined
\begin{equation}
 \Sigma\equiv1+2\sum_{n=1}^\infty \left[1-\frac{1}{3n}+2n-2n(n+1)\ln\Bigl(\frac{n+1}{n}\Bigr)\right]=0.6052253730\ldots\,. \label{eq:infser}
\end{equation}
Notably the infinite sum~\eqref{eq:infser} converges, such that the results obtained in Eqs.~\eqref{eq:Reeta0}-\eqref{eq:Reeta2} are manifestly finite.
In summary, the real part of the photon polarization tensor~\eqref{eq:PI_comp_largeB} in the strong magnetic field limit can conveniently be represented as
\begin{multline}
\Re\left\{
 \begin{array}{c}
 \!\Pi_{\parallel}\!\\
 \!\Pi_{\perp}\!\\
 \!\Pi_{0}\!
 \end{array}
\right\}
=\frac{\alpha}{2\pi}\Biggl[
\left(eB-\frac{k_\perp^2}{2}\right)\Re\int_{-1}^1\frac{{\rm d}\nu}{2}\frac{(1-\nu^2)}{\phi_0^\parallel}
\left\{
 \begin{array}{c}
 \!k_\parallel^2\! \\
 0 \\
 0
 \end{array}
\right\} \\
+
\frac{2}{3}\left[\ln\biggl(\frac{m^2}{2eB}\biggr)+\gamma\right]k^2+
\Sigma
\left\{
 \begin{array}{c}
 \!k_\perp^2\! \\
 \!k^2\! \\
 \!k^2\!
 \end{array}
\right\}
+\left[\frac{2}{3}+\frac{m^2}{eB}\ln\left(\frac{m^2}{2eB}\right)\right]
\left\{
 \begin{array}{c}
 \!k_\perp^2\! \\
 \!k_\parallel^2\! \\
 \!k^2\!
 \end{array}
\right\}
+{\cal O}\Bigl(\tfrac{1}{eB}\Bigr)\Biggr]. \label{eq:PisfB_Re}
\end{multline}
The contributions to \Eqref{eq:PisfB_Re} at ${\cal O}(eB)$ have already been determined by \cite{Skobelev:1975,Shabad:1976,Melrose:1976dr}.
While the polarization tensor for the $\parallel$ mode features a term which depends linearly on the magnetic field strength, the corresponding expressions for the other modes start contributing only at order ${\cal O}(\ln(eB))$ and ${\cal O}((eB)^0)$.
Particularly the term $\sim eB$ has various important phenomenological consequences: It leads to essential deviations of the photon dispersion law from $k^2=0$; cf. \cite{Shabad:2003}. Moreover, it is responsible for modifications of the Coulomb potential in the presence of a strong magnetic field, cf. \cite{Shabad:2007xu,Sadooghi:2007ys,Machet:2010yg}, and gives rise to screening effects, cf. \cite{Shabad:2007xu,Shabad:2007zu,Godunov:2011aa}.
All terms compatible with $n=l=0$ in Eqs.~\eqref{eq:sf_block1_2}-\eqref{eq:intN2_s}  can be traced back to the function $N_1$, giving rise to imaginary parts in Eqs.~\eqref{eq:sf_block2_2} and \eqref{eq:intN1_s} only.
With the help of \Eqref{eq:ImpartsBf} it is straightforward to derive
\begin{equation}
 \Im\int_{-1}^1\frac{{\rm d}\nu}{2}\,\eta^\parallel_1(B)=-eB\,\frac{4m^2}{k_\parallel^2}\frac{2\pi}{\sqrt{(k_\parallel^2+4m^2)k_\parallel^2}}\,\Theta(-k_\parallel^2-4m^2)
\end{equation}
and
\begin{equation}
 \Im\int_{-1}^1\frac{{\rm d}\nu}{2}\int\limits_{0}^{\infty-{\rm i}\eta}\frac{{\rm d} s}{s}{\rm e}^{-{\rm i}\phi_0^\parallel s}\left({\rm e}^{-{\rm i}n_2k_\perp^2s}-1\right)N_1(eBs)=-\left(1-{\rm e}^{-\frac{k_\perp^2}{2eB}}\right)\Im\int_{-1}^1\frac{{\rm d}\nu}{2}\,\eta^\parallel_1(B)\,.
\end{equation}
All other terms do not feature an imaginary part in the strong field limit.
Correspondingly the full imaginary part of the photon polarization tensor~\eqref{eq:PI_comp_largeB} in the strong magnetic field limit is given by [cf. \Eqref{eq:Pieta} and below \Eqref{eq:valin_k^2=0}]
\begin{equation}
 \Im\{\Pi_\parallel\}={\rm e}^{-\frac{k_\perp^2}{2eB}}\,\Im\{\left.\Pi_\parallel\right|_{k_\perp^2=0}\}=-\alpha\,eB\,{\rm e}^{-\frac{k_\perp^2}{2eB}}\,\frac{4m^2}{\sqrt{(k_\parallel^2+4m^2)k_\parallel^2}}\,\Theta(-k_\parallel^2-4m^2)\,, \label{eq:Im_sf}
\end{equation}
while $\Im\{\Pi_\perp\}=\Im\{\left.\Pi_\perp\right|_{k^2_\perp=0}\}=\Im\{\Pi_0\}=\Im\{\left.\Pi_0\right|_{k^2_\perp=0}\}=0$. Thus, also the imaginary part of the photon polarization tensor for the $\parallel$ mode features a linear magnetic field dependence. 
The exponential suppression $\sim\exp\{k_\perp^2/(2eB)\}$, indicates the nonperturbative nature of \Eqref{eq:Im_sf}.
We emphasize that Eqs.~\eqref{eq:PisfB_Re} and \eqref{eq:Im_sf}, constituting the photon polarization tensor in the strong magnetic field limit, are one of our main results in this section.
The other main result is an analogous expression of the polarization tensor in the limit of strong electric fields; see \Eqref{eq:PI_sfE} below.  

Even though it is well known, and has been discussed in detail in the literature (cf., e.g., \cite{Shabad:2003}), that the dispersion law for photons propagating in such strong electromagnetic fields deviates significantly from the dispersion law at zero field, $k^2=0$, we subsequently state the results for the quantities $\kappa_p$ and $n_p$ as defined in Eqs.~\eqref{eq:np} and \eqref{eq:kappa_p}, involving an explicit constraint to $k^2=0$.
We do this solely for the sake of an easier comparison with previous results derived in the literature. Of course, these quantities can no longer be identified with the physical indices of refraction and the physical absorption coefficients.
Equation~\eqref{eq:Im_sf} results in
\begin{equation}
\left\{
 \begin{array}{c}
 \!\kappa_{\parallel}\!\\
 \!\kappa_{\perp}\!
 \end{array}
\right\}=\alpha\,\sin\theta\,\left\{
 \begin{array}{c}
 \!1\! \\
 \!0\!
 \end{array}
\right\}
\frac{eB}{\omega^2\sin^2\theta}\,{\rm e}^{-\frac{\omega^2\sin^2\theta}{2eB}}\,\frac{4m^2}{\sqrt{\omega^2\sin^2\theta-4m^2}}\,\Theta(\omega\sin\theta-2m)\,,  \label{eq:kappa_sfB}
\end{equation}
where we made use of the fact that $k^2=0$ of course implies $k_\perp^2=-k_\parallel^2=\omega^2\sin^2\theta$ [cf. \Eqref{eq:olcd}].
Moreover, keeping terms at order $\omega^2$ only,
we exactly reproduce the expression as derived by Tsai and Erber for the case of very-low-energy photons and $\frac{eB}{m^2}\gg1$, given in Eq.~(38) of \cite{Tsai:1975iz} and written in a slightly different representation,
\begin{equation}
\left\{
 \begin{array}{c}
 \!n_{\parallel}\!\\
 \!n_{\perp}\!
 \end{array}
\right\}=
1+\frac{\alpha}{4\pi}\sin^2\theta\Biggl[
\left(\frac{2}{3}\frac{eB}{m^2}-\Sigma\right)
\left\{
 \begin{array}{c}
 \!1\! \\
 \!0\!
 \end{array}
\right\}
-\left[\frac{2}{3}+\frac{m^2}{eB}\ln\left(\frac{m^2}{2eB}\right)\right]
\left\{
 \begin{array}{c}
 \!1 \\
 \!-1\!
 \end{array}
\right\}
+{\cal O}\Bigl(\tfrac{1}{eB}\Bigr)+{\cal O}(\omega^2)
\Biggr]. \label{eq:np_sf}
\end{equation}
Notably, \Eqref{eq:np_sf} has also been derived by \cite{Heyl:1997hr} pursuing an alternative approach.
By a comparison of \Eqref{eq:np_sf} with Eq.~(38) of \cite{Tsai:1975iz}, we find the following identity,
\begin{equation}
 \Sigma=8L_1-\frac{2\gamma}{3}-1\,, \label{eq:SigmaL1}
\end{equation}
where the constant $L_1$ can be obtained from the Raabe integral \cite{Tsai:1975iz},
\begin{equation}
 L_1=\frac{1}{3}+\int_0^1{\rm d}\chi\,\ln\Gamma_1(1+\chi)=0.248754477\ldots\,,
\end{equation}
with the logarithm of the generalized Gamma function given by \cite{Tsai:1975iz,Bendersky:1933}
\begin{equation}
 \ln\Gamma_1(\chi)=\int_0^\chi{\rm d}t\,\ln\Gamma(t)+\frac{\chi}{2}(\chi-1)-\frac{\chi}{2}\ln(2\pi)\,.
\end{equation}
Correspondingly, \Eqref{eq:SigmaL1} provides us with an explicit integral representation of the infinite sum~\eqref{eq:infser}.
Alternatively, the constant $L_1$ -- and thus the sum~\eqref{eq:infser} -- can also be expressed in terms of the first derivative of the Riemann $\zeta$-function \cite{Dittrich:2000zu},
\begin{equation}
 L_1=\frac{1}{12}-\zeta'(-1)\,. 
\end{equation}

Let us finally note that the corresponding true physical refractive indices and absorption coefficients could be obtained form Eqs.~\eqref{eq:PisfB_Re} and \eqref{eq:Im_sf} by taking into account the {\it modified light cone condition} in the {\it polarized quantum vacuum} $k^2+\Pi_p=0$ for each polarization mode $p$.

For an electric field the relevant $\nu$ integrals follow from \Eqref{eq:basicnuint_E}.
As already noted below \Eqref{eq:basicnuint_E}, in case of an electric field there is no threshold condition for pair production and the respective integrals are genuinely complex.
The expression for the photon polarization tensor in the strong electric field limit then reads
\begin{multline}
\left\{
 \begin{array}{c}
 \!\Pi_{\perp}\!\\
 \!\Pi_{\parallel}\!\\
 \!\Pi_{0}\!
 \end{array}
\right\}
=\frac{\alpha}{2\pi}\Biggl[
-\left({\rm i}eE+\frac{k_\parallel^2}{2}\right)\int_{-1}^1\frac{{\rm d}\nu}{2}\frac{(1-\nu^2)}{\phi_0^\perp}
\left\{
 \begin{array}{c}
 \!k_\perp^2\! \\
 0 \\
 0
 \end{array}
\right\} \\
+
\frac{2}{3}\left[\ln\biggl(\frac{m^2}{2eE}\biggr)+\gamma+{\rm i}\frac{\pi}{2}\right]k^2+
\Sigma
\left\{
 \begin{array}{c}
 \!k_\parallel^2\! \\
 \!k^2\! \\
 \!k^2\!
 \end{array}
\right\}
+\left[\frac{2}{3}+{\rm i}\frac{m^2}{eE}\ln\left(\frac{m^2}{2eE}\right)\right]
\left\{
 \begin{array}{c}
 \!k_\parallel^2\! \\
 \!k_\perp^2\! \\
 \!k^2\!
 \end{array}
\right\}
+{\cal O}\Bigl(\tfrac{1}{eE}\Bigr)\Biggr]. \label{eq:PI_sfE}
\end{multline}
Note that \Eqref{eq:PI_sfE} implies a strong absorption $\sim eE$ of photons polarized in the $\perp$ mode and propagating perpendicular to the electric field.
Contrarily, the real part starts contributing only at ${\cal O}((eE)^0)$.
In contrast to \Eqref{eq:Im_sf}, which basically holds throughout the strong magnetic field regime,
the imaginary part of the terms written explicitly in \Eqref{eq:PI_sfE} -- like the respective real part -- receives corrections at ${\cal O}(\frac{1}{eE})$.
Again, this is an explicit manifestation of the discrete kinematic threshold structure governing pair creation in a magnetic field, as compared to the possibility of pair creation for arbitrary kinematics in the presence of an electric field.

\section{Conclusions}\label{sec:conclusions}

In this paper, we have studied in detail the photon polarization tensor in the presence of a homogeneous, purely magnetic and electric field, respectively.
The simultaneous study of the magnetic field case together with the electric field case was strongly motivated by the electric-magnetic duality~\eqref{eq:trafo}.

While the similarities between the (proper time) expressions of the photon polarization tensor for purely magnetic and electric fields were recognized from the days of its very first derivation, 
and the formal correspondence $B\leftrightarrow{\rm i}E$ and $\parallel\leftrightarrow\perp$ [cf.~\Eqref{eq:formcorresp}] even employed in Urrutia's derivation of the polarization tensor {\it in parallel homogeneous electric and magnetic fields} \cite{Urrutia:1977xb}, we are not aware of a discussion of the electric-magnetic duality~\eqref{eq:trafo} in the context of the photon polarization tensor. 
As detailed in this paper, e.g., in the special case of a magnetic field but $k_\perp^2=0$, the electric-magnetic duality allows us to explicitly perform the propertime integration, irrespective of the kinematics conditions
-- particularly also above the pair creation threshold.

Our main focus was on explicit results for various well-specified physical parameter regimes.
After shortly reviewing the perturbative weak field expansion, focusing on its basic structure and regime of applicability, we studied in great detail 
\begin{itemize}
 \item approximations \'{a} la Tsai and Erber:
While we retraced the original approach of Tsai and Erber to study {\it the propagation of photons in homogeneous magnetic fields} \cite{Tsai:1974fa,Tsai:1975iz},
we did not restrict ourselves to on-the-light-cone dynamics from the outset and in particular kept track of all the contributions arising in this type of expansion.
In this context, we have in particular obtained new analytical results for the imaginary part of the photon polarization tensor at leading order in a $1/\xi$ expansion (cf. Sec.~\ref{subsec:xitoinfty}),
leading to particularly handy expressions for the photon absorption coefficient $\kappa_p$ of photons polarized in mode $p=\{\parallel,\perp\}$ in both pure magnetic and electric fields.
We have compared these expressions with both the original expressions derived by Tsai and Erber \cite{Tsai:1974fa,Tsai:1975iz}, and results obtained with the method of stationary phase by \cite{Baier:2007dw,Baier:2009it,Dunne:2009gi}.
Moreover, in Sec.~\ref{subsec:xito0} we have highlighted the problems of an expansion in $\xi\to0$, and have explicitly studied the restrictions to be imposed on the parameter regime for the leading contribution to scale $\sim(ef)^{2/3}$, with $f\in\{B,E\}$,
both in case of on- and off-the-light-cone dynamics.
\end{itemize}
and the
\begin{itemize}
 \item strong field limit:
In Sec.~\ref{sec:strongfield} we have obtained analytical insights into the strong field limit, $ef\gg\{m^2,|k_\parallel^2|,k_\perp^2\}$, by performing formal expansions in both $y={\rm e}^{-{\rm i}z}$ and $k_\perp^2/(eB)$ for magnetic fields,
and analogously $y={\rm e}^{-z'}$ and $k_\parallel^2/(eE)$ for electric fields. Having implemented the above expansions,
and resorting to a different representation of the original contact term, more amenable to a strong field expansion,
the propertime integrals could be carried out straightforwardly.
In turn, also the remaining parameter integral over $\nu$ could be performed explicitly. Expanding the results of these integrations in powers of the inverse field strength $1/(ef)$,
we have obtained closed form analytical expressions of the photon polarization tensor in the strong field limit. Here we have neglected all terms suppressed by at least the inverse field strength, i.e., terms of ${\cal O}(1/(ef))$.
Noteworthy, the strong field expansion naturally induces logarithmic contributions of the structure $\sim(ef)^{-n}\ln(ef)$, with $n\in\mathbb{N}_0$. As such contributions can be large for very strong fields,
we have also explicitly accounted for the corresponding leading logarithmic corrections.
\end{itemize}

Besides providing for reliable analytical results into various physical parameter regimes for homogeneous electric and magnetic fields, our study is also relevant beyond the constant field limit,
namely for inhomogeneous field configurations that may locally be approximated by a constant: For inhomogeneities with a typical scale of variation $w$ much larger than the Compton wavelength of the virtual particles, $w\gg\lambda_c=1/m$,
using the constant field expressions locally is well justified \cite{Gies:2013yxa}.
Of course, also the photon polarization tensor for other field configurations and in scalar QED \cite{Witte:1990,Schubert:2000yt} could be studied along the very same lines.

\section*{Acknowledgments}

It is a great pleasure to thank H.~Gies for many enlightening and stimulating discussions and support, as well as for valuable comments on the manuscript. Moreover, I would like to thank B.~D\"obrich for various helpful conversations and discussions.
Motivating and entertaining conversations with M.~Zepf and M.~Tingu are gratefully acknowledged.
I am grateful to S.~Gavrilov and A.~Shabad for helpful remarks on the manuscript.
Finally, I would like to thank G.~Raffelt for bringing a misprint in Eqs.~\eqref{eq:Tsai_smallxi} and \eqref{eq:soho1} in a previous version of this paper to my attention.

\pagebreak
\appendix

\section{Useful identities}

In this appendix we collect useful identities and intermediate steps omitted in the main text.

To arrive at \Eqref{eq:vac0} from \Eqref{eq:PI_zerofield}, and again in the derivation of \Eqref{eq:PI_tsai}, we made use of the following identity,
\begin{multline}
 \int_{-1}^{1}\frac{{\rm d}\nu}{2}(1-\nu^2){\rm e}^{-{\rm i}k^2\frac{1-\nu^2}{4}s-{\rm i}a(1-\nu^2)^2} \\
 =\int_{-1}^{1}\frac{{\rm d}\nu}{2}(1-\nu^2)+\frac{\rm i}{2}\int_{-1}^{1}\frac{{\rm d}\nu}{2}\left(\frac{\nu^2}{3}-1\right)\nu^2\left[k^2s+8a (1-\nu^2)\right]{\rm e}^{-{\rm i}k^2\frac{1-\nu^2}{4}s-{\rm i}a(1-\nu^2)^2}\,, \label{eq:partint}
\end{multline}
which is obtained straightforwardly by employing integrations by parts.
In the first case we needed \Eqref{eq:partint} with $a=0$, while in the second case we had $a=k_\perp^2(eB)^2s^3/48$.

\subsection{Explicit evaluation of the $\nu$ integrals in \Eqref{eq:nuintegral}} \label{app:Bint}

The integrals~\eqref{eq:nuintegral} with $j=1,2$ can be expressed in terms of the most basic one with $j=0$ by employing the following identities,
\begin{align}
 \int{\rm d}\nu\,\frac{\nu}{a+2b\nu+c\nu^2}&=\frac{1}{2c}\int{\rm d}\nu\,\partial_{\nu}\ln\left(a+2b\nu+c\nu^2\right)-\frac{b}{c}\int{\rm d}\nu\,\frac{1}{a+2b\nu+c\nu^2}\,, \label{eq:lala} \\ 
 \int{\rm d}\nu\,\frac{\nu^2}{a+2b\nu+c\nu^2}&=\int{\rm d}\nu\,\partial_{\nu}\biggl[\frac{\nu}{c}-\frac{b}{c^2}\ln\left(a+2b\nu+c\nu^2\right)\biggr]+\frac{2b^2-ac}{c^2}\int{\rm d}\nu\,\frac{1}{a+2b\nu+c\nu^2}\,. \label{eq:lalala}
\end{align}
Here and in the remainder of this section we omit the integration constants for indefinite integrals.
In consequence of the ${\rm i}\epsilon$ contained in the parameter $a$ [cf. \Eqref{eq:abc}], for a magnetic field the logarithm in Eqs.~\eqref{eq:lala} and \eqref{eq:lalala} can always be rewritten in the following form:
\begin{equation}
 \ln\left(a+2b\nu+c\nu^2\right)=\ln\left|a+2b\nu+c\nu^2\right|-{\rm i}\pi\Theta\left(-a-2b\nu-c\nu^2\right), \label{eq:lalalala}
\end{equation}
and thus is purely real valued for $\nu=\pm1$.

In case of a magnetic field, the integral~\eqref{eq:nuintegral} with $j=0$ can be performed with the help of formula 2.103.4 of \cite{Gradshteyn},
\begin{align}
 \int{\rm d}\nu\,\frac{1}{a+2b\nu+c\nu^2}&=\frac{1}{\sqrt{ac-b^2}}\arctan\biggl(\frac{c\nu+b}{\sqrt{ac-b^2}}\biggr), \label{eq:basicnuint_below}
\end{align}
which holds for $ac>b^2$, or equivalently, expressed in terms of physical parameters [cf. \Eqref{eq:abc}],
\begin{equation}
 \left(\sqrt{m^2+2eBn}-\sqrt{m^2+2(n+l)eB}\right)^2<-k_\parallel^2<\left(\sqrt{m^2+2eBn}+\sqrt{m^2+2(n+l)eB}\right)^2. \label{eq:ac>b^2}
\end{equation}
Under the above conditions, \Eqref{eq:basicnuint_below} is purely real valued.
By analytical continuation, it is straightforward to  derive results for the complementary regime, $ac<b^2$, and more generally for genuinely complex parameters $a$, $b$ and $c$ from \Eqref{eq:basicnuint_below} also. 
For magnetic fields, $B\geq0$,
and $ac<b^2$ we obtain
\begin{multline}
 \int{\rm d}\nu\,\frac{1}{a+2b\nu+c\nu^2}=
\frac{1}{2}\frac{1}{\sqrt{b^2-ac}}\biggl\{\ln\biggl|\frac{\sqrt{b^2-ac}-c\nu-b}{\sqrt{b^2-ac}+c\nu+b}\biggr|  -{\rm i}\pi\,{\rm sign}\bigl[k_\parallel^2(c\nu+b)\bigr]\Theta\bigl(|c\nu+b|-\sqrt{b^2-ac}\bigr)\biggr\}. \label{eq:Bsfint_abovethresh}
\end{multline}
This expression develops an imaginary part if the following condition is met,
\begin{equation}
 |c\nu+b|-\sqrt{b^2-ac}\geq0 \quad\leftrightarrow\quad (a+2b\nu+c\nu^2)c\geq 0\,, \label{eq:imagpartB_ineq}
\end{equation}
and for $\nu=\pm1$ simply implies an imaginary part for $-k_\parallel^2\geq0$.
Taking into account this condition, the imaginary part only survives the integration of \Eqref{eq:Bsfint_abovethresh} from $\nu=-1$ to $\nu=+1$, if
\begin{equation}
 {\rm sign}(-c-b)-{\rm sign}(c-b)\neq0 \quad\leftrightarrow\quad -k_\parallel^2>2eBl\,. \label{eq:t1}
\end{equation}
By expanding out the right-hand side, one can straightforwardly prove that also the following inequality holds,
\begin{equation}
 2eBl\geq\left(\sqrt{m^2+2eBn}-\sqrt{m^2+2(n+l)eB}\right)^2\,,
\end{equation}
which, in combination with \Eqref{eq:t1}, rules out an imaginary part of the photon polarization tensor in the regime [cf. \Eqref{eq:ac>b^2}]
\begin{equation}
 -k_\parallel^2\leq\left(\sqrt{m^2+2eBn}-\sqrt{m^2+2(n+l)eB}\right)^2. \label{eq:Bpairthresh}
\end{equation}
Correspondingly, the imaginary part of \Eqref{eq:Bsfint_abovethresh} integrated over the full $\nu$ interval $\nu\in[-1\ldots1]$
can be written as
\begin{equation}
 \Im\int_{-1}^1{\rm d}\nu\,\frac{1}{a+2b\nu+c\nu^2}=2\pi\Theta\Bigl(-k_\parallel^2-\left(\sqrt{m^2+2eBn}+\sqrt{m^2+2(n+l)eB}\right)^2\Bigr).
\end{equation}

\section{Series representations of the scalar function $N_i(z)$ and $n_2(z)$}\label{app:seriesNn}

With the definition $\csc z=(\sin z)^{-1}$, the scalar functions in~\Eqref{eq:scalarfcts_B} featuring a $z$ dependence can be rewritten as follows,
\begin{align}
 N_0(z)&=z\left[\csc z-\left(\partial_z\csc z\right)\partial_z\right]\left(\cos \nu z -\cos z\right)\,, \nonumber \\
 N_1(z)&=z\left(1-\nu^2\right)\cot z\,, \nonumber\\
 N_{2}(z)&=z\left(\cos \nu z -\cos z\right)\left(1+\partial_z^2\right)\csc z\,, \nonumber\\
 n_{2}(z)&=(2z)^{-1}\csc z\left(\cos{\nu z}-\cos{z}\right)\,.  \label{eq:scalarfcts_B_anders}
\end{align}
For completeness, also note the following relation,
\begin{equation}
 \cos{\nu z}-\cos{z}=2\sin\left(\tfrac{1+\nu}{2}z\right)\sin\left(\tfrac{1-\nu}{2}z\right).
\end{equation}
Here we aim at series representations of the form~\eqref{eq:Ni},
\begin{equation}
 N_i(z)=\sum_{n=0}^{\infty}N_i^{(2n)}z^{2n} \quad\quad{\rm and}\quad\quad n_2(z)=\sum_{n=0}^{\infty}n_2^{(2n)}z^{2n}\,, \label{eq:Ni_app}
\end{equation}
with $i\in\{0,1,2\}$.
Employing the series representation,
\begin{equation}
 \cos{\nu z}-\cos{z}=\sum_{j=0}^\infty\frac{(-1)^j}{[2(j+1)]!}\left(1-\nu^{2(j+1)}\right)z^{2(j+1)}\,,
\end{equation}
as well as formulae 1.411.7 and 1.411.11 of \cite{Gradshteyn}, both valid for $|z|<\pi$,
\begin{equation}
  \csc(z)=\sum_{n=0}^{\infty}(-1)^{n-1}\,\frac{(2^{2n}-2){\cal B}_{2n}}{(2n)!}\,z^{2n-1}\,, \quad\quad
 \cot(z)=\sum_{n=0}^{\infty}(-1)^{n}\,\frac{2^{2n}{\cal B}_{2n}}{(2n)!}\,z^{2n-1}\,, \label{eq:csc&cot}
\end{equation}
and using the Cauchy product, we obtain
\begin{align}
 N_0^{(2n)}&=(-1)^n\sum_{j=0}^n\frac{(2^{2j}-2)\,{\cal B}_{2j}}{[2(n-j)]!\,(2j)!}
 \left[\left(1-\nu^{2(n-j)}\right)+\frac{2j-1}{2(n-j)+1}\left(1-\nu^{2(n-j+1)}\right)\right], \nonumber \\
 N_1^{(2n)}&=\left(1-\nu^2\right)(-1)^n\frac{2^{2n}\,{\cal B}_{2n}}{(2n)!}\,, \nonumber\\
 N_{2}^{(2n)}&=(-1)^n\sum_{j=0}^n\frac{(2^{2j}-2)\,{\cal B}_{2j}}{[2(n-j)]!\,(2j)!}
 \left[\left(1-\nu^{2(n-j)}\right)-\frac{j-1}{n-j+1}\frac{2j-1}{2(n-j)+1}\left(1-\nu^{2(n-j+1)}\right)\right], \nonumber \\
 n_2^{(2n)}&=\frac{(-1)^{n-1}}{2}\sum_{j=0}^n\frac{(2^{2j}-2){\cal B}_{2j}}{[2(n-j+1)]!(2j)!}\left(1-\nu^{2(n-j+1)}\right), \label{eq:koeffs}
\end{align}
where ${\cal B}_{i}$ ($i\in\mathbb{N}_0$) denote Bernoulli numbers.
As all the functions $N^{(2n)}_i$ with $i\in\{0,1,2\}$, and $n_2^{(2n)}$ contain either a factor of $\cot(z)$ and $\csc(z)$ or its derivatives [cf. \Eqref{eq:scalarfcts_B_anders}], the series representations in \Eqref{eq:Ni_app}
are valid for $|z|<\pi$ [cf. \Eqref{eq:csc&cot}].
Equation~\eqref{eq:koeffs} in particular implies
\begin{gather}
N_0^{(0)}=N_1^{(0)}=N_2^{(0)}=1-\nu^2\,, \\
 N_0^{(2)}=\frac{(1-\nu^2)^2}{6}\,,\quad\quad N_1^{(2)}=-\frac{1-\nu^2}{3}\,, \quad\quad N_2^{(2)}=\frac{(1-\nu^2)(5-\nu^2)}{12}\,,
\end{gather}
as well as
\begin{align}
 n_2^{(0)}&=\frac{1-\nu^{2}}{4}\,, \quad\quad
 n_2^{(2)}=\frac{1}{3}\left(\frac{1-\nu^2}{4}\right)^2\,,\quad\quad\text{and}\quad\quad
 n_2^{(4)}=\frac{3-\nu^2}{90}\left(\frac{1-\nu^2}{4}\right)^2\,.
\end{align}

Note that for $n\in\mathbb{N}$, $n_2^{(2n)}$ is at least of ${\cal O}\left((1-\nu^2)^2\right)$, i.e., an overall factor $\sim(1-\nu^2)^2$ can be factored out.
To see this, we substitute $\nu^2\equiv1-\tilde u$ in the expression for $n_2^{(2n)}$ as provided in \Eqref{eq:koeffs} and make use of the Binomial theorem to rewrite
\begin{equation}
 1-\nu^{2(n-j+1)}=1-(1-\tilde u)^{n-j+1}=-\sum_{l=1}^{n-j+1}(-1)^l\binom{n-j+1}{l}\,\tilde u^l\,, \label{eq:binom}
\end{equation}
resulting in
\begin{equation}
 n_2^{(2n)}=\frac{(-1)^{n}}{2}\sum_{j=0}^n\frac{(2^{2j}-2){\cal B}_{2j}}{[2(n-j+1)]!(2j)!}\sum_{l=1}^{n-j+1}(-1)^l\binom{n-j+1}{l}\,\tilde u^l\,,
\end{equation}
and focus on the contribution linear in $u$, given by
\begin{equation}
 \left.n_2^{(2n)}\right|_{\sim\tilde u}=-\frac{(-1)^{n}}{4}\sum_{j=0}^n\frac{(2^{2j}-2){\cal B}_{2j}}{(2n-2j+1)!(2j)!}\,\tilde u\,.
\end{equation}
By comparison with \Eqref{eq:koeffs}, we infer
\begin{equation}
 \left.n_2^{(2n)}\right|_{\sim\tilde u}=-\left.\partial_{\nu}n_2^{(2n)}\right|_{\nu=1}\frac{\tilde u}{2}\,.
\end{equation}
From $\left.\partial_{\nu}n_2\right|_{\nu=1}=-\frac{1}{2}$ [cf. \Eqref{eq:scalarfcts_B_anders}], it is obvious that there is a nonvanishing contribution for $n=0$ only, such that
\begin{equation}
 \left.n_2^{(2n)}\right|_{\sim\tilde u}=
  \begin{cases}
    \frac{\tilde u}{4}\quad&\text{for}\quad n=0\,, \\
    0 \quad&\text{for}\quad n\in\mathbb{N}\,,
  \end{cases}
\end{equation}
and correspondingly
\begin{equation}
 n_2^{(2n)}=
 \begin{cases}
  \frac{1-\nu^2}{4}\quad&\text{for}\quad n=0\,, \\
  {\cal O}\left((1-\nu^2)^2\right) \quad&\text{for}\quad n\in\mathbb{N}\,.
 \end{cases}
\end{equation}

Substituting $\nu^2\equiv1-\tilde u$ in the other expressions in \Eqref{eq:koeffs} and employing \Eqref{eq:binom}, it is immediately obvious that $N_{0}^{(2n)}$, $N_{1}^{(2n)}$ and $N_{2}^{(2n)}$ share an overall factor of $(1-\nu^2)$ to be factored out.

\section{Identities and series representations relevant for Section~\ref{sec:TsaiErber}}\label{app:AiryBessel}

The Airy functions can be represented in terms of Bessel functions \cite{Gradshteyn}. For $\Re(\chi)\geq0$,
\begin{align}
 {\rm Ai}(\chi)&=\frac{\sqrt{\chi}}{3}\left[{\rm I}_{-1/3}\!\left(\tfrac{2}{3}\chi^{3/2}\right)-{\rm I}_{1/3}\!\left(\tfrac{2}{3}\chi^{3/2}\right)\right]=\frac{1}{\pi}\sqrt{\frac{\chi}{3}}\;{\rm K}_{1/3}\!\left(\tfrac{2}{3}\chi^{3/2}\right), \label{eq:AiIK+} \\
 {\rm Bi}(\chi)&=\sqrt{\frac{\chi}{3}}\left[{\rm I}_{-1/3}\!\left(\tfrac{2}{3}\chi^{3/2}\right)+{\rm I}_{1/3}\!\left(\tfrac{2}{3}\chi^{3/2}\right)\right],
\end{align}
and for $\Re(\chi)\leq0$:
\begin{align}
 {\rm Ai}(\chi)&=\frac{\sqrt{-\chi}}{3}\left[{\rm J}_{-1/3}\!\left(\tfrac{2}{3}(-\chi)^{3/2}\right)+{\rm J}_{1/3}\!\left(\tfrac{2}{3}(-\chi)^{3/2}\right)\right], \label{eq:AiIK-} \\
 {\rm Bi}(\chi)&=\sqrt{\frac{-\chi}{3}}\left[{\rm J}_{-1/3}\!\left(\tfrac{2}{3}(-\chi)^{3/2}\right)-{\rm J}_{1/3}\!\left(\tfrac{2}{3}(-\chi)^{3/2}\right)\right],
\end{align}
with ${\rm J}_\kappa(\chi)$ the Bessel function of the first kind, and ${\rm I}_\kappa(\chi)$ and ${\rm K}_\kappa(\chi)$ the modified Bessel functions of the first kind and second kind, respectively.

Given that $0\leq\nu<1$ and $|B^2k_{\perp}^2|\neq0$, the condition $\Re({\rm sign}(B^2k_{\perp}^2)\tilde\xi^{2/3})\geq0$ is equivalent to $\Re(B^2k_{\perp}^2\phi_0)\geq0$.
Specifying \Eqref{eq:Gi} to $\chi=\pm{\rm sign}(B^2k_{\perp}^2)(\frac{3}{2}\tilde\xi)^{2/3}$ and employing definition~\eqref{eq:xi!} as well as the following identities (cf. formulae 6.511.3 and 6.511.11 of \cite{Gradshteyn}),
\begin{align}
 \int_0^{{\rm sign}(B^2k_{\perp}^2)\left(\tfrac{3}{2}\tilde\xi\right)^{2/3}}{\rm d}t\,\sqrt{t}\,{\rm I}_{\pm1/3}\!\left(\tfrac{2}{3}t^{3/2}\right)&=\int_0^{\xi_+}{\rm d}t\,{\rm I}_{\pm1/3}\!\left(t\right)=2\sum_{n=0}^{\infty}(-1)^n\,{\rm I}_{2n+1\pm1/3}\!\left(\xi_+\right), \label{eq:intI} \\
 \int_0^{-{\rm sign}(B^2k_{\perp}^2)\left(\tfrac{3}{2}\tilde\xi\right)^{2/3}}{\rm d}t\,\sqrt{t}\,{\rm J}_{\pm1/3}\!\left(\tfrac{2}{3}t^{3/2}\right)&=\int_0^{\xi_-}{\rm d}t\,{\rm J}_{\pm1/3}\!\left(t\right)=2\sum_{n=0}^{\infty}{\rm J}_{2n+1\pm1/3}\!\left(\xi_-\right), \label{eq:intJ}
\end{align}
also ${\rm Gi}\left({\rm sign}(B^2k_{\perp}^2)(\frac{3}{2}\tilde\xi)^{2/3}\right)$ can be expressed in terms of Bessel functions; for $\Re(B^2k_{\perp}^2\phi_0)\geq0$,
\begin{multline}
 {\rm Gi}\left({\rm sign}(B^2k_{\perp}^2)(\tfrac{3}{2}\tilde\xi)^{2/3}\right)=\frac{1}{3\sqrt{3}}\left[{\rm sign}(B^2k_\perp^2)\left(\tfrac{3}{2}\tilde\xi\right)^{2/3}\right]^{\frac{1}{2}}\biggl\{{\rm I}_{-1/3}\!\left(\xi_+\right)+{\rm I}_{1/3}\!\left(\xi_+\right)\\ +4\sum_{n=0}^{\infty}(-1)^n
 \bigl[{\rm I}_{-1/3}\!\left(\xi_+\right){\rm I}_{2n+4/3}\!\left(\xi_+\right)-{\rm I}_{1/3}\!\left(\xi_+\right){\rm I}_{2n+2/3}\!\left(\xi_+\right)\bigr]\biggr\},
\end{multline}
and for $\Re(B^2k_{\perp}^2\phi_0)\leq0$,
\begin{multline}
 {\rm Gi}\left({\rm sign}(B^2k_{\perp}^2)(\tfrac{3}{2}\tilde\xi)^{2/3}\right)=\frac{1}{3\sqrt{3}}\left[-{\rm sign}(B^2k_\perp^2)\left(\tfrac{3}{2}\tilde\xi\right)^{2/3}\right]^{\frac{1}{2}}\biggl\{{\rm J}_{-1/3}(\xi_-)-{\rm J}_{1/3}(\xi_-)\\ +4\sum_{n=0}^{\infty}
 \bigl[{\rm J}_{-1/3}(\xi_-){\rm J}_{2n+4/3}(\xi_-)-{\rm J}_{1/3}(\xi_-){\rm J}_{2n+2/3}(\xi_-)\bigr]\biggr\}.
\end{multline}
In turn, \Eqref{eq:tsaielm} can be written as an infinite sum of Bessel functions; for $\Re(B^2k_{\perp}^2\phi_0)\geq0$,
\begin{multline}
\int_{0}^{\infty}{\rm d} s\,{\rm e}^{-{\rm i}\phi_0(k^2)s - {\rm i} k_{\perp}^2\frac{(1-\nu^2)^2}{48}(eB)^2s^3}
 =\frac{\pi}{2}\frac{\tilde\xi^{2/3}}{\phi_0}\xi_+^{1/3}\Biggl\{\frac{\sqrt{3}}{\pi}{\rm K}_{1/3}(\xi_+)
-{\rm sign}(B^2k_\perp^2)\frac{{\rm i}}{\sqrt{3}} \\
\times\biggr[{\rm I}_{-1/3}(\xi_+)+{\rm I}_{1/3}(\xi_+)+4\sum_{n=0}^{\infty}(-1)^n\bigl[{\rm I}_{-1/3}(\xi_+)\,{\rm I}_{2n+4/3}(\xi_+)-{\rm I}_{1/3}(\xi_+)\,{\rm I}_{2n+2/3}(\xi_+)\bigr]\biggr]\Biggr\}, \label{eq:Tsai_int_s0+}
\end{multline}
and for $\Re(B^2k_{\perp}^2\phi_0)\leq0$,
\begin{multline}
\int_{0}^{\infty}{\rm d} s\,{\rm e}^{-{\rm i}\phi_0(k^2)s - {\rm i} k_{\perp}^2\frac{(1-\nu^2)^2}{48}(eB)^2s^3}
 =\frac{\pi}{2}\frac{\tilde\xi^{2/3}}{\phi_0}\xi_-^{1/3}\Biggl\{{\rm J}_{-1/3}(\xi_-)+{\rm J}_{1/3}(\xi_-) -{\rm sign}(B^2k_{\perp}^2)\frac{{\rm i}}{\sqrt{3}} \\
\times\biggl[{\rm J}_{-1/3}(\xi_-)-{\rm J}_{1/3}(\xi_-)
+4\sum_{n=0}^{\infty}\bigl[{\rm J}_{-1/3}(\xi_-)\,{\rm J}_{2n+4/3}(\xi_-)-{\rm J}_{1/3}(\xi_-)\,{\rm J}_{2n+2/3}(\xi_-)\bigr]\biggr]\Biggr\}. \label{eq:Tsai_int_s0-}
\end{multline}
Trading the derivatives with respect to $\phi_0$ for derivatives with respect to $\xi_\pm$ [cf. Eqs.~\eqref{eq:xi} and \eqref{eq:xi!}],
\begin{equation}
 \frac{\partial}{\partial\phi_0}=\frac{3}{2}\frac{\tilde\xi^{2/3}}{\phi_0}\left(\pm{\rm sign}(B^2k_\perp^2)\xi_\pm^{1/3}\frac{\partial}{\partial\xi_\pm}\right), \label{eq:diffphi0nachxi}
\end{equation}
and employing
\begin{equation}
 \frac{\partial}{\partial\phi_0}\biggl(\frac{\tilde\xi^{2/3}}{\phi_0}\biggr)=0\,,
\end{equation}
it is then straightforward to derive the corresponding expressions for the generic propertime integral~\eqref{eq:gentsaierberint}, as provided in Eqs.~\eqref{eq:Tsai_int_sj+} and \eqref{eq:Tsai_int_sj-}.
With the help of formula 8.406.1 of \cite{Gradshteyn},
\begin{equation}
 {\rm I}_{\kappa}(\chi)={\rm e}^{-{\rm i}\frac{\pi}{2}\kappa}{\rm J}_{\kappa}({\rm e}^{{\rm i}\frac{\pi}{2}}\chi)\,, \quad{\rm for}\quad -\pi<{\rm arg}(\chi)\leq\frac{\pi}{2}\,, \label{eq:BesselI->J}
\end{equation}
also \Eqref{eq:Tsai_int_s0+} can be entirely expressed in terms of Bessel functions of the first kind,

\subsection{Very weak fields - large momentum: The limit $\xi\to\infty$}

Taking into account \Eqref{eq:BesselI->J}, the behavior of Eqs.~\eqref{eq:Tsai_int_sj+} and \eqref{eq:Tsai_int_sj-} in the limit $\xi\to\infty$
can be inferred from the Bessel function of the first kind.
Its asymptotic expansion for large values of $\chi$, while $|{\rm arg}(\chi)|<\pi$, reads (cf. formula 8.451.1 of \cite{Gradshteyn})
\begin{multline}
 {\rm J}_{\pm\kappa}(\chi)=\sqrt{\frac{2}{\pi\chi}}\Biggl\{\cos\left(\chi\mp\tfrac{\pi}{2}\kappa-\tfrac{\pi}{4}\right)\Biggl[\sum_{k=0}^{l-1}c^{(2k)}_{\kappa}\frac{(-1)^k}{(2\chi)^{2k}}+{\cal O}\left((\tfrac{1}{\chi})^{2l}\right)\Biggr] \\
-\sin\left(\chi\mp\tfrac{\pi}{2}\kappa-\tfrac{\pi}{4}\right)\Biggl[\sum_{k=0}^{l-1}c^{(2k+1)}_{\kappa}\frac{(-1)^k}{(2\chi)^{2k+1}}+{\cal O}\left((\tfrac{1}{\chi})^{2l+1}\right)\Biggr]\Biggr\}, \label{eq:BesselJassympt}
\end{multline}
where we introduced
\begin{equation}
 c^{(n)}_{\kappa}=\frac{\Gamma(\kappa+n+\tfrac{1}{2})}{n!\,\Gamma(\kappa-n+\tfrac{1}{2})}\,. \label{eq:c_kappa^n}
\end{equation}
Equation~\eqref{eq:c_kappa^n} is even in $\kappa$, as for $n\in\mathbb{N}$ (formula 8.339.4 of \cite{Gradshteyn}),
\begin{equation}
 \frac{\Gamma(\kappa+n+\tfrac{1}{2})}{\Gamma(\kappa-n+\tfrac{1}{2})}=\frac{(4\kappa^2-1^2)(4\kappa^2-3^2)\ldots[4\kappa^2-(2n-1)^2]}{2^{2n}}\,. \label{eq:GammabyGamma}
\end{equation}

From Eqs.~\eqref{eq:BesselI->J} and \eqref{eq:BesselJassympt} it is straightforward to derive the asymptotic expansion of the modified Bessel function of the second kind (cf. also formula 8.451.6 of \cite{Gradshteyn}). Specifying to $\kappa=1/3$, one finds
\begin{equation}
 {\rm K}_{1/3}(\chi)=\sqrt{\frac{\pi}{2\chi}}\,{\rm e}^{-\chi}\Biggl[\sum_{k=0}^{l-1}\frac{1}{(2\chi)^k}\frac{\Gamma(\frac{5}{6}+k)}{k!\,\Gamma(\frac{5}{6}-k)}+{\cal O}\left((\tfrac{1}{\chi})^l\right)\Biggr]. \label{eq:assymptK}
\end{equation}
Moreover, note that
\begin{multline}
 {\rm J}_{-1/3}(\chi)+{\rm J}_{1/3}(\chi)=\sqrt{\frac{3}{\pi\chi}}\Biggl\{\cos(\chi)\left[\sum_{k=0}^{l-1}(-1)^k\left(c^{(2k)}_{1/3}\frac{1}{(2\chi)^{2k}}+c^{(2k+1)}_{1/3}\frac{1}{(2\chi)^{2k+1}}\right)+{\cal O}\left((\tfrac{1}{\chi})^{2l}\right)\right] \\
+\sin(\chi)\left[\sum_{k=0}^{l-1}(-1)^k\left(c^{(2k)}_{1/3}\frac{1}{(2\chi)^{2k}}-c^{(2k+1)}_{1/3}\frac{1}{(2\chi)^{2k+1}}\right)+{\cal O}\left((\tfrac{1}{\chi})^{2l}\right)\right]\Biggr\}.  \label{eq:assymptJ+J}
\end{multline}
For estimates of the remainders of the asymptotic series in Eqs.~\eqref{eq:BesselJassympt}, \eqref{eq:assymptK} and \eqref{eq:assymptJ+J} we refer the reader to \cite{Gradshteyn}.

\subsection{Weak fields - large momentum, and momentum dominance: The limit $\xi\to0$}\label{app:xi->0}

Aiming at results for $\xi_\pm\to0$, it is helpful to note that the (modified) Bessel function of the first kind has an exact series representation, formulae 8.440 and 8.445 of \cite{Gradshteyn},
\begin{equation}
 {\rm J}_{\kappa}(\chi)=\sum_{j=0}^{\infty}\frac{(-1)^j}{j!\,\Gamma(j+1+\kappa)}\left(\frac{\chi}{2}\right)^{2j+\kappa}\quad {\rm for}\quad |{\rm arg}(\chi)|<\pi\,, \quad
 {\rm and} \quad {\rm I}_{\kappa}(\chi)=\sum_{j=0}^{\infty}\frac{1}{j!\,\Gamma(j+1+\kappa)}\left(\frac{\chi}{2}\right)^{2j+\kappa}. \label{eq:Iseries}
\end{equation}
Given the structure of Eqs.~\eqref{eq:Tsai_int_sj+} and \eqref{eq:Tsai_int_sj-}, also the following identity, formula 8.442 of \cite{Gradshteyn},
\begin{equation}
 {\rm J}_{\kappa}(\chi){\rm J}_{\lambda}(\chi)=\sum_{j=0}^\infty\frac{(-1)^j}{j!}\frac{\frac{\Gamma(2j+1+\kappa+\lambda)}{\Gamma(j+1+\kappa+\lambda)}}{\Gamma(j+1+\kappa)\Gamma(j+1+\lambda)}\left(\frac{\chi}{2}\right)^{2j+\kappa+\lambda}, \label{eq:JJ}
\end{equation}
is useful. An analogous expression for modified Bessel functions can be obtained by employing \Eqref{eq:BesselI->J} in \Eqref{eq:JJ}.  

\subsection{Building blocks of the polarization tensor in the limit $\xi\to0$}\label{sec:basicbuild}

Here we provide more explicit expressions for the basic building blocks of \Eqref{eq:PI_tsai} in the limit $\xi\to0$, relevant for the discussion in Sec.~\ref{subsec:xito0}.
These expressions in particular confirm that the assumptions invoked in \Eqref{eq:c} give rise to a well-defined expansion scheme in terms of the small quantities $\frac{1}{\rho}$, $\frac{1}{c}$ and $\frac{c}{\rho}$.
With the help of Eqs.~\eqref{eq:Tsai_smallxiLO_1} and \eqref{eq:Tsai_smallxiLO_2}, we obtain
\begin{multline}
 \int_0^{c/\rho}\frac{{\rm d}{\tilde u}}{\sqrt{1-{\tilde u}}} \,\int_{0}^{\infty}{\rm d} s\,{\rm e}^{-{\rm i}\phi_0s - {\rm i} \frac{\tilde u^2}{48}k_{\perp}^2(eB)^2s^3}\,
 \left\{\begin{array}{c}
  (k^2)^2s^0\left[1+{\cal O}({\tilde u})\right]\\
 k_p^2\frac{z^{2n}}{s}\,{\tilde u}\left[1+{\cal O}({\tilde u})\right]\\
  k^2z^2k_{\perp}^2{\tilde u}\left[1+{\cal O}({\tilde u})\right]\\
 \end{array}\right\}\\
\sim\left\{\begin{array}{c}
  k^2\frac{c}{\rho}\left\{\frac{k^2}{c^{2/3}\phi_0(c/\rho)}\left[1+{\cal O}(\tfrac{c}{\rho})+{\cal O}(\tfrac{1}{c^{2/3}})\right]+\frac{k^2}{m^2}{\cal O}(1)\right\}\\
  k_p^2\left(\frac{c}{\rho}\right)^2\left\{\left(\frac{eB}{c^{2/3}\phi_0(c/\rho)}\right)^{2n}\left[1+{\cal O}(\tfrac{c}{\rho})+{\cal O}(\tfrac{1}{c^{2/3}})\right]\right\}\\
  k^2\left[1+{\cal O}(\tfrac{c}{\rho})+{\cal O}(\tfrac{1}{c^{2/3}})\right] \\
 \end{array}\right\}, \label{eq:x1to0_1}
\end{multline}
and
\begin{multline}
\int_0^{c/\rho}\frac{{\rm d}{\tilde u}}{\sqrt{1-{\tilde u}}} \, \int_{0}^{\infty}\frac{{\rm d} s}{s}\,{\rm e}^{-{\rm i}\phi_0s - {\rm i} \frac{\tilde u^2}{48}k_{\perp}^2(eB)^2s^3}k_p^2{\tilde u}\,\bigl\{-{\rm i}k_{\perp}^2s\,{\tilde u}^2\left[1+{\cal O}({\tilde u})\right]\bigr\}^{j}\!z^{2(n+2j)} \\
\sim k_p^2\left(\frac{c}{\rho}\right)^2\left(\frac{eB}{c^{2/3}\phi_0(c/\rho)}\right)^{2(n+j)}\left[1+{\cal O}(\tfrac{c}{\rho})+{\cal O}(\tfrac{1}{c^{2/3}})\right]  , \label{eq:x1to0_2}
\end{multline}
as well as
\begin{multline}
 \int_{c/\rho}^1\frac{{\rm d}{\tilde u}}{\sqrt{1-{\tilde u}}} \,\int_{0}^{\infty}{\rm d} s\,{\rm e}^{-{\rm i}\phi_0s - {\rm i} \frac{\tilde u^2}{48}k_{\perp}^2(eB)^2s^3}\,
 \left\{\begin{array}{c}
  (k^2)^2s^0\left[1+{\cal O}({\tilde u})\right]\\
 k_p^2\frac{z^{2n}}{s}\,{\tilde u}\left[1+{\cal O}({\tilde u})\right]\\
  k^2z^2k_{\perp}^2{\tilde u}\left[1+{\cal O}({\tilde u})\right]\\
 \end{array}\right\}\\
\sim\left\{\begin{array}{c}
  k^2\frac{k^2}{\phi_0(c/\rho)}\frac{1}{\rho^{2/3}}\left\{{\cal O}(1)+{\cal O}(\tfrac{1}{\rho^{2/3}})-\left(\frac{c}{\rho}\right)^{\frac{1}{3}}\left[1+{\cal O}(\tfrac{c}{\rho})+{\cal O}(\tfrac{1}{c^{2/3}})\right]\right\}\\
  k_p^2\left(\frac{eB}{\phi_0(c/\rho)}\right)^{2n}\left\{\left(\frac{1}{\rho^{2/3}}\right)^{2n}\left[{\cal O}(1)+{\cal O}(\tfrac{1}{\rho^{2/3}})\right]-\left(\frac{c}{\rho}\right)^2\left(\frac{1}{c^{2/3}}\right)^{2n}\left[1+{\cal O}(\tfrac{c}{\rho})+{\cal O}(\tfrac{1}{c^{2/3}})\right]\right\}\\
  k^2\left[1+{\cal O}(\tfrac{c}{\rho})+{\cal O}(\tfrac{1}{c^{2/3}})+{\cal O}(\tfrac{1}{\rho^{2/3}})\right]\\
 \end{array}\right\}, \label{eq:x1to0_1b}
\end{multline}
and
\begin{align}
&\int_{c/\rho}^1\frac{{\rm d}{\tilde u}}{\sqrt{1-{\tilde u}}} \, \int_{0}^{\infty}\frac{{\rm d} s}{s}\,{\rm e}^{-{\rm i}\phi_0s - {\rm i}\frac{\tilde u^2}{48} k_{\perp}^2(eB)^2s^3}k_p^2{\tilde u}\,\bigl\{-{\rm i}k_{\perp}^2s\,{\tilde u}^2\left[1+{\cal O}({\tilde u})\right]\bigr\}^{j}\!z^{2(n+2j)} \label{eq:x1to0_2b}\\
&\sim k_p^2\left(\frac{eB}{\phi_0(c/\rho)}\right)^{2(n+j)}\left\{\left(\frac{1}{\rho^{2/3}}\right)^{2(n+j)}\left[{\cal O}(1)+{\cal O}(\tfrac{1}{\rho^{2/3}})\right]-\left(\frac{c}{\rho}\right)^2\left(\frac{1}{c^{2/3}}\right)^{2(n+j)}\left[1+{\cal O}(\tfrac{c}{\rho})+{\cal O}(\tfrac{1}{c^{2/3}})\right]\right\},  \nonumber
\end{align}
with $p\in\{\parallel,\perp\}$ and $\{n,j\}\in\mathbb{N}$.
Note that the structure of the contributions in the second line on the right-hand side of \Eqref{eq:x1to0_1} and the right-hand side of \Eqref{eq:x1to0_2} is quite similar.
The same is true for the second line of \Eqref{eq:x1to0_1b} and \Eqref{eq:x1to0_2b}.
The result of an integration over the full $\nu$ (or equivalently $\tilde u$) interval is obtained by adding \Eqref{eq:x1to0_1} to \Eqref{eq:x1to0_1b} and \Eqref{eq:x1to0_2} to \Eqref{eq:x1to0_2b}.

\section{Results from the method of stationary phase} \label{sec:methstat}

References \cite{Baier:2007dw,Baier:2009it} and \cite{Dunne:2009gi} have employed the method of stationary phase to approximate the imaginary part of the photon polarization tensor for on-the-light-cone dynamics, i.e., $k^2=0$.
This approach is limited to weak fields, $ef/m^2\ll1$, with $f=\{E,B\}$. It constitutes a reliable approximation
\begin{itemize}
 \item to the regime of relatively low photon energy, characterized by $\frac{k_\perp^2}{4m^2}\sim1$, while $\frac{k_\perp^2}{4m^2}-1\gg \frac{eB}{m^2}$ or $\frac{k_\perp^2}{4m^2}\gg \frac{eE}{m^2}$, respectively,
 \item and, given that $\beta_f(r)\gg\frac{ef}{m^2}$ (see below), also at high photon energy, $\frac{k_\perp^2}{4m^2}\gg1$.
\end{itemize}
In order to keep this paper self-contained, we reproduce their results here.
In our conventions, the result of \cite{Baier:2007dw} for a magnetic field (cf. in particular their Appendix~B) reads
\begin{equation}
 \Im(\Pi_\parallel)=-\alpha eB\,\sqrt{\frac{r}{(r-1)l_B(r)\beta_B(r)}}\,{\rm e}^{-\frac{m^2}{eB}\beta_B(r)}\,, \quad\quad
 \Im(\Pi_\perp)=\frac{r-1}{2r}\,\Im(\Pi_\parallel)\,, \label{eq:methstat_B}
\end{equation}
with the definitions $r=\frac{k_\perp^2}{4m^2}$, $l_B(r)=\ln\frac{\sqrt{r}+1}{\sqrt{r}-1}$ and $\beta_B(r)=2\sqrt{r}-(r-1)l_B(r)$.
The corresponding result for electric fields was first derived by \cite{Dunne:2009gi}, and also obtained by \cite{Baier:2009it}. It reads
\begin{equation}
 \Im(\Pi_\parallel)=-\alpha eE\,\frac{1}{2}\sqrt{\frac{r+1}{r\,l_E(r)\beta_E(r)}}\,{\rm e}^{-\frac{m^2}{eE}\beta_E(r)}\,,  \quad\quad
 \Im(\Pi_\perp)=\frac{2r}{r+1}\,\Im(\Pi_\parallel)\,, \label{eq:methstat_E}
\end{equation}
with $l_E(r)=2\arctan\frac{1}{\sqrt{r}}$ and $\beta_E(r)=(r+1)l_E(r)-2\sqrt{r}$.
For completeness, note that Baier and Katkov \cite{Baier:2009it} have also determined the correction term to the very last equation in the limit $r\ll1$:
The corresponding result is obtained by a multiplication with the factor $\left(1+\frac{1}{\pi}\frac{eE}{k_\perp^2}\right)$.

\section{The photon polarization tensor in the strong electric field limit} \label{app:strongEfield}

Here we detail the derivation of the photon polarization tensor in the strong electric field limit, \Eqref{eq:PI_sfE}.
Following the derivation of the analogous expression for strong magnetic fields in Sec.~\ref{sec:strongfield},
we first provide the results for the various $\nu$ integrals encountered in the derivation [cf. \Eqref{eq:nuintegral}].
Specializing to an electric field, the coefficients $a$, $b$ and $c$ as introduced in the main text now read [cf. \Eqref{eq:abc}]
\begin{equation}
  a=m^2-{\rm i}\epsilon-{\rm i}eE(2n+l)+\frac{k_\perp^2}{4}\,, \quad b=-{\rm i}\frac{eEl}{2}\,, \quad c=-\frac{k_\perp^2}{4}\,. \label{eq:abc2}
\end{equation}
For $\{n,l\}>0$ we get
\begin{equation}
 \int_{-1}^1{\rm d}\nu\,\frac{1}{a+2b\nu+c\nu^2}
\left\{\begin{array}{c}
1\\ \nu\\ \nu^2
\end{array}\right\}
=
\frac{\rm i}{leE}
\left\{\begin{array}{c}
\ln\left(\frac{n+l}{n}\right)\\ 2-\left(1+\frac{2n}{l}\right)\ln\left(\frac{n+l}{n}\right)\\ \left(1+\frac{2n}{l}\right)\left[-2+\left(1+\frac{2n}{l}\right)\ln\left(\frac{n+l}{n}\right)\right]
\end{array}\right\}
+{\cal O}\Bigl(\frac{1}{(eE)^{2}}\Bigr), \label{eq:strongE1}
\end{equation}
while for $l=0$, $n>0$, we find
\begin{equation}
 \int_{-1}^1{\rm d}\nu\,\frac{1}{a+2b\nu+c\nu^2}
\left\{\begin{array}{c}
1\\ \nu\\ \nu^2
\end{array}\right\}
=
\frac{\rm i}{neE}\left\{\begin{array}{c}
1\\ 0\\ 1/3
\end{array}\right\}
+{\cal O}\Bigl(\frac{1}{(eE)^{2}}\Bigr).
\end{equation}
For $n=0$, $l>0$, we obtain
\begin{multline}
 \int_{-1}^1{\rm d}\nu\,\frac{1}{a+2b\nu+c\nu^2}
\left\{\begin{array}{c}
1\\ \nu\\ \nu^2
\end{array}\right\} 
=
\frac{\rm i}{leE}\left[
\left\{\begin{array}{c}
\ln\left(\frac{2leE}{m^2}\right)-{\rm i}\frac{\pi}{2}\\ 2-\ln\left(\frac{2leE}{m^2}\right)+{\rm i}\frac{\pi}{2}\\ \ln\left(\frac{2leE}{m^2}\right)-{\rm i}\frac{\pi}{2}-2
\end{array}\right\}\right.\\
\left.+\frac{\rm i}{leE}\left\{\begin{array}{c}
k_\perp^2+\frac{1}{2}m^2-\frac{1}{2}k_\perp^2\left[\ln\left(\frac{2leE}{m^2}\right)-{\rm i}\frac{\pi}{2}\right]\\ \left(\frac{1}{2}k_\perp^2-m^2\right)\left[\ln\left(\frac{2leE}{m^2}\right)-{\rm i}\frac{\pi}{2}\right]-\frac{3}{2}k_\perp^2-\frac{1}{2}m^2\\ \frac{5}{3}k_\perp^2-\frac{3}{2}m^2 + \left(2m^2-\frac{1}{2}k_\perp^2\right)\left[\ln\left(\frac{2leE}{m^2}\right)-{\rm i}\frac{\pi}{2}\right]
\end{array}\right\}\right]
+{\cal O}\Bigl(\frac{1}{(eE)^{3}}\Bigr), \label{eq:strongE3}
\end{multline}
and in particular [cf. \Eqref{eq:strongE3}]
\begin{multline}
 \int_{-1}^1{\rm d}\nu\,\frac{1}{a+2b\nu+c\nu^2}
\left\{\begin{array}{c}
1+\nu \\ 1-\nu^2
\end{array}\right\} 
=
\frac{\rm i}{leE}\left[\left\{\begin{array}{c}
2 \\ 2
\end{array}\right\} + \frac{\rm i}{leE}\left\{\begin{array}{c}
-\frac{1}{2}k_\perp^2-m^2\left[\ln\left(\frac{2leE}{m^2}\right)-{\rm i}\frac{\pi}{2}\right] \\ 2m^2-\frac{2}{3}k_\perp^2-2m^2\left[\ln\left(\frac{2leE}{m^2}\right)-{\rm i}\frac{\pi}{2}\right]
\end{array}\right\}\right]
+{\cal O}\Bigl(\frac{1}{(eE)^{3}}\Bigr). \label{eq:strongE3b}
\end{multline}
Note that the left-hand sides of Eqs.~\eqref{eq:strongE1}-\eqref{eq:strongE3b} can alternatively be obtained from Eqs.~~\eqref{eq:i_start}-\eqref{eq:1+nu/-nu^2}, relabeling $\parallel\leftrightarrow\perp$ and substituting $B\to{\rm e}^{-{\rm i}\frac{\pi}{2}}E$ [cf. \Eqref{eq:trafo}].
Utilizing the above identities in Eqs.~\eqref{eq:sf_block1_2}-\eqref{eq:sf_block3_2}, with $\parallel\leftrightarrow\perp$ and $B\to-{\rm i}E$, we obtain
\begin{equation}
 \int_{-1}^1\frac{{\rm d}\nu}{2}\int\limits_{0}^{\infty-{\rm i}\eta}\frac{{\rm d} s}{s}{\rm e}^{-{\rm i}\phi_0^\perp s}\left({\rm e}^{-{\rm i}n_2k_\parallel^2s}-1\right)N_1(-{\rm i}eEs)=-\frac{k_\parallel^2}{2}\,\int_{-1}^1\frac{{\rm d}\nu}{2}\,\frac{1-\nu^2}{\phi_0^\perp}
+ {\cal O}\Bigl(\tfrac{1}{eE}\Bigr),
\end{equation}
while for $i\in\{0,2\}$ we find
\begin{equation}
 \int_{-1}^1\frac{{\rm d}\nu}{2}\int\limits_{0}^{\infty-{\rm i}\eta}\frac{{\rm d} s}{s}{\rm e}^{-{\rm i}\phi_0^\perp s}\left({\rm e}^{-{\rm i}n_2k_\parallel^2s}-1\right)N_i(-{\rm i}eEs)={\cal O}\Bigl(\tfrac{1}{eE}\Bigr).
\end{equation}
Analogously, Eqs.~\eqref{eq:intN0_s}-\eqref{eq:intN2_s}, with $\parallel\leftrightarrow\perp$ and $B\to-{\rm i}E$, result in
\begin{align}
 \int_{-1}^1\frac{{\rm d}\nu}{2}\,\eta_0(-{\rm i}E)\
&=\left(\frac{2}{3}+{\rm i}\frac{m^2}{eE}\right)\ln\biggl(\frac{m^2}{2eE}\biggr)+\frac{2\gamma+{\rm i}\pi}{3}+\frac{2}{3}+\Sigma+{\cal O}\Bigl(\tfrac{1}{eE}\Bigr), \\
 \int_{-1}^1\frac{{\rm d}\nu}{2}\,\eta_1(-{\rm i}E)
&=-\left({\rm i}eE+\frac{k_\parallel^2}{2}\right)\int_{-1}^1\frac{{\rm d}\nu}{2}\frac{1-\nu^2}{\phi_0^\perp}+\frac{2}{3}\ln\biggl(\frac{m^2}{2eE}\biggr)+\frac{2\gamma+{\rm i}\pi}{3}+{\cal O}\Bigl(\tfrac{1}{eE}\Bigr), \\
 \int_{-1}^1\frac{{\rm d}\nu}{2}\,\eta_2(-{\rm i}E)
&=\frac{2}{3}\ln\biggl(\frac{m^2}{2eE}\biggr)+\frac{2\gamma+{\rm i}\pi}{3}+\Sigma+{\cal O}\Bigl(\tfrac{1}{eE}\Bigr).
\end{align}


\begin{thebibliography}{10}\setlength{\itemsep}{-0.5mm}

\bibitem{Toll:1952}
J.~S.~Toll,
Ph.D. thesis, Princeton Univ., 1952 (unpublished).

\bibitem{Baier}
R.~Baier and P.~Breitenlohner, 
{\it Act.~Phys.~Austriaca} {\bf 25}, 212 (1967); 
{\it Nuov.~Cim.~B}\ {\bf 47} 117 (1967).

\bibitem{BialynickaBirula:1970vy}
  Z.~Bialynicka-Birula and I.~Bialynicki-Birula,
  {\it Phys.\ Rev.\  D}\ {\bf 2}, 2341 (1970).

\bibitem{Adler:1971wn}
  S.~L.~Adler,
  {\it Annals Phys.}\  {\bf 67}, 599 (1971).

\bibitem{BatShab}
  I.~A.~Batalin and A.~E.~Shabad,
  Zh.\ Eksp.\ Teor.\ Fiz.\  {\bf 60}, 894 (1971)
  [Sov.\ Phys.\ JETP\ {\bf 33}, 483 (1971)].

\bibitem{Urrutia:1977xb}
  L.~F.~Urrutia,
  {\it Phys.\ Rev.\  D}\ {\bf 17}, 1977 (1978).

\bibitem{Dittrich:2000zu}
  W.~Dittrich and H.~Gies,
  {\it Springer Tracts Mod.\ Phys.}\  {\bf 166}, 1 (2000).

\bibitem{Schubert:2000yt}
  C.~Schubert,
  {\it Nucl.\ Phys.\  B}\ {\bf 585}, 407 (2000)
  [arXiv:hep-ph/0001288].

\bibitem{Tsai:1974ap}
  W.~y.~Tsai,
  Phys.\ Rev.\  D {\bf 10}, 2699 (1974).

\bibitem{physics/0605038} 
  W.~Heisenberg and H.~Euler,
  {\it Z.\ Phys.}\ {\bf 98}, 714  (1936); 
  an English translation is available at [physics/0605038].

\bibitem{Jentschura:2001qr}
  U.~D.~Jentschura, H.~Gies, S.~R.~Valluri, D.~R.~Lamm and E.~J.~Weniger,
  Can.\ J.\ Phys.\  {\bf 80} 267 (2002)
  [arXiv:hep-th/0107135].

\bibitem{Shabad:1975ik}
  A.~E.~Shabad,
  Annals Phys.\  {\bf 90}, 166 (1975).

\bibitem{Melrose:1976dr} 
  D.~B.~Melrose and R.~J.~Stoneham,
  Nuovo Cim.\ A {\bf 32}, 435 (1976).

\bibitem{Melrose:1977} 
  D.~B.~Melrose and R.~J.~Stoneham,
J.\ Phys.A:Math.Gen. {\bf 10}, 1211 (1977).

\bibitem{Witte:1990}
N.~S.~ Witte,
J.\ Phys.A:Math.Gen. {\bf 23}, 5257 (1990).

\bibitem{Hattori:2012je} 
  K.~Hattori and K.~Itakura,
  Annals Phys.\  {\bf 330}, 23 (2013)
  [arXiv:1209.2663 [hep-ph]].

\bibitem{Hattori:2012ny} 
  K.~Hattori and K.~Itakura,
  Annals Phys.\  {\bf 334}, 58 (2013)
  [arXiv:1212.1897 [hep-ph]].

\bibitem{Ishikawa:2013fxa} 
  K.~-I.~Ishikawa, D.~Kimura, K.~Shigaki and A.~Tsuji,
  Int.\ J.\ Mod.\ Phys.\ A {\bf 28}, 1350100 (2013)
  [arXiv:1304.3655 [hep-ph]].

\bibitem{Nikishov:1964zza} 
  A.~I.~Nikishov and V.~I.~Ritus,
  Zh.\ Eksp.\ Teor.\ Fiz.\  {\bf 46}, 776 (1964)
  [Sov.\ Phys.\ JETP {\bf 19}, 529 (1964)].
  
\bibitem{Nikishov:1964zz} 
  A.~I.~Nikishov and V.~I.~Ritus,
  Zh.\ Eksp.\ Teor.\ Fiz.\  {\bf 46}, 1768 (1964)
  [Sov.\ Phys.\ JETP {\bf 19}, 1191 (1964)].

\bibitem{narozhnyi:1968}
  N.~B.~Narozhnyi,
 Zh.\ Eksp.\ Teor.\ Fiz.\ {\bf 55}, 714 (1968)
 [Sov.\ Phys.\ JETP \textbf{28}, 371 (1969)].
  
\bibitem{ritus:1972}
 V.~I.~Ritus,
 Ann.\ Phys.\ {\bf 69}, 555 (1972).
  
\bibitem{constcrossedfields:1976}
  V.~N.~Ba\u{\i}er, A.~I.~Mil'shte\u{\i}n and V.~M.~Strakhovenko,
  Zh.\ Eksp.\ Teor.\ Fiz.\ {\bf 69}, 1893 (1975)
 [Sov.\ Phys.\ JETP {\bf 42}, 961 (1976)].

\bibitem{Heinzl:2006pn} 
  T.~Heinzl and O.~Schroeder,
  J.\ Phys.\ A {\bf 39}, 11623 (2006)
  [hep-th/0605130].
  
\bibitem{Marklund:2008gj} 
  M.~Marklund and J.~Lundin,
  Eur.\ Phys.\ J.\ D {\bf 55}, 319 (2009)
  [arXiv:0812.3087 [hep-th]].

\bibitem{Dunne:2008kc} 
  G.~V.~Dunne,
  Eur.\ Phys.\ J.\ D {\bf 55}, 327 (2009)
  [arXiv:0812.3163 [hep-th]].
  
\bibitem{DiPiazza:2011tq} 
  A.~Di Piazza, C.~Muller, K.~Z.~Hatsagortsyan and C.~H.~Keitel,
  Rev.\ Mod.\ Phys.\  {\bf 84}, 1177 (2012)
  [arXiv:1111.3886 [hep-ph]].
  
\bibitem{Tsai:1974fa}
  W.~y.~Tsai and T.~Erber,
  {\it Phys.\ Rev.\  D}\ {\bf 10}, 492 (1974).

\bibitem{Tsai:1975iz}
  W.~y.~Tsai and T.~Erber,
  {\it Phys.\ Rev.\  D}\ {\bf 12}, 1132 (1975).

\bibitem{Baier:2007dw}
  V.~N.~Baier and V.~M.~Katkov,
  Phys.\ Rev.\  D {\bf 75}, 073009 (2007)
  [arXiv:hep-ph/0701119].

\bibitem{Baier:2009it}
  V.~N.~Baier and V.~M.~Katkov,
  Phys.\ Lett.\  A {\bf 374}, 2201 (2010)
  [arXiv:0912.5250 [hep-ph]].

\bibitem{Schwinger:1951nm}
  J.~S.~Schwinger,
  Phys.\ Rev.\  {\bf 82}, 664 (1951).

\bibitem{Dittrich:1985yb}
  W.~Dittrich and M.~Reuter,
  Lect.\ Notes Phys.\  {\bf 220}, 1 (1985).

\bibitem{Gradshteyn}
I.~S.~Gradshteyn and I.~M.~Ryzhik, \textit{Table of Integrals, Series, and Products}, Fifth Edition, Academic Press, UK (1994).

\bibitem{Cover:1974ij}
  R.~A.~Cover, G.~Kalman,
  {\it Phys.\ Rev.\ Lett.}\  {\bf 33}, 1113 (1974).

\bibitem{Tsai:1975tw}
  W.~Y.~Tsai and T.~Erber,
 {\it Acta Phys.\ Austriaca}\ {\bf 45}, 245 (1976).

\bibitem{Bakshi:1976vd} 
  P.~Bakshi, G.~Kalman and R.~A.~Cover,
  Phys.\ Rev.\ D {\bf 14}, 2532 (1976).
  
\bibitem{Cover:1979zz} 
  R.~A.~Cover, G.~Kalman and P.~Bakshi,
  Phys.\ Rev.\ D {\bf 20}, 3015 (1979).

\bibitem{Dunne:2004uk}
  G.~V.~Dunne, C.~Schubert,
  [math/0406610 [math-nt]].

\bibitem{Dyson:1952tj}
  F.~J.~Dyson,
  Phys.\ Rev.\  {\bf 85}, 631 (1952).

\bibitem{Dunne:2004nc}
  G.~V.~Dunne,
  In *Shifman, M. (ed.) et al.: From fields to strings, vol. 1* 445-522.
  [hep-th/0406216].

\bibitem{Dunne:2002rq}
  G.~V.~Dunne,
  [hep-th/0207046].

\bibitem{Abram}
M. Abramowitz and I.E. Stegun (ed.),   
\textit{Handbook of Mathematical Functions With Formulas, Graphs and Mathematical Tables}.
National Bureau of Standards, Washington (1964).

\bibitem{Fradkin:1981sc} 
  E.~S.~Fradkin and D.~M.~Gitman,
  Fortsch.\ Phys.\  {\bf 29}, 381 (1981).

\bibitem{Barashev:1985zm} 
  V.~P.~Barashev, A.~E.~Shabad and S.~.M.~Shvartsman,
  Sov.\ J.\ Nucl.\ Phys.\  {\bf 43}, 617 (1986)
  [Yad.\ Fiz.\  {\bf 43}, 964 (1986)].
  
\bibitem{Gavrilov:1987}
  S.~P.~Gavrilov, D.~M.~Gitman and E.~S.~Fradkin, 
  Sov.\ J.\ Nucl.\ Phys. (USA) \textbf{46}, 107 (1987)
  [Yad.\ Fiz.\ {\bf 46}, 172 (1987)].

\bibitem{Fradkin:1991zq}
  E.~S.~Fradkin, D.~M.~Gitman and S.~M.~Shvartsman,
  \emph{Quantum Electrodynamics with Unstable Vacuum}, Springer-Verlag, Berlin (1991).

\bibitem{Gavrilov:2007hq} 
  S.~P.~Gavrilov and D.~M.~Gitman,
  Phys.\ Rev.\ D {\bf 78}, 045017 (2008)
  [arXiv:0709.1828 [hep-th]].
  
\bibitem{Dunne:2009gi} 
  G.~V.~Dunne, H.~Gies and R.~Schutzhold,
  Phys.\ Rev.\ D {\bf 80}, 111301 (2009)
  [arXiv:0908.0948 [hep-ph]].

\bibitem{shabad}
A.~E.~Shabad and V.~V.~Usov,
Astrophysics and Space Science {\bf 102}, 2 (1984).

\bibitem{Shabad:1972rg}
  A.~E.~Shabad,
  Lett.\ Nuovo Cim.\  {\bf 3}, 457 (1972).

\bibitem{Daugherty:1984tr} 
  J.~K.~Daugherty and A.~K.~Harding,
  Astrophys.\ J.\  {\bf 273}, 761 (1983).

\bibitem{Ahlers:2006iz} 
  M.~Ahlers, H.~Gies, J.~Jaeckel and A.~Ringwald,
  Phys.\ Rev.\ D {\bf 75}, 035011 (2007)
  [hep-ph/0612098].

\bibitem{Dobrich:2012sw} 
  B.~Dobrich, H.~Gies, N.~Neitz and F.~Karbstein,
  Phys.\ Rev.\ Lett.\  {\bf 109}, 131802 (2012)
  [arXiv:1203.2533 [hep-ph]].
  
\bibitem{Dobrich:2012jd} 
  B.~Dobrich, H.~Gies, N.~Neitz and F.~Karbstein,
  Phys.\ Rev.\ D {\bf 87}, 025022 (2013)
  [arXiv:1203.4986 [hep-ph]].
  
\bibitem{Karbstein:2011ja} 
  F.~Karbstein, L.~Roessler, B.~Dobrich and H.~Gies,
  Int.\ J.\ Mod.\ Phys.\ Conf.\ Ser.\  {\bf 14}, 403 (2012)
  [arXiv:1111.5984 [hep-ph]].

\bibitem{Skobelev:1975}
 V.~V.~Skobelev,
 Sov.\ Phys.\ J. {\bf 18} 1481 (1975)
 [Izv. Vyssh. Uchebn. Zaved., Fiz. {\bf 10} 142, (1975)].

\bibitem{Shabad:1976}
 A.~E.~Shabad,
 Sov.\ Phys., Lebedev\ Inst.\ Rep. \textbf{3}, 13 (1976).

\bibitem{Shabad:2003} 
  A.~E.~Shabad,
  Sov.\ Phys.\ JETP {\bf 98}, 186 (2004)
  [Zh.\ Eksp.\ Teor.\ Fiz.\  {\bf 125}, 210 (2004)].

\bibitem{Shabad:2007xu} 
  A.~E.~Shabad and V.~V.~Usov,
  Phys.\ Rev.\ Lett.\  {\bf 98}, 180403 (2007)
  [arXiv:0704.2162 [astro-ph]].
  
\bibitem{Sadooghi:2007ys} 
  N.~Sadooghi and A.~Sodeiri Jalili,
  Phys.\ Rev.\ D {\bf 76}, 065013 (2007)
  [arXiv:0705.4384 [hep-th]].
  
\bibitem{Machet:2010yg} 
  B.~Machet and M.~I.~Vysotsky,
  Phys.\ Rev.\ D {\bf 83}, 025022 (2011)
  [arXiv:1011.1762 [hep-ph]].

\bibitem{Shabad:2007zu} 
  A.~E.~Shabad and V.~V.~Usov,
  Phys.\ Rev.\ D {\bf 77}, 025001 (2008)
  [arXiv:0707.3475 [astro-ph]].
  
\bibitem{Godunov:2011aa} 
  S.~I.~Godunov, B.~Machet and M.~I.~Vysotsky,
  Phys.\ Rev.\ D {\bf 85}, 044058 (2012)
  [arXiv:1112.1891 [hep-ph]].

\bibitem{Heyl:1997hr} 
  J.~S.~Heyl and L.~Hernquist,
  J.\ Phys.\ A {\bf 30}, 6485 (1997)
  [hep-ph/9705367].
 
\bibitem{Bendersky:1933}
 L.~Bendersky,
 Acta\ Math.\ {\bf 61}, 263 (1933).

\bibitem{Gies:2013yxa} 
  H.~Gies, F.~Karbstein and N.~Seegert,
  New J.\  Phys.\  {\bf 15}, 083002 (2013)
  [arXiv:1305.2320 [hep-ph]].
  
\end{thebibliography}
\end{document}